\documentclass[journal,a4paper,onecolumn,final]{IEEEtran}
\usepackage{cite}
\usepackage[cmex10]{amsmath}
\usepackage{algorithm}
\usepackage{algorithmic}
\usepackage{amssymb,amsbsy,amsthm}
\usepackage{algorithmic}
\usepackage{bm}
\usepackage{psfrag}
\usepackage{fixmath}
\usepackage{subfigure}
\usepackage[usenames]{color}
\usepackage{pstool}

\newcommand{\norm}[1]{\left\lVert{#1}\right\rVert}

\newcommand{\eff}{\text{eff}}

\newcommand{\mse}{\mathsf{mse}}

\newcommand{\mmse}{\mathsf{mmse}}

\newcommand{\Tx}{\text{T}}
\newcommand{\Rx}{\text{R}}

\newcommand{\maximum}{\text{max}}
\newcommand{\minimum}{\text{min}}

\newcommand{\Ba}{\B{a}}
\newcommand{\Bb}{\B{b}}

\newcommand{\Bd}{\B{d}}
\newcommand{\Be}{\B{e}}

\newcommand{\Bh}{\B{h}}

\newcommand{\Bp}{\B{p}}
\newcommand{\Bq}{\B{q}}
\newcommand{\Br}{\B{r}}
\newcommand{\Bs}{\B{s}}
\newcommand{\Bt}{\B{t}}
\newcommand{\Bu}{\B{u}}
\newcommand{\Bv}{\B{v}}
\newcommand{\Bw}{\B{w}}
\newcommand{\Bx}{\B{x}}
\newcommand{\By}{\B{y}}
\newcommand{\Bz}{\B{z}}
\newcommand{\BA}{\B{A}}
\newcommand{\BB}{\B{B}}

\newcommand{\BD}{\B{D}}

\newcommand{\BF}{\B{F}}
\newcommand{\BG}{\B{G}}
\newcommand{\BH}{\B{H}}

\newcommand{\BK}{\B{K}}
\newcommand{\BL}{\B{L}}

\newcommand{\BP}{\B{P}}
\newcommand{\BQ}{\B{Q}}
\newcommand{\BR}{\B{R}}
\newcommand{\BS}{\B{S}}
\newcommand{\BT}{\B{T}}
\newcommand{\BU}{\B{U}}

\newcommand{\BW}{\B{W}}
\newcommand{\BX}{\B{X}}
\newcommand{\BY}{\B{Y}}
\newcommand{\BZ}{\B{Z}}

\newcommand{\Balpha}{\bm{\alpha}}
\newcommand{\Bbeta}{\bm{\beta}}

\newcommand{\Blambda}{\bm{\lambda}}

\newcommand{\Bxi}{\bm{\xi}}

\newcommand{\Bsigma}{\bm{\sigma}}

\newcommand{\Bupsilon}{\bm{\upsilon}}

\newcommand{\Bpsi}{\bm{\psi}}
\newcommand{\Bomega}{\bm{\omega}}

\newcommand{\BLambda}{\bm{\Lambda}}
\newcommand{\BXi}{\bm{\Xi}}
\newcommand{\BPi}{\bm{\Pi}}

\newcommand{\BPsi}{\bm{\Psi}}

\newcommand{\BUpsilon}{\bm{\Upsilon}}

\newcommand{\myid}{\mathbf{I}} %identity matrix
\newcommand{\mynull}{\mathbf{0}} %null vector or null matrix
\newcommand{\myone}{\mathbf{1}} %all-ones vector or all-ones matrix

\DeclareMathAlphabet{\mathbit}{OML}{cmr}{bx}{it}
\DeclareMathAlphabet{\mathgoth}{U}{ygoth}{m}{n}
\DeclareMathAlphabet{\mathfrak}{U}{yfrak}{m}{n}
\DeclareMathAlphabet{\mathswab}{U}{yswab}{m}{n}

\newcommand{\B}[1]{\mathbit{#1}}
\newtheorem{mythm}{Theorem}[section]

\newtheorem{mycor}{Corollary}[section]
\newtheorem{mylem}{Lemma}[section]
\newtheorem{mydef}{Definition}[section]

\DeclareMathOperator*{\argmax}{argmax}

\DeclareMathOperator{\Prob}{\mathsf{Pr}}
\DeclareMathOperator{\Exp}{\mathbb{E}}

\newcommand{\diffd}{{\operatorname{d}}}
\newcommand{\e}{{\operatorname{e}}}
\DeclareMathOperator{\diag}{diag}

\DeclareMathOperator{\trace}{tr}

\DeclareMathOperator{\interior}{int}

\DeclareMathOperator{\vectorize}{vec}
\DeclareMathOperator{\rank}{rank}

\DeclareMathOperator{\range}{range}
\DeclareMathOperator{\col}{\mathcal{C}}

\DeclareMathOperator{\Transpose}{T}
\DeclareMathOperator{\Hermitian}{\dagger}
\newcommand{\Tr}{{\Transpose}}
\newcommand{\He}{{\Hermitian}}

\newcommand{\dB}{{\operatorname{dB}}}

\newcommand{\Frob}{\text{F}}

\begin{document}

\title{A Framework for Joint Design\\of Pilot Sequence and Linear Precoder}

\author{
\IEEEauthorblockN{
Adriano~Pastore,~\IEEEmembership{Student Member,~IEEE},
Michael~Joham,~\IEEEmembership{Member,~IEEE},\\
and Javier~Rodr\'{i}guez~Fonollosa,~\IEEEmembership{Senior Member,~IEEE}
}
}

\maketitle

\begin{abstract}
%\boldmath
Most performance measures of pilot-assisted multiple-input multiple-output (MIMO) systems are functions that depend on both the linear precoding filter and the pilot sequence. A framework for the optimization of these two parameters is proposed, based on a matrix-valued generalization of the concept of effective signal-to-noise ratio (SNR) introduced in a famous work by Hassibi and Hochwald \cite{HaHo03}. The framework applies to a wide class of utility functions of said effective SNR matrix, most notably a well-known mutual information expression for Gaussian inputs, an upper bound on the minimum mean-square error (MMSE), as well as approximations thereof. The approach consists in decomposing the joint optimization problem into three subproblems: first, we describe how to reformulate the optimization of the linear precoder subject to a fixed pilot sequence as a convex problem. Second, we do likewise for the optimization of the pilot sequence subject to a fixed precoder. Third, we describe how to generate pairs of precoders and pilot sequences that are Pareto optimal in the sense that they attain the Pareto boundary of the set of feasible effective SNR matrices. By combining these three optimization problems into an iteration, we obtain an algorithm which allows to compute jointly optimal pairs of precoders and pilot sequences with respect to some generic utility function of the effective SNR.
%This Pareto optimization is studied in the case where the budgets for time, training energy and transmit power can be freely allocated between the tasks of channel estimation and data transmission, as well as in the more stringent case in which there are two separate constraints imposed on the training energy and on the transmit power, respectively.
%An iterative algorithm for concave utility functions is proposed, which performs the joint design of pilot symbols and of the linear precoder, and which achieves the joint optimum if an appropriate initialization is chosen.
\renewcommand{\thefootnote}{}
  \footnote{This work was supported by the Spanish Science and Technology Commissions and FEDER funds from the EC (TEC2010-19171/TCM and CONSOLIDER INGENIO CSD2008-00010 COMONSENS), and 2009SGR-1236 of the Catalan government.

Parts of this work were published in \cite{PaJoFo11_isit}.}
\end{abstract}
\setcounter{footnote}{0}


\IEEEpeerreviewmaketitle
\section{Introduction}

\IEEEPARstart{W}{hen} the receiver has no genie-provided knowledge of the fading gains, a common approach is to incorporate a pre-agreed pattern of {\em training} or {\em pilot} symbols into the transmitted signal. The receiver first exploits the training observation to generate an estimate of the fading gains, and then uses this channel estimate to decode the transmitted message. This two-stage approach is suboptimal compared to optimal full-blown maximum-likelihood decoding, but it drastically reduces decoding complexity while maintaining near-optimal performance results. We shall consider a narrow-band MIMO channel with time-duplexed training, that is, certain time slots are reserved exclusively for transmitting pilot symbols, while other time slots are reserved for data symbols. However, much of what is described in this article applies as well to wideband channels with frequency-duplexed training (pilot tones).

In our time-discrete Rayleigh-distributed block-fading model, the statistics of the channel gains are fully described by their second-order moments (the covariance of the fading coefficients) and by the fading-block length, also called the {\em coherence time}. Therefore, when the feedback is limited to being statistical (as in the scenario we shall consider), the pilot sequence and the precoder can only be designed based on these two statistical channel parameters.

%This optimization effort is justified by the fact that, in the long run, said statistical channel properties may vary, prompting for a readjustment of both the precoder and the pilot sequence in order to maintain optimal system operation.
A frequent yet suboptimal choice in the literature is that of generic orthonormal pilot symbols. Besides, many publications focus on distortion measures like the mean-square error when designing the pilot sequence (e.g., \cite{ToSaDo04}, \cite{BiGe06}), while focusing on other measures such as bit-error rate or mutual information when designing the precoder. In light of this situation, it is of both practical and theoretical interest to examine what performance gains can potentially be achieved by {\em jointly} designing the pilot sequence and the precoder, based on statistical channel knowledge and with respect to a single system performance metric.

In the present article, the metric of choice will be a well-known expression for the Gaussian-input mutual information between the channel input and the ouput of a mismatched decoder, which takes the channel estimate as if it were the true channel gain, and seeks to minimize the expected Euclidian distance between the received signal and the expected output that would have been produced by the candidate codeword. This so-called nearest-neighbor decoder was studied in \cite{LaSh02} in the single-antenna setting and in \cite{WeStSh04} for the general multi-antenna case. Predating these publications, the mutual information achieved by this decoding scheme was also well-known (in a less general interpretation) as a lower bound on the mutual information between the channel input and the raw receiver observation, in systems with imperfect channel-state information at the receiver. In this weaker formulation, it was originally proposed by M\'edard \cite{Me00} and later generalized to MIMO training-based systems by Hassibi and Hochwald \cite{HaHo03}. In numerous variations and different settings, the essence of the bounding technique proposed in \cite{Me00} has been extensively used in subsequent works on transmission with imperfect channel-state information (e.g. in \cite{BaFoMe01,YoYoGo04,MuDoNaAg05,Lo08,SoUl10a,DiBl10}, to cite only a few), most often as a performance metric for system design.

In \cite{HaHo03}, the problem was considered of finding the optimal time share between training and transmission, as well as optimally balancing the training and transmit power levels. Later works have followed a similar approach: in \cite{YoYoGo04}, the optimal transmit covariance was shown to be diagonal, and its eigenvalues would turn out to be solutions to a convex problem. However, in both \cite{HaHo03} and \cite{YoYoGo04}, all results were derived exclusively for {\em uncorrelated} fading. When facing the more difficult---yet more realistic---situation of {\em correlated} fading, the question of joint optimality of pilot sequence and precoder is much more involved. For example, one can intuit that the number of pilot symbols and the number of data streams will depend, among other things, on the conditioning of the channel's correlation structure. The authors of \cite{SoUl10a} went about this problem by designing the pilot sequence so as to minimize the variance of the channel estimation error by a waterfilling-type algorithm. But evidently, this approach is merely heuristic. The present work proposes a framework to tackle this problem {\em optimally}.

We consider a single-user multiple-input multiple-output (MIMO) link and assume a highly scattering environment at the receiver---as is the case in many downlink scenarios---so that the fading is correlated only at the transmitter side. This encompasses the important special case of fully correlated multiple-input single-output (MISO) links. Additionally, the main results of the present work can be generalized straightforwardly to MIMO multiple-access channels with transmit-side correlation, when viewing the multiple-access channel as a large MIMO channel with additional block-diagonality constraints on the channel correlation, the transmit covariance and the pilot sequence.

In \cite{HaHo03}, the concept of {\em effective signal-to-noise ratio} (SNR) was introduced to designate an SNR that accounts for the imperfection of channel state information (CSI) at the receiver. The mutual information for Gaussian inputs (which in \cite{HaHo03} is interpreted as a capacity lower bound) is an increasing function of this effective SNR, which thus serves as the figure of merit to be maximized. In the present work, we follow a similar line of thought, however, we treat the more general case of correlated fading, for which the definition of the effective SNR needs to be extended from a scalar to a matrix-valued quantity. Hence, the concept of effective SNR maximization needs to be extended accordingly to a Pareto optimization. Among all Pareto optimal solutions, the optimum will be determined by the specific choice of the utility function.

A procedure is proposed by which the non-convex joint pilot-precoder optimization problem is decomposed into three subproblems, each of which can be cast into a convex optimization problem. An iteration cycles through these three optimization steps to compute the joint optimum: the first step consists in optimizing the precoder while keeping the pilot sequence fixed, the second step consists in optimizing the pilot sequence while keeping the precoder fixed, and the third step adjusts the pilot-precoder pair so as to be Pareto optimal in terms of the matrix-valued effective SNR.

One main result in the analysis of the joint optimization problem is that the left singular vectors of the precoder and of the pilot sequence matrix must be eigenvectors of the channel covariance matrix. Loosely speaking, this means that the training symbols and the multiple beamforming vectors should be aligned in direction of the channel eigenmodes.

The article is structured as follows: Section~\ref{sec:notation} defines notation; Section~\ref{sec:system_model} describes the system model; Section~\ref{sec:utility_functions} defines and motivates the class of utility functions considered in the optimization framework; Section~\ref{sec:problem_statement} states the optimization problem to consider; Sections~\ref{sec:Precoder_Design_for_Prescribed_Pilots} and \ref{sec:Pilot_Design_for_a_Prescribed_Precoder} describe how the optimization of the precoder (resp. pilot sequence) subject to a fixed pilot sequence (resp. precoder) is cast into a convex problem; Section~\ref{sec:Joint_Pilot_and_Precoder_Optimization} specifies the jointly optimal training and transmit directions and shows how the residual problem of computing pilot-precoder power loading vectors that are Pareto optimal in terms of the effective SNR, can be formulated as a quasi-convex problem; Section~\ref{sec:iterative_joint_design} assembles the findings from Sections~\ref{sec:Precoder_Design_for_Prescribed_Pilots}, \ref{sec:Pilot_Design_for_a_Prescribed_Precoder} and \ref{sec:Joint_Pilot_and_Precoder_Optimization} into an iterative algorithm that achieves the jointly optimal pilot-precoder design.
%Section~\ref{sec:low_snr_analysis} analyzes what benefits are attained at low SNR by joint optimization as compared to partial optimization or no optimization.

\section{Notation}   \label{sec:notation}

The operators $(\bullet)^\Tr$, $(\bullet)^*$ and $(\bullet)^\He$ denote the transpose, the complex conjugate, and the conjugate transpose (Hermitian adjoint) of a matrix, respectively. Matrix square roots are denoted as $(\bullet)^\frac{1}{2}$ and are assumed to be Hermitian. The Moore-Penrose pseudoinverse $\BA^+$ of a matrix $\BA$ is uniquely defined by the four identities
\begin{align*}
	\BA\BA^+\BA &= \BA	&	\BA^+\BA\BA^+ &= \BA^+ \nonumber\\
	(\BA\BA^+)^\He &= \BA\BA^+	&	(\BA^+\BA)^\He &= \BA^+\BA.
\end{align*}
The range of a matrix $\BA$, denoted as $\range(\BA)$, shall be the linear space spanned by its columns. The set of columns of a matrix $\BA$ is denoted as $\col(\BA)$.

The trace and determinant of a square matrix are written as $\trace(\bullet)$ and $\det(\bullet)$, respectively.

We will occasionally use the entrywise comparison $\Ba \leq \Bb$ between two real-valued vectors $\Ba$ and $\Bb$, of $i$-th entries $a_i$ and $b_i$, defined as $\Ba \leq \Bb \Leftrightarrow \forall i \colon a_i \leq b_i$. If $\BA$ and $\BB$ denote Hermitian matrices, $\BA \preceq \BB$ (resp. $\BA \prec \BB$) means that $\BA-\BB$ is positive semidefinite (resp. positive definite).

For a set of real vectors $\mathcal{X} \subset \mathbb{R}^n$, the so-called Pareto border $\partial^+\mathcal{X} \subseteq \mathcal{X}$ contains those points from $\mathcal{X}$ which are not dominated by any other point from $\mathcal{X}$, in the sense that for any point $\Bx^+ \in \partial^+\mathcal{X}$, there is no $\Bx \in \mathcal{X}$ distinct from $\Bx^+$ such that $\Bx \geq \Bx^+$.

The expectation of a random variable is denoted by $\Exp[\bullet]$. If the distribution of a complex random vector $\Bx$ is proper Gaussian, we write $\Bx \sim \mathcal{N}_\mathbb{C}(\bar{\Bx},\BR_\Bx)$, where $\bar{\Bx} = \Exp[\Bx]$ and $\BR_\Bx = \Exp[(\Bx-\bar{\Bx})(\Bx-\bar{\Bx})^\He]$ stand for the mean and the covariance of $\Bx$, respectively.

We denote by $\mathbb{U}^{m \times n} \subset \mathbb{C}^{m \times n}$ the set of (sub-)unitary complex matrices defined by
\begin{itemize}
	\item $\BU\BU^\He = \myid$ if $m \leq n$
	\item $\BU^\He\BU = \myid$ if $m \geq n$.
\end{itemize}
We denote by $\mathbb{P}^n = \mathbb{U}^{n \times n} \cap \{0,1\}^{n \times n}$ the symmetric group of permutation matrices.

Unless stated otherwise, $\BU_\BA$ denotes the reduced left singular basis of a matrix $\BA$. For $\BA$ Hermitian, $\BU_\BA$ is thus the reduced eigenbasis, with the number of columns equal to the rank of $\BA$.

The cone of positive definite (resp. positive semidefinite) matrices from $\mathbb{C}^{n \times n}$ is denoted $\mathbb{C}_{++}^{n \times n}$ (resp. $\mathbb{C}_+^{n \times n}$), and is formally defined as:
\begin{align*}
	\mathbb{C}_+^{n \times n}
	&= \bigl\{ \BA \in \mathbb{C}^{n \times n} \, \big| \, \forall \Bx \in \mathbb{C}^n \colon \Bx^\He\BA\Bx \geq 0 \bigr\} \\
	\mathbb{C}_{++}^{n \times n}
	&= \bigl\{ \BA \in \mathbb{C}^{n \times n} \, \big| \, \forall \Bx \in \mathbb{C}^n \colon \Bx^\He\BA\Bx > 0 \bigr\}.
\end{align*}

Subsets of the non-negative orthant whose elements sum up to a value not larger than $a$, will be denoted $\mathcal{D}(a)$:
\begin{align*}
	\mathcal{D}(a) = \bigl\{ \Bx \in \mathbb{R}_+^n \; \big| \; \myone^\Tr\Bx \leq a \bigr\}.
\end{align*}
The dimension of $\mathcal{D}(a)$ ($n$ in the above case) will be clear from the context. Usually, the dimension will be equal to the system's number of transmit antennas.

The vectorization operator $\vectorize(\BA)$ takes a matrix $\BA = [\Ba_1, \Ba_2, \dotsc ]$ as argument, and returns a vector $\vectorize(\BA) = [\Ba_1^\Tr, \Ba_2^\Tr , \dotsc]^\Tr$ containing the columns of $\BA$ stacked on top of each other.

\section{System Model}   \label{sec:system_model}

Our system consists of a standard single-user MIMO link with an $N_\Rx \times N_\Tx$ channel matrix $\BH$ expressible as
\begin{align}   \label{H}
	\BH = \BW\BR^{\frac{1}{2}},
\end{align}
where the entries of $\BW \in \mathbb{C}^{N_\Rx \times N_\Tx}$ are independent and identically distributed (i.i.d.) zero-mean circularly-symmetric unit-variance complex Gaussian, i.e., $\vectorize(\BW) \sim \mathcal{N}_\mathbb{C}(\mynull,\myid)$. The matrix $\BW$ is the white random component of the channel matrix, whereas $\BR = \frac{1}{N_\Rx}\Exp[\BH^\He\BH]$ is the deterministic component and represents the transmit-side correlation. The latter is assumed as full-rank, since we ignore keyhole effects. This correlation model is valid in setups where numerous scatterers are located in the vicinity of the transmitter, and notably subsumes the case of correlated multiple-input single-output (MISO) channels, which are particularly relevant in wireless downlinks.

The channel remains constant for a duration $T$ called the channel coherence time, after which it changes to a new realization that is independent of all previous ones (block-fading). Within every such fading block, we reserve $T_\tau$ time slots to transmit a sequence of pilot symbols known at the receiver, while the data is transmitted during the remaining $T-T_\tau$ time slots. Without loss of generality, we can accommodate the pilot symbols into the first $T_\tau$ time slots of each fading block. During data transmission phases, the received signal at time instant $k$ is
\begin{align}   \label{system_equation}
	\By^{(k)} = \BH\BF\Bx^{(k)} + \Bz^{(k)},
	\qquad k = T_\tau+1, \dotsc, T
\end{align}
where $\Bx^{(k)} \sim \mathcal{N}_\mathbb{C}(\mynull,\myid_r)$ is a $r \times 1$ vector containing Gaussian inputs multiplexed into $r$ independent substreams, $\BF \in \mathbb{C}^{N_\Tx \times r}$ is the linear precoder, and $\Bz^{(k)} \sim \mathcal{N}_\mathbb{C}(\mynull,\myid)$ is a normalized independent additive Gaussian noise term. The Gram matrix $\BQ = \BF\BF^\He$ represents the covariance of the transmit signal $\BF\Bx^{(k)}$, and is thus called the {\em transmit covariance}. We assume that the $\Bx^{(k)}$ and $\Bz^{(k)}$ are i.i.d. across the time index $k$. During training phases, a sequence $\BT = \left[ \Bt^{(1)}, \Bt^{(2)}, \dotsc \right] \in \mathbb{C}^{N_\Tx \times T_\tau}$ of $T_\tau$ pilot symbols is sent. At time instant $k$, the receiver observes
\begin{equation}   \label{training_equation}
	\By^{(k)} = \BH\Bt^{(k)} + \Bz^{(k)},
	\qquad k = 1, \dotsc, T_\tau.
\end{equation}
The noisy training observations $\By^{(k)}$ are stored in a matrix $\BY_\tau = \left[ \By^{(1)}, \By^{(2)}, \dotsc \right]$. It can be shown that the MMSE channel estimate $\hat{\BH}$ is obtained by right-multiplying $\BY_\tau$ with the estimator matrix $\BG = (\BT^\He\BR\BT+\myid)^{-1}\BT^\He\BR$:
\begin{align}
	\hat{\BH} = \BY_\tau\BG.
\end{align}
As a consequence of the correlation models for $\BH$ and $\Bz$, the respective marginal distributions of the estimate $\hat{\BH}$ and of the estimation error $\tilde{\BH} = \BH - \hat{\BH}$ turn out to be $\hat{\Bh} = \vectorize(\hat{\BH}) \sim \mathcal{N}_\mathbb{C}(\mynull,\hat{\BR}^\Tr \otimes \myid)$ and $\tilde{\Bh} = \vectorize(\tilde{\BH}) \sim \mathcal{N}_\mathbb{C}(\mynull,\tilde{\BR}^\Tr \otimes \myid)$ with transmit-side covariances
\begin{subequations}   \label{R_definitions}
\begin{align}
	\hat{\BR} &= \tfrac{1}{N_\Rx} \Exp[\hat{\BH}^\He\hat{\BH}] = \BR - \tilde{\BR}   \label{R_hat} \\
	\tilde{\BR} &= \tfrac{1}{N_\Rx} \Exp[\tilde{\BH}^\He\tilde{\BH}] = (\BR^{-1} + \BP)^{-1},   \label{R_tilde}
\end{align}
\end{subequations}
where $\BP = \BT\BT^\He$ denotes the Gram matrix of the pilot sequence matrix $\BT$, and shall from now on be called the {\em pilot Gram}.
Note that we can write
\begin{align}   \label{H_hat_and_H_tilde}
	\hat{\BH} &= \hat{\BW} \hat{\BR}^{\frac{1}{2}}
	&
	\tilde{\BH} &= \tilde{\BW} \tilde{\BR}^{\frac{1}{2}}
\end{align}
with $\vectorize(\hat{\BW}) \sim \mathcal{N}_\mathbb{C}(\mynull,\myid)$ and $\vectorize(\tilde{\BW}) \sim \mathcal{N}_\mathbb{C}(\mynull,\myid)$.

The error covariance $\tilde{\BR}$ is non-singular by construction, whereas for $\hat{\BR}$, the following always holds:
\begin{align}   \label{rank_equality}
	\rank(\hat{\BR}) = \rank(\BP).
\end{align}
This rank equality is easily seen by application of the matrix inversion lemma: denoting by $\BP = \BU_\BP \BLambda_\BP \BU_\BP^\He$ the reduced eigendecomposition of $\BP$, where $\BLambda_\BP$ is diagonal full-rank of dimension $\rank(\BP) \times \rank(\BP)$, and $\BU_\BP \in \mathbb{C}^{N_\Tx \times \rank(\BP)}$ has orthonormal columns, we have
\begin{align}   \label{mil}
	\hat{\BR}
	&= \BR - (\BR^{-1} + \BU_\BP\BLambda_\BP\BU_\BP^\He)^{-1} \nonumber\\
	&= \BR\BU_\BP(\BLambda_\BP^{-1} + \BU_\BP^\He\BR\BU_\BP)^{-1}\BU_\BP^\He\BR.
\end{align}
Since $\BR\BU_\BP$ has full column rank, it becomes manifest that we always have
\begin{align}
	\rank(\hat{\BR})
	= \rank\bigl((\BLambda_\BP^{-1} + \BU_\BP^\He\BR\BU_\BP)^{-1}\bigr)
	= \rank(\BP).
\end{align}

\section{Utility Functions}   \label{sec:utility_functions}

\subsection{Matrix-valued effective SNR}

Omitting the time index $k$ for notational concision, we rewrite the system equation \eqref{system_equation} as
\begin{align}   \label{system_equation_2}
	\By = \hat{\BH}\BF\Bx + \underbrace{\tilde{\BH}\BF\Bx + \Bz}_{\displaystyle \Bz_\eff}
\end{align}
The first term $\hat{\BH}\BF\Bx$ represents the useful signal portion of the observation $\By$, while the remaining term $\Bz_\eff = \tilde{\BH}\BF\Bx + \Bz$ is a non-Gaussian noise term, uncorrelated with (but not independent of) the input $\Bx$. This noise term is commonly called {\em effective noise} \cite{HaHo03}. By treating the effective noise as if it were independent of the input, we incur into a suboptimality in estimating/decoding $\Bx$. Note that the effective noise has a covariance
\begin{align}
	\Exp[\Bz_\eff\Bz_\eff^\He]
	= (1+\trace(\BF^\He\tilde{\BR}\BF)) \myid_{N_\Rx}
	\triangleq \sigma_\eff^2 \myid_{N_\Rx}.
\end{align}
We whiten the random channel and normalize the transmit signal and effective noise by scaling and rewriting \eqref{system_equation_2} as
\begin{align}
	\sqrt{\frac{1}{\sigma_\eff^2}} \By
	= \hat{\BW}\BK\Bx + \bar{\Bz}_\eff,
\end{align}
where $\bar{\Bz}_\eff$ is defined as $\bar{\Bz}_\eff = \Bz_\eff \sqrt{1/\sigma_\eff^2}$ and the matrix $\BK$ is defined as
\begin{align}
	\BK
	= \frac{\hat{\BR}^{\frac{1}{2}}\BF}{\sqrt{1+\trace(\BF^\He\tilde{\BR}\BF)}}.
\end{align}
If the receiver attempts to generate an estimate $\hat{\Bx}$ of the transmit symbols $\Bx$, he may do so by minimizing the mean-square error (MSE) conditioned on the receiver side information $\BY_\tau$ and on the observation $\BY$. This MSE is the trace of the MSE matrix
\begin{align}
	\mse[g]
	= \Exp\bigl[(\hat{\Bx}-\Bx)(\hat{\Bx}-\Bx)^\He \big| \BY_\tau, \BY \bigr],
\end{align}
wherein the estimate $\hat{\Bx} = g(\BY_\tau,\BY)$ is some deterministic function of the side information $\BY_\tau$ and observation $\BY$. 
It is well known that this MSE functional takes its minimum (the minimum MSE, in short MMSE) when $g$ is the conditional mean estimator (CME), i.e.,
\begin{align}
	\hat{\Bx}
	&= g_\text{CME}(\BY_\tau,\BY)
	\triangleq \Exp\bigl[\Bx\big|\BY_\tau,\BY\bigr].
\end{align}
But this estimate being difficult to compute exactly in our channel model, we content ourselves with the (suboptimal) linear minimum mean-square estimate (LMMSE)
\begin{align}   \label{LMMSE_symbol_estimate}
	\hat{\Bx}
	= g_\text{LMMSE}(\BY_\tau,\BY)
	= \BG_\text{LMMSE}(\BY_\tau) \By
\end{align}
wherein the linear estimator $\BG_\text{LMMSE}(\BY_\tau)$ reads as
\begin{align}
	\BG_\text{LMMSE}(\BY_\tau)
	&= \Exp\bigl[ \Bx\By^\He \big| \BY_\tau \bigr] \Exp\bigl[ \By\By^\He \big| \BY_\tau \bigr]^{-1} \nonumber\\
	&= \BK^\He\hat{\BW}^\He \bigl(\myid + \hat{\BW}\BS\hat{\BW}^\He\bigr)^{-1},
\end{align}
where the matrix $\BS$, which is the Gram matrix of $\BK$, i.e.,
\begin{align}   \label{S}
	\BS = \BK\BK^\He = \frac{\hat{\BR}^{\frac{1}{2}}\BQ\hat{\BR}^{\frac{1}{2}}}{1+\trace(\BQ\tilde{\BR})},
\end{align}
represents the matrix-valued {\em effective SNR} \cite{HaHo03}.
The above LMMSE estimate \eqref{LMMSE_symbol_estimate} is suboptimal in the sense that it yields an MSE that is larger than the actual MMSE, i.e., $\mse[g_\text{LMMSE}] \succeq \mse[g_\text{CME}] = \mmse$, and reads as (cf. Utility 9 in Table~\ref{table:utilities}, Appendix~\ref{app:examples_of_utilities})
\begin{align}   \label{LMMSE_matrix}
	\mse[g_\text{LMMSE}]
	&= \myid_r - \BK^\He\hat{\BW}^\He \bigl(\myid+\hat{\BW}\BS\hat{\BW}^\He\bigr)^{-1} \hat{\BW}\BK.
\end{align}
The average scalar mean-square error achieved with said LMMSE symbol estimator is thus
\begin{align}   \label{MMSE_upper}
	\Exp\bigl[ \lVert \Bx - \hat{\Bx} \rVert_2^2 \bigr]
	&= \Exp\trace(\mse[g_\text{LMMSE}]) \nonumber\\
	&= r - \trace\Exp\bigl[ (\myid + \hat{\BW}\BS\hat{\BW}^\He)^{-1} \hat{\BW}\BS\hat{\BW}^\He \bigr] \nonumber\\
	&= r - N_\Rx + \trace\Exp\bigl[ (\myid + \hat{\BW}\BS\hat{\BW}^\He)^{-1} \bigr].
\end{align}
This upper bound on the average data symbol MMSE constitutes a basic performance metric of the considered MIMO system.

Another important figure of merit, besides this MMSE bound, is the input-output mutual information of the channel. Denoting the differential entropy as $h(\bullet)$ and following the same lines as in the derivation found in \cite{VoSc06}, the input-output mutual information $I(\Bx;\By|\BY_\tau)$ can be lower bounded as
\begin{align}   \label{I_lower_derivation}
	I(\Bx;\By|\BY_\tau)
	&\geq I(\Bx;\BG_\text{LMMSE}(\BY_\tau)\By|\BY_\tau) \nonumber\\
	&= h(\Bx) - h(\Bx|\BG_\text{LMMSE}(\BY_\tau)\By,\BY_\tau) \nonumber\\
	&= h(\Bx) - h(\Bx - \BG_\text{LMMSE}(\BY_\tau)\By|\BY_\tau) \nonumber\\
	&\geq h(\Bx) - \Exp\log\det(\pi\e\,\mse[g_\text{LMMSE}]).
\end{align}
Here, the first inequality is the data processing inequality, while the second inequality comes from upper-bounding the entropy $h(\Bx - \BG_\text{LMMSE}(\BY_\tau)\By|\BY_\tau)$ by the entropy of a Gaussian variable of same covariance. By inserting \eqref{LMMSE_matrix} into \eqref{I_lower_derivation}, this mutual information lower bound reads as
\begin{align}
	\Exp\log\det(\myid + \hat{\BW}\BS\hat{\BW}^\He)
	\leq I(\Bx;\By|\hat{\BH}).
\end{align}
This bound has been widely used and studied in the literature, e.g., \cite{Me00}, \cite{HaHo03}, \cite{YoGo06}, \cite{PaJo09_asilomar}. It was generalized in \cite{LaSh02} (for the single-antenna case) and \cite{WeStSh04} (for the multiple-antenna case), where the authors showed that this lower bound is in fact the mutual information between the input signal and the output of a nearest-neighbor decoder based on the Euclidian distance metric in the received signal space. Henceforth, we shall denote by $I$ this mutual information, or by $I(\BS)$ whenever we interpret it as a function of the effective SNR:
\begin{align}   \label{I_lower}
	I(\BS)
	&= \Exp\log\det\bigl(\myid + \hat{\BW}\BS\hat{\BW}^\He\bigr).
\end{align}
This mutual information will be the main figure of merit that we seek to maximize.

Notice that both figures of merit presented above, namely the MMSE bound \eqref{MMSE_upper} and the mutual information \eqref{I_lower}, depend only on $\BS$, which in turn depends on the linear precoder $\BF$ and on the training sequence $\BT$ via their Gram forms alone, that is, the transmit covariance $\BQ = \BF\BF^\He$ and the pilot Gram $\BP = \BT\BT^\He$.
Thus, we may occasionally write $\BS = \BS(\BP,\BQ)$ to emphasize this dependency.
The matrix $\BS$ plays a central role in all subsequent considerations, since it concentrates all system parameters (the channel covariance $\BR$, the pilot Gram $\BP$ and the transmit covariance $\BQ$) into a single matrix.

Said matrix $\BS$ constitutes a matrix-valued generalization of the scalar {\em effective SNR} introduced by Hassibi and Hochwald in \cite{HaHo03}. Much in the same way as the authors do in \cite{HaHo03}, we will seek to maximize this effective SNR (in a Pareto sense, to be specified later). Evidently, $\BS$ increases in the sense of matrix monotonicity when scaling up the pilot energy \eqref{pilot_scaling} or the transmit power \eqref{precoder_scaling}:
\begin{subequations}   \label{power_monotonicity}
\begin{align}
	0 \leq k < k'
	&\Rightarrow \BS(k\BP,\BQ) \prec \BS(k'\BP,\BQ)   \label{pilot_scaling} \\
	0 \leq k < k'
	&\Rightarrow \BS(\BP,k\BQ) \prec \BS(\BP,k'\BQ).   \label{precoder_scaling}
\end{align}
\end{subequations}

\begin{IEEEproof}
See Appendix~\ref{app:proof:power_monotonicity}.
\end{IEEEproof}

The monotonicity in $\BP$ even holds in the stronger sense
\begin{align}
	\mynull \preceq \BP \prec \BP'
	&\Rightarrow \BS(\BP,\BQ) \prec \BS(\BP',\BQ),
\end{align}
yet this is not true for the monotonicity in $\BQ$.

\subsection{General utility functions}

For a matrix $\BX$, let $\Blambda(\BX)$ denote the vector of non-increasingly ordered eigenvalues of $\BX$. We call $\Blambda(\BX)$ the {\em (eigenvalue) profile} of $\BX$. Since $\hat{\BW}$ has i.i.d. circularly-symmetric complex Gaussian entries, it is invariant against unitary rotations, i.e., $\hat{\BW}$ and $\hat{\BW}\BU$ have the same marginal distribution for any unitary $\BU$. Therefore, the function $I$ is invariant against unitary transformations:
\begin{align}
	I(\BS)
	= I(\BU^\He\BS\BU).
\end{align}
It is thus a symmetric function of the profile of $\BS$, which we shall denote as $\Bs = \Blambda(\BS)$. Henceforth, we may write $I(\Bs)$ or $I(\BS)$ without distinction.

There exists a number of other physically meaningful examples of utilities besides the mutual information $I(\Bs)$ that are functions of the profile $\Bs$. Such functions constitute a class $\mathcal{F}$ of utilities (formally defined below) and can result from different design goals, optimization criteria, asymptotic or heuristic approximations of utilities, etc. Any utility function $F \in \mathcal{F}$ shares two essential properties with $I$, namely, that it should be matrix-monotonic and invariant against unitary transformations, as put forth in the formal definition below.

\begin{mydef}   \label{def:class_F_matrix}
A function $F \colon \mathbb{C}_+^{n \times n} \to \mathbb{R}$ belongs to the class $\mathcal{F}$ if it is matrix-monotonic and invariant against unitary transformations, i.e., if the following two conditions are met:
\begin{align}   \label{f_properties_matrix}
\begin{cases}
	\mynull \preceq \BS \preceq \BS' \Rightarrow F(\BS) \leq F(\BS'), \\
	\forall \BU \in \mathbb{U}^{n \times n} \colon F(\BU\BS\BU^\He) = F(\BS).
\end{cases}
\end{align}
\end{mydef}

Note that the invariance against unitary transformations implies that functions from the class $\mathcal{F}$ are actually functions of the set of eigenvalues of $\BS$ alone, the eigenbasis of $\BS$ being irrelevant. Therefore, instead of defining the class $\mathcal{F}$ based on matrix-to-scalar functions, we can equivalently define the class $\mathcal{F}$ based on vector-to-scalar functions.

\begin{mydef}   \label{def:class_F_vector}
A function $f \colon \mathbb{R}_+^n \to \mathbb{R}$ belongs to the class $\mathcal{F}$ if it is vector-monotonic and symmetric (permutation-invariant), i.e., if
\begin{align}   \label{f_properties_vector}
\begin{cases}
	\mynull \leq \Bs \leq \Bs' \Rightarrow f(\Bs) \leq f(\Bs'), \\
	\forall \BPi \in \mathbb{P}^n \colon f(\BPi\Bs) = f(\Bs).
\end{cases}
\end{align}
\end{mydef}

Both definitions \ref{def:class_F_matrix} and \ref{def:class_F_vector} provably characterize the same set of functions. Though the two definitions apply to different types of functions (matrix-to-scalar vs. vector-to-scalar), we use the same letter $\mathcal{F}$ to denote both sets. Which of the two is meant will always be clear from the context.

Besides $I(\BS)$, two other simple examples of utility functions from the class $\mathcal{F}$ are
\begin{align}
	\det(\BS)
	&= \prod_{i=1} s_i
	&\trace(\BS)
	&= \sum_{i=1} s_i,
\end{align}
where the $s_i$ denote the entries of $\Bs$. More examples are given in Table~\ref{table:utilities} in Appendix~\ref{app:examples_of_utilities}. Below the table are included some brief explanations that motivate the use of most of the utilities listed.

\section{Problem Statement}   \label{sec:problem_statement}

For a fixed coherence time $T$ and training duration $T_\tau$, let us define the compact set of admissible values of the pilot-precoder pair $(\BP,\BQ)$ as
\begin{align}   \label{admissible_set}   %ADJUST
	\mathcal{PQ}
	= \Bigl\{ (\BP,\BQ) \in \mathbb{C}_+^{N_\Tx \times N_\Tx} \times \mathbb{C}_+^{N_\Tx \times N_\Tx} \Big| \trace(\BP) + (T-T_\tau)\trace(\BQ) \leq T\mu \Bigr\}.
\end{align}
Here, the pilot energy $\trace(\BP)$ and the transmit power $\trace(\BQ)$ are related via the energy conservation equation
\begin{align}   \label{energy_conservation}
	\trace(\BP) + (T-T_\tau) \trace(\BQ) \leq T\mu,
\end{align}
where the scalar $\mu$ stands for the maximum average energy consumption per time unit of the system.

If the training duration $T_\tau$ is also subject to optimization, the full-fledged problem of joint pilot and precoder optimization reads in its most general formulation as
\begin{align}   \label{joint_problem_full_fledged}
	\max_{T_\tau \in \{1,\dotsc,T-1\}} \max_{(\BP,\BQ) \in \mathcal{PQ}} \frac{T-T_\tau}{T} f(\Bs(\BP,\BQ)).
\end{align}
where the output value of the function $f$ represents a utility per {\em data} (non-training) channel use, and the factor $\frac{T-T_\tau}{T}$ accounts for the loss due to the time invested in channel estimation. Accordingly, the quantity $\frac{T-T_\tau}{T} f(\Bs(\BP,\BQ))$ represents the average utility per channel use.

In \cite{HaHo03}, the authors postulate for a similar setup that the receiver should have a representative estimate of the complete channel state, described by $N_\Tx N_\Rx$ fading coefficients. Therefore, they assume that the training duration $T_\tau$ should be at least the number of transmit antennas $N_\Tx$, so as to generate at least as many observables as there are coefficients to estimate. However, in the case where only a limited number of data streams are to be precoded, it might be more economic to only estimate a properly chosen subspace of the channel covariance spanned by the stronger eigenmodes. In fact, since $T_\tau$ is defined as the number of columns of the pilot matrix $\BT$, and given that all utility functions and constraints depend on $\BT$ only via its Gram matrix $\BP = \BT\BT^\He$, we can set the training duration equal to the rank of $\BP$, i.e., $T_\tau = \rank(\BP) \leq N_\Tx$, and accordingly reduce the search interval in \eqref{joint_problem_full_fledged} from $\{1,\dotsc,T-1\}$ down to $\{1,\dotsc,\min(T-1,N_\Tx)\}$. In any case, the optimization over $T_\tau$ is over a finite set and can be solved by an exhaustive search. Therefore, we will leave this problem aside until Section~\ref{sec:iterative_joint_design}, and focus in the meantime on the inner problem:
\begin{align}   \label{joint_problem_generic}
	\max_{(\BP,\BQ) \in \mathcal{PQ}} f(\Bs(\BP,\BQ)).
\end{align}
In the next two sections, based on Problem~\eqref{joint_problem_generic}, we will treat the partial problems that consist in optimizing one among the two variables $\BP$ and $\BQ$, while the other variable has a constant value. These individual optimizations will be two components of an algorithmic approach that aims to solve the joint problem \eqref{joint_problem_generic}. However, they may also be considered as two stand-alone problems in their own right.

\section{Precoder Design for Prescribed Pilots}   \label{sec:Precoder_Design_for_Prescribed_Pilots}

In this section, we consider the optimization of the transmit covariance $\BQ$ alone, while the pilot Gram $\BP$ has a fixed value. In a first approach, we will keep the matrix notation $F(\BS)$ instead of the equivalent vector notation $f(\Bs)$, as we will first investigate the problems in the matrix domain. The problem at hand reads as
\begin{align}   \label{marginal_problem_Q}
	\BQ^\star(\BP)
	&= \argmax_{\BQ \in \mathcal{Q}} F(\BS(\BP,\BQ))
%	&= \argmax_{\BS \in \BS(\BP,\mathcal{Q})} F(\BS)
\end{align}
where the search set $\mathcal{Q}$ is bounded by a trace constraint
\begin{align}   \label{set_Q}
	\mathcal{Q}
	&= \left\{ \BQ \in \mathbb{C}_+^{N_\Tx \times N_\Tx} \colon \trace(\BQ) \leq \mu_\mathcal{Q} \right\}.
\end{align}
The constant $\mu_\mathcal{Q}$ may be computed from the energy conservation relation \eqref{energy_conservation} as $\mu_\mathcal{Q} = \frac{T\mu - \trace(\BP)}{T-T_\tau}$. It may as well be considered as some arbitrary constant.

\subsection{Preliminaries}

Prior to delving into analytical derivations, it is instructive to take a glance at how $\BS(\BP,\BQ)$ depends on its second argument $\BQ$, in order to get to grips with the optimization problem at hand. In the expression of the matrix-to-matrix function
\begin{align}   \label{S(P,Q)}
	\BQ \mapsto \BS(\BP,\BQ)
	= \frac{\hat{\BR}^{\frac{1}{2}}\BQ\hat{\BR}^{\frac{1}{2}}}{1+\trace(\BQ\tilde{\BR})},
\end{align}
we see that the argument $\BQ$ appears in the matrix-valued numerator, and inside a trace operator in the denominator. This function $\BQ \mapsto \BS(\BP,\BQ)$ is thus reminiscent of fractions of monomials such as $q \mapsto \frac{aq}{1+bq}$, except that it is defined for matrices. In fact, the function $\BQ \mapsto \BS(\BP,\BQ)$ pertains to what can be defined in the following Definition~\ref{def:linear_fractional} as a generalization of linear fractional functions. The latter are commonly defined for the scalar case (e.g., \cite[Sec.~2.3.3]{BoVa04}).

\begin{mydef}   \label{def:linear_fractional}
Let $\mathcal{X} \subset \mathbb{C}^{n \times n}$ denote a set of Hermitian matrices of size $n \times n$ whose elements $\BX \in \mathcal{X}$ satisfy $\trace(\BB\BX) \neq -1$ with some given Hermitian matrix $\BB \in \mathbb{C}^{n \times n}$. A function $\BX \mapsto \phi(\BX;\BA,\BB)$ that is defined as
\begin{align}
	\mathcal{X} \ &\to \ \mathbb{C}^{m \times m}, \ 
	\BX \ \mapsto \ \phi(\BX;\BA,\BB) = \frac{\BA\BX\BA^\He}{1 + \trace(\BB\BX)}
\end{align}
shall be called a {\em linear fractional} function with parameters $\BA \in \mathbb{C}^{m \times n}$ and $\BB \in \mathbb{C}^{n \times n}$.
\end{mydef}

Note that the Hermitianity of $\BB$ and of the argument $\BX$ ensures the Hermitianity of the image $\phi(\BX;\BA,\BB)$. Linear fractional functions may or may not be injective functions, depending on the properties of the parameter $\BA$.

\begin{mylem}   \label{lem:linear_fractional}
The linear fractional function $\BX \mapsto \phi(\BX;\BA,\BB)$ from Definition~\ref{def:linear_fractional} is injective (one-to-one) if one at least of the following two conditions apply:
\begin{enumerate}
	\item	The parameter $\BA$ has full column rank
	\item	The parameter $\BA$ has full row rank and the domain $\mathcal{X}$ is such that $\forall \BX \in \mathcal{X} \colon \range(\BX)=\range(\BA^\He)$.\footnote{In case $\BA$ has neither full column nor full row rank, one can bring the problem back to one of the two considered cases by an appropriate rank reduction.}
\end{enumerate}
In these two respective cases, its inverse function $\phi^{-1} \colon \phi(\mathcal{X};\BA,\BB) \to \mathcal{X}, \BY \mapsto \phi^{-1}(\BY;\BA,\BB)$ is
\begin{enumerate}
	\item	linear fractional with parameters $\BA^\sharp$ and $-\BA^{\sharp\He}\BB\BA^\sharp$, where $\BA^\sharp = (\BA^\He\BA)^{-1}\BA^\He$ denotes the left pseudoinverse of $\BA$, i.e., $\phi^{-1}(\bullet;\BA,\BB) = \phi(\bullet;\BA^\sharp,-\BA^{\sharp\He}\BB\BA^\sharp)$.
	\item	linear fractional with parameters $\BA^\flat$ and $-\BA^{\flat\He}\BB\BA^\flat$, where $\BA^\flat = \BA^\He(\BA\BA^\He)^{-1}$ denotes the right pseudoinverse of $\BA$, i.e., $\phi^{-1}(\bullet;\BA,\BB) = \phi(\bullet;\BA^\flat,-\BA^{\flat\He}\BB\BA^\flat)$.
\end{enumerate}
\end{mylem}
\begin{IEEEproof}
See Appendix~\ref{proof:lem:linear_fractional}.
\end{IEEEproof}

In the following we will optimize $\BS(\BP,\BQ)$ rather than $\BQ$ directly, and thus the above Lemma~\ref{lem:linear_fractional} can be used for computing the optimal transmit covariance $\BQ$ from the optimal $\BS$, by means of the appropriate inverse linear fractional function.

\subsection{Convexity of the set of feasible $\BS$ for prescribed pilots}

Prescribing the pilot Gram $\BP$ means that the matrices $\tilde{\BR} = (\BR^{-1}+\BP)^{-1}$ and $\hat{\BR} = \BR - \tilde{\BR}$ are prescribed. Therefore, the function $\BQ \mapsto \BS(\BP,\BQ)$ as given in \eqref{S(P,Q)} is linear fractional with parameters $\BA = \hat{\BR}^{\frac{1}{2}}$ and $\BB = \tilde{\BR}$, i.e.,
\begin{align}
	\BS(\BP,\BQ)
	= \phi(\BQ;\hat{\BR}^{\frac{1}{2}},\tilde{\BR}).
\end{align}
With this new notation, the problem \eqref{marginal_problem_Q} reads as
\begin{align}   \label{marginal_problem_Q_2}
	\BQ^\star(\BP)
	&= \argmax_{\BQ \in \mathcal{Q}} F(\phi(\BQ;\hat{\BR}^{\frac{1}{2}},\tilde{\BR})).
\end{align}
The key property of linear fractional functions that we need for understanding Problem~\eqref{marginal_problem_Q_2} is that they preserve the linearity of segments.

\begin{mylem}   \label{lem:segment_preservation}
An injective linear fractional function $\varphi(\bullet) = \phi(\bullet;\BA,\BB)$ with some given parameters $\BA$ and $\BB$ uniquely maps linear segments onto linear segments in a one-to-one manner, i.e.,
%\begin{align}   \label{segment_preservation}   %adjust
%	&\forall (\BX_1,\BX_2,\alpha) \in \mathcal{X}^2 \times [0;1], \exists \beta \in [0;1] \colon \nonumber\\
%	&\ \ \varphi(\alpha\BX_1 + (1-\alpha)\BX_2) = \beta\varphi(\BX_1) + (1-\beta)\varphi(\BX_2).
%\end{align}
\begin{align}   \label{segment_preservation}
	\forall (\BX_1,\BX_2,\alpha) \in \mathcal{X}^2 \times [0;1], \exists \beta \in [0;1] \colon \varphi(\alpha\BX_1 + (1-\alpha)\BX_2) = \beta\varphi(\BX_1) + (1-\beta)\varphi(\BX_2).
\end{align}
\end{mylem}

\begin{IEEEproof}
This is readily verified by inserting the explicit value
\begin{align}
	\beta = \frac{\alpha(1 + \trace(\BB\BX_1))}{1 + \alpha\trace(\BB\BX_1) + (1-\alpha)\trace(\BB\BX_2)}
\end{align}
into the equality \eqref{segment_preservation}.
\end{IEEEproof}

\begin{figure}[htb]
\begin{center}
\includegraphics[width=.5\columnwidth]{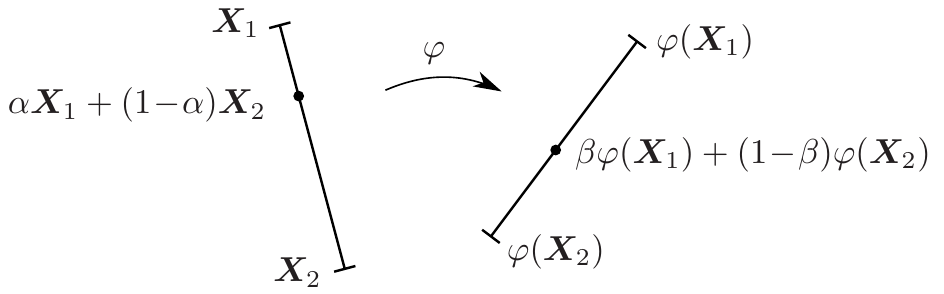}
\caption{Linear fractional functions preserve segments}
\label{fig:segments}
\end{center}
\end{figure}

Figure~\ref{fig:segments} symbolically depicts the behavior of linear fractional functions: a convex combination of two points is mapped onto a convex combination of the respective images of said points, thus preserving segments. They are not linear functions though, because $\alpha$ and $\beta$ can be different.

\begin{mycor}   \label{cor:set_convexity}
Linear fractional mappings preserve set convexity.
\end{mycor}

\begin{IEEEproof}
Take a pair $(\BX_1,\BX_2) \in \mathcal{X}^2$ with a convex $\mathcal{X}$. According to Lemma~\ref{lem:segment_preservation}, any convex combination of $\BX_1$ and $\BX_2$ is mapped onto a convex combination of $\varphi(\BX_1)$ and $\varphi(\BX_2)$. Therefore, the codomain $\varphi(\mathcal{X})$ is convex.
\end{IEEEproof}

As a consequence, $\BS(\BP,\mathcal{Q})$ is a convex set because $\mathcal{Q}$ is convex [cf.~\eqref{set_Q}]. So if a utility $F$ is concave in $\BS$, then Problem \eqref{marginal_problem_Q}, which may be rewritten in the $\BS$-domain as
\begin{align}   \label{marginal_problem_Q_matricial_in_S_domain}
	\BS^\star(\BP)
	= \argmax_{\BS \in \BS(\BP,\mathcal{Q})} F(\BS),
\end{align}
is convex. The optimal transmit covariance $\BQ^\star(\BP)$ is then computed from $\BS^\star(\BP)$ by means of the appropriate inverse linear fractional function (cf. Lemma~\ref{lem:linear_fractional}). More generally speaking, if $F$ is quasi-concave in $\BS$, then the problem \eqref{marginal_problem_Q} can be recast into a convex problem by an appropriate transformation. Even if $F$ is only unimodal on $\BS(\BP,\mathcal{Q})$---that is, it has a single local maximum on the convex compact $\BS(\BP,\mathcal{Q})$---one can still optimize it efficiently via bisection. The mutual information $I$ is one example of a concave utility. Other examples of concave or log-concave (quasi-concave) utilities are given in Table~\ref{table:utilities} in Appendix~\ref{app:examples_of_utilities}.

The next theorem specifies an important property of the range space of the optimal $\BQ^\star(\BP)$.

\begin{mythm}   \label{thm:Q_rank}
For any utility $F \in \mathcal{F}$ and a prescribed pilot Gram $\BP$, the range space of the optimal transmit covariance $\BQ^\star(\BP)$ must be contained in the range space of the channel estimate covariance $\hat{\BR}$:
\begin{align}   \label{range_space_inclusion}
	\range(\BQ^\star(\BP)) \subseteq \range(\hat{\BR}).
\end{align}
\end{mythm}
\begin{IEEEproof}
See Appendix~\ref{proof:thm:Q_rank}.
\end{IEEEproof}

Note that, together with Identity~\eqref{rank_equality}, Theorem~\ref{thm:Q_rank} directly implies the rank inequality
\begin{align}   \label{Q_rank_inequality}
	\rank(\BQ^\star(\BP)) \leq \rank(\hat{\BR}) = \rank(\BP),
\end{align}
or in words,
\begin{align}   \label{Q_rank_inequality_in_words}
	\textbf{number of streams $\leq$ number of pilot symbols}
\end{align}
The idea behind the proof of Theorem~\ref{thm:Q_rank} is that, if $\BQ^\star(\BP)$ had eigenvectors (transmit directions) lying outside the range space of the estimate covariance $\hat{\BR}$, then the transmitter would be radiating some of its transmit power into channel directions of which the receiver has no estimate (and thus cannot detect coherently), thus incurring a waste of power. As a particular consequence, \eqref{Q_rank_inequality} tells us that the number of precoded streams should never exceed the number of training symbols.

\subsection{Convexity of the set of feasible $\Bs$ for prescribed pilots}   \label{ssec:Convexity_of_the_set_of_feasible_profiles_s_for_prescribed_P}

By virtue of the equivalence of Definitions~\ref{def:class_F_matrix} and \ref{def:class_F_vector}, we may rewrite Problem~\eqref{marginal_problem_Q} as
\begin{align}   \label{marginal_problem_Q_vectorial}
	\BQ^\star(\BP)
	&= \argmax_{\BQ \in \mathcal{Q}} f(\Bs(\BP,\BQ)),
\end{align}
or alternatively, in the $\Bs$-domain [compare with \eqref{marginal_problem_Q_matricial_in_S_domain}] as
\begin{align}
	\Bs^\star(\BP)
	&= \argmax_{\Bs \in \Bs(\BP,\mathcal{Q})} f(\Bs).
\end{align}
We are now focusing on the eigenvalue profile $\Bs$ instead of the matrix $\BS$, though both problem formulations [matrix-based \eqref{marginal_problem_Q} and vector-based \eqref{marginal_problem_Q_vectorial}] are in fact equivalent. In the previous subsection, we have shown that the set $\BS(\BP,\mathcal{Q})$ is convex. Note that this convexity, however, does not generally imply (nor is implied by) the convexity of the set of eigenvalue profiles $\Bs(\BP,\mathcal{Q})$. Nevertheless, it turns out that $\Bs(\BP,\mathcal{Q})$ is also convex and has a simplex shape, whose vertices are characterized by Theorem~\ref{thm:prescribed_pilot_region} below.

Let $\omega_i$ denote the non-increasingly ordered eigenvalues of the generalized eigenvalue problem
\begin{align}   \label{generalized_evp}
	\hat{\BR} \Bv_i = \omega_i \bigl(\mu_\mathcal{Q}^{-1}\myid + \tilde{\BR}\bigr) \Bv_i.
\end{align}
Due to $\rank(\hat{\BR}) = \rank(\BP)$ [cf.~\eqref{rank_equality}], only the first $r_\BP = \rank(\BP)$ eigenvalues $\omega_i$ are different from zero.

\begin{mythm}   \label{thm:prescribed_pilot_region}
The set [cf.~\eqref{set_Q}, \eqref{S(P,Q)}]
\begin{align}   \label{simplex_set_written_out}
	\Bs(\BP,\mathcal{Q})
	= \left\{ \Blambda\left( \frac{\hat{\BR}^{\frac{1}{2}}\BQ\hat{\BR}^{\frac{1}{2}}}{1+\trace(\BQ\tilde{\BR})} \right) \ \middle| \ \BQ \in \mathbb{C}_+^{N_\Tx \times N_\Tx}, \ \trace(\BQ) \leq \mu_\mathcal{Q} \right\}
\end{align}
is a simplex given by the convex hull of the origin $\Bsigma^{(0)} \triangleq \mynull$ and of the $r_\BP$ linearly independent points
\begin{align}   \label{simplex_vertices}
	\Bsigma^{(n)}
	= \mathcal{H}(\omega_1,\dotsc,\omega_n) \sum_{j=1}^n \Be_j,
	\quad n \in \{1,\dotsc,r_\BP\}
\end{align}
where $[\Be_1,\dotsc,\Be_{N_\Tx}] = \myid$ is the canonical basis, and $\mathcal{H}(x_1,\dotsc,x_n) = (\sum_{i=1}^n x_i^{-1})^{-1}$ with $n$ arguments $x_1,\dotsc,x_n$ denotes the harmonic mean thereof, divided by $n$.
\end{mythm}

\begin{IEEEproof}
See Appendix~\ref{proof:thm:prescribed_pilot_region}.
\end{IEEEproof}

\begin{figure}[ht]
\begin{center}
\includegraphics[width=.45\columnwidth]{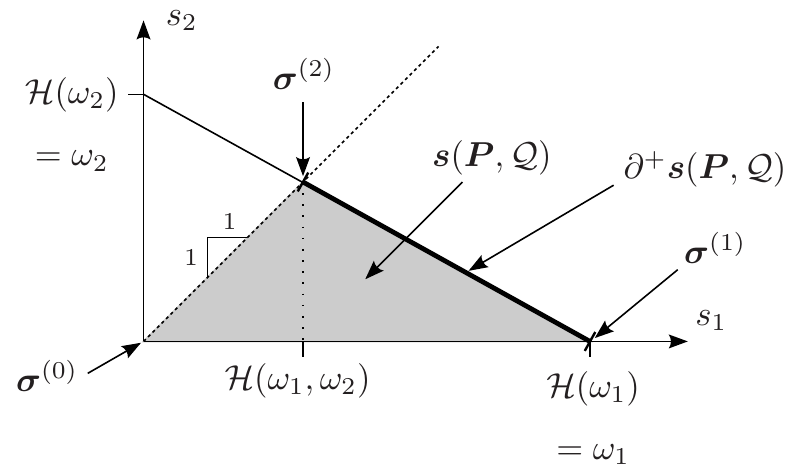}
\caption{Sketch of a simplex set $\Bs(\BP,\mathcal{Q})$. The so-called Pareto border $\partial^+\Bs(\BP,\mathcal{Q})$ contains those points from $\Bs(\BP,\mathcal{Q})$ that are not dominated by any other point from $\Bs(\BP,\mathcal{Q})$, and is the convex hull of $\Bsigma^{(n)}$ for $n \in \{1,\dotsc,r_\BP\}$ (excluding the origin).}
\label{fig:simplex_shape}
\end{center}
\end{figure}

As a byproduct, the proof of Theorem~\ref{thm:prescribed_pilot_region} reveals that if the set of eigenvectors of $\hat{\BR}$ is contained in the set of eigenvectors of $\tilde{\BR}$, i.e., $\col(\BU_{\hat{\BR}}) \subseteq \col(\BU_{\tilde{\BR}})$, then it is optimal with respect to any utility $F \in \mathcal{F}$ that the eigenbasis $\BU_{\BQ^\star(\BP)}$ of the optimal matrix $\BQ^\star(\BP)$ be chosen such that as
\begin{align}    \label{alignment_of_Q}
	\col(\BU_\BQ) \subseteq \col(\BU_{\hat{\BR}}).
\end{align}
Note that this requirement is stronger than the range space inclusion property of Theorem~\ref{thm:Q_rank} [cf.~\eqref{range_space_inclusion}].
This particular situation of eigenbasis alignment $\col(\BU_{\hat{\BR}}) \subseteq \col(\BU_{\tilde{\BR}})$ occurs, for example, when
\begin{itemize}
	\item	using $N_\Tx$ unitary pilots (i.e., $\BP = \frac{\trace(\BP)}{N_\Tx}\myid_{N_\Tx}$ is a scaled identity matrix)
	\item	the channel gains are independently and identically distributed ($\BR = \myid$)
	\item	the channel estimation error vanishes ($\tilde{\BR} = \mynull$, $\hat{\BR} = \BR$)
	\item	the pilots are aligned with the channel covariance, i.e., $\col(\BU_\BP) \subseteq \col(\BU_\BR)$.
\end{itemize}
As we shall see later in Section~\ref{sec:Joint_Pilot_and_Precoder_Optimization}, the latter condition $\col(\BU_\BP) \subseteq \col(\BU_\BR)$ is in fact necessary for joint optimality of $\BP$ and $\BQ$.

\section{Pilot Design for a Prescribed Precoder}   \label{sec:Pilot_Design_for_a_Prescribed_Precoder}

To complement the previous Section~\ref{sec:Precoder_Design_for_Prescribed_Pilots}, we will now swap the roles of $\BP$ and $\BQ$ and consider the optimization of the pilot Gram $\BP$ under a trace constraint, while the transmit covariance $\BQ$ has a fixed value. This problem reads as
\begin{align}   \label{marginal_problem_P}
	\BP^\star(\BQ)
	&= \argmax_{\BP \in \mathcal{P}} F(\BS(\BP,\BQ))
\end{align}
with a search set
\begin{align}   \label{set_P}
	\mathcal{P}
	&= \left\{ \BP \in \mathbb{C}_+^{N_\Tx \times N_\Tx} \colon \trace(\BP) \leq \mu_\mathcal{P} \right\},
\end{align}
or alternatively, in the $\BS$-domain,
\begin{align}   \label{marginal_problem_P_in_S_domain}
	\BS^\star(\BQ)
	&= \argmax_{\BS \in \BS(\mathcal{P},\BQ)} F(\BS).
\end{align}
The constant $\mu_\mathcal{P}$ may be computed from the energy conservation relation \eqref{energy_conservation} as $\mu_\mathcal{P} = T\mu - (T-T_\tau) \trace(\BQ)$, or it may be considered as some given constant.

Finally, in analogy to the rank inequality \eqref{Q_rank_inequality} between $\BP$ and $\BQ^\star(\BP)$, which follows from Theorem~\ref{thm:Q_rank} and applies to Problem~\eqref{marginal_problem_Q}, we also have a corresponding rank inequality for Problem~\eqref{marginal_problem_P}.
\begin{mythm}   \label{thm:P_rank}
For any utility $F \in \mathcal{F}$ and a prescribed transmit covariance $\BQ$, the rank of the optimal pilot Gram $\BP^\star(\BQ)$ is not larger than the rank of $\BQ$:
\begin{align}   \label{P_rank_inequality}
	\rank(\BP^\star(\BQ)) \leq \rank(\BQ).
\end{align}
\end{mythm}
\begin{IEEEproof}
See Appendix~\ref{proof:thm:P_rank}.
\end{IEEEproof}

In words, we can state this as [compare with \eqref{Q_rank_inequality_in_words}]
\begin{align}   \label{P_rank_inequality_in_words}
	\textbf{number of streams $\geq$ number of pilot symbols}
\end{align}
The interpretation behind this rank inequality is that, if there were more orthogonal training directions than there are data streams precoded, we would necessarily be wasting some pilot energy into directions that are not used for transmission anyway.

Next, we will show that the set $\BS(\mathcal{P},\BQ)$ is convex. We write out $\tilde{\BR}$ as $\BR - \hat{\BR}$, then $\BS$ reads as [cf.~\eqref{S}]
\begin{align}
	\BS
	= \frac{\hat{\BR}^{\frac{1}{2}}\BQ\hat{\BR}^{\frac{1}{2}}}{1+\trace(\BQ\BR)-\trace(\BQ\hat{\BR})},
\end{align}
which is unitarily equivalent to
\begin{align}   \label{S_prime}
	\BS'
	= \frac{\BQ^{\frac{1}{2}}\hat{\BR}\BQ^{\frac{1}{2}}}{1+\trace(\BQ\BR)-\trace(\BQ\hat{\BR})},
\end{align}
since the Hermitian matrices $\hat{\BR}^{\frac{1}{2}}\BQ\hat{\BR}^{\frac{1}{2}}$ and $\BQ^{\frac{1}{2}}\hat{\BR}\BQ^{\frac{1}{2}}$ have the same eigenvalues because of the identity $\Blambda(\BA\BB) = \Blambda(\BB\BA)$.
Due to the invariance property in Definition~\ref{def:class_F_matrix}, the matrices $\BS$ and $\BS'$ yield the same utility, i.e., $F(\BS) = F(\BS')$ for any $F \in \mathcal{F}$, so they can be used interchangeably. Let us further abbreviate $1 + \trace(\BQ\BR)$ as $\tau$ so we get
\begin{align}   \label{S_prime_2}
	\BS'
	= \frac{\BQ^{\frac{1}{2}}\hat{\BR}\BQ^{\frac{1}{2}}}{\tau-\trace(\BQ\hat{\BR})}.
\end{align}
By comparing Expression~\eqref{S_prime_2} with with the definition of linear fractional functions (cf. Definition~\ref{def:linear_fractional}), we identify $\BS'$ as a linear fractional function of $\hat{\BR}$ with parameters $\BA = \frac{1}{\sqrt{\tau}}\BQ^{\frac{1}{2}}$ and $\BB = -\frac{1}{\tau}\BQ$, i.e.,
\begin{align}
	\BS'(\BP,\BQ)
	= \phi\Bigl(\hat{\BR};\tfrac{1}{\sqrt{\tau}}\BQ^{\frac{1}{2}},-\tfrac{1}{\tau}\BQ\Bigr).
\end{align}
In Appendix~\ref{app:R_hat_convexity}, we show that the set of feasible $\hat{\BR}$ is convex, from which follows immediately with Corollary~\ref{cor:set_convexity} that $\BS'(\mathcal{P},\BQ)$ is a convex set. Provided that the utility $F$ is concave, quasi-concave or unimodal, Problem \eqref{marginal_problem_P_in_S_domain} is convex in the domain of $\BS'$, and as such, can be solved efficiently with convex optimization methods.
\footnote{As regards the optimization $\BP^\star(\BQ)$ in the {\em eigenvalue} domain, it turns out that, unlike for the set $\Bs(\BP,\mathcal{Q})$ studied in the previous section \ref{sec:Precoder_Design_for_Prescribed_Pilots} and characterized as a simplex in Theorem~\ref{thm:prescribed_pilot_region}, there does not seem to exist a comparably simple analytic characterization of the set $\Bs(\mathcal{P},\BQ)$.}

\section{Jointly Pareto Optimal Pilot-Precoder Pairs}   \label{sec:Joint_Pilot_and_Precoder_Optimization}

\subsection{Problem statement}

When restated in the domain of feasible profiles $\Bs$ [cf.~\eqref{f_properties_vector}], the original problem~\eqref{joint_problem_generic} reads as
\begin{align}   \label{joint_problem_generic_2}
	\max_{\Bs \in \Bs(\mathcal{PQ})} f(\Bs),
\end{align}
where the feasible set $\Bs(\mathcal{PQ})$ is
\begin{align}
	\Bs(\mathcal{PQ})
	= \Bigl\{ \Bs(\BP,\BQ) \in \mathbb{R}_+^{N_\Tx} \Bigm| (\BP,\BQ) \in \mathcal{PQ} \Bigr\}
\end{align}
and $\mathcal{PQ}$ was defined in \eqref{admissible_set}. Furthermore, we can exploit the monotonicity of utilities $f \in \mathcal{F}$ [cf.~\eqref{power_monotonicity}] to restrict the search set $\Bs(\mathcal{PQ})$ to its Pareto border alone:
\begin{align}   \label{joint_problem_generic_3}
	\max_{\Bs \in \partial^+\Bs(\mathcal{PQ})} f(\Bs).
\end{align}
Said Pareto border $\partial^+\Bs(\mathcal{PQ})$ consists of Pareto optimal points, i.e., points in $\Bs(\mathcal{PQ})$ that are not dominated by any other point in $\Bs(\mathcal{PQ})$, or in mathematical notation:
\begin{align}   %adjust
	\partial^+\Bs(\mathcal{PQ})
	= \Bigl\{ \Bs' \in \Bs(\mathcal{PQ}) \Bigm| \nexists \Bs'' \in \Bs(\mathcal{PQ}) \colon \Bs'' \geq \Bs' \text{ with } \Bs'' \neq \Bs' \Bigr\}.
\end{align}

\begin{figure}[htb]
\begin{center}
\includegraphics[width=.35\columnwidth]{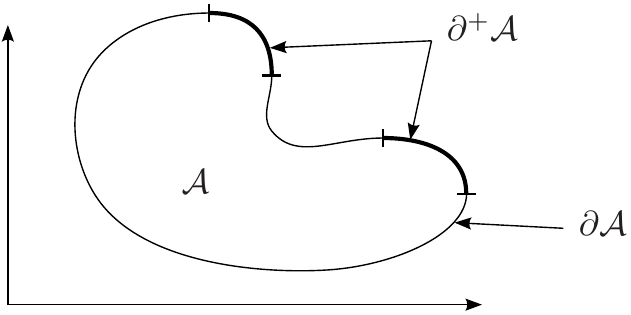}
\caption{Pareto border $\partial^+\mathcal{A}$ of a compact set $\mathcal{A} \subset \mathbb{R}^2$}
\label{fig:Pareto}
\end{center}
\end{figure}

The practical computation of the joint global optimum \eqref{joint_problem_generic_2} depends mainly on the properties of the utility function $f$ that we are considering. In fact, whether the problem at hand is convex, non-convex, quasi-convex, etc., depends on the function $f$ and possibly also on the values of $\BR$ and $T$, hence there cannot be a generic optimization procedure that is guaranteed to converge to the global joint optimum. Instead, the problem must be analyzed case-by-case for every utility function and set of parameters. However, there exists an important subproblem of \eqref{joint_problem_generic_2} that is common to all utility functions of the class $\mathcal{F}$ and can be generally solved, as we shall see: the computation of the search set $\partial^+\Bs(\mathcal{PQ})$. The present Section~\ref{sec:Joint_Pilot_and_Precoder_Optimization} deals with this problem.

\subsection{Number of pilot symbols and number of streams}   \label{sec:Number of pilot symbols and number of streams}

The joint problem~\eqref{joint_problem_generic_3} can be decomposed in an outer optimization (which we shall call {\em energy boost}\footnote{In the literature, the optimal balancing between pilot/data symbol {\em powers} under an overall average power constraint and for fixed time fractions assigned to training and data transmission, is sometimes referred to as {\em power boost} (e.g., \cite{Lo08}). Our setup is different: the training duration $T_\tau$ is not fixed, but is given by the inner optimization via $T_\tau = \rank(\BP)$. The constraint for the outer optimization is not on powers, but on the sum of pilot {\em energy} $\mu_\mathcal{P}$ and data symbol {\em energy} $(T-T_\tau)\mu_\mathcal{Q}$. This is why we talk about {\em energy boost}.}) that consists in finding the optimal balance between the pilot symbol energy and the data symbol power, and an inner optimization of $\Bs$ over a set $\Bs(\mathcal{P},\mathcal{Q})$:
\begin{align}   \label{joint_problem_decomposed}
	\max_{ \substack{ \mu_\mathcal{P}, \mu_\mathcal{Q} \geq 0 \\ \mu_\mathcal{P} + (T-T_\tau)\mu_\mathcal{Q} = T\mu } } \left\{ \max_{\Bs \in \partial^+\Bs(\mathcal{P},\mathcal{Q})} f(\Bs) \right\}.
\end{align}
Note that the inner optimization is over the set $\partial^+\Bs(\mathcal{P},\mathcal{Q})$ and not over the set $\partial^+\Bs(\mathcal{PQ})$ as in \eqref{joint_problem_generic_3}: here, the sets $\mathcal{P}$ and $\mathcal{Q}$ are understood to be the trace-constrained sets as defined in \eqref{set_P} and \eqref{set_Q}, respectively. The inner optimization inside the braces of \eqref{joint_problem_decomposed} can as well be written as
\begin{align}   \label{innermost_problem}
	\max_{(\BP,\BQ) \in \mathcal{P} \times \mathcal{Q}} f\bigl(\Bs(\BP,\BQ)\bigr)
	&= \max_{\BP \in \mathcal{P}} \left\{ f\bigl(\Bs(\BP,\BQ^\star(\BP))\bigr) \right\} \nonumber\\
	&= \max_{\BQ \in \mathcal{Q}} \left\{ f\bigl(\Bs(\BP^\star(\BQ),\BQ)\bigr) \right\} %\nonumber\\
\end{align}
with $\BQ^\star(\BP)$ and $\BP^\star(\BQ)$ defined by \eqref{marginal_problem_Q} and \eqref{marginal_problem_P}, respectively. The jointly optimal pilot-precoder pair $(\BP^\star,\BQ^\star)$ must therefore simultaneously fulfill the rank inequalities \eqref{Q_rank_inequality} and \eqref{P_rank_inequality} (the latter being set forth by Theorem~\ref{thm:P_rank}), from which follows that $\BP^\star$ and $\BQ^\star$ must have equal rank at the joint optimum. Since this rank equality holds regardless of the value of the pair $(\mu_\mathcal{P},\mu_\mathcal{Q})$, it also generally holds for the optimal pair $(\BP^\star,\BQ^\star)$ in Problem \eqref{joint_problem_decomposed}. Since said rank equality holds also independently of the value of $T_\tau$, it also holds for the full-fledged problem \eqref{joint_problem_full_fledged} (with training overhead taken into account), so that we can state that [compare with \eqref{Q_rank_inequality_in_words}, \eqref{P_rank_inequality_in_words}]
\begin{align}
	\textbf{number of streams $=$ number of pilot symbols}
\end{align}
is a necessary condition for a pilot-precoder pair $(\BP,\BQ)$ to be jointly optimal for Problems~\eqref{innermost_problem}, \eqref{joint_problem_generic_2}, and \eqref{joint_problem_full_fledged}.

\subsection{Jointly optimal transmit and training directions}

A fortunate circumstance when treating the joint problem \eqref{joint_problem_generic}/\eqref{joint_problem_generic_2} is that the jointly optimal transmit and training directions have a very simple and intuitive characterization, enunciated in Theorem~\ref{thm:joint_basis} below.
Let us rewrite Problem~\eqref{joint_problem_generic_3} like in \eqref{joint_problem_decomposed}, and only consider the inner optimization problem inside the curly braces \eqref{joint_problem_decomposed}, namely
\begin{align}   \label{joint_problem_decomposed_revisited_inner}
	\max_{\Bs \in \partial^+\Bs(\mathcal{P},\mathcal{Q})} f(\Bs),
\end{align}
which is the pilot-precoder joint optimization problem {\em without} energy boost.

Let the channel covariance $\BR$, the pilot Gram $\BP$ and the transmit covariance $\BQ$ have the following (reduced) eigendecompositions:
\begin{align*}
	\BR &= \BU_\BR\BLambda_\BR\BU_\BR^\He, &
	\BP &= \BU_\BP\BLambda_\BP\BU_\BP^\He, &
	\BQ &= \BU_\BQ\BLambda_\BQ\BU_\BQ^\He.
\end{align*}
Without loss of generality, we assume that the eigenvalues of $\BR$ are arranged in non-increasing order on the diagonal positions of $\BLambda_\BR$, whereas the eigenvalues of $\BLambda_\BP$ and $\BLambda_\BQ$ are not sorted in any specific order.

\begin{mythm}   \label{thm:joint_basis}
For any utility $f \in \mathcal{F}$, in the joint optimization problem \eqref{joint_problem_decomposed_revisited_inner}, there is no loss of optimality in setting the eigenvectors of the pilot Gram $\BP$ (i.e., the left singular vectors of the pilot sequence $\BT$) and the eigenvectors of the transmit covariance $\BQ$ (i.e., the left singular vectors of the precoder $\BF$) to be a common subset of the eigenvectors of the channel covariance $\BR$ corresponding to the largest eigenvalues of $\BR$. Formally, this is to say that the (reduced) eigenbases $\BU_\BP$ and $\BU_\BQ$ should satisfy [cf. Section~\ref{sec:notation}]
\begin{align}   \label{eigenbasis_alignment}
	\col(\BU_\BP)
	= \col(\BU_\BQ)
	= \{\Bu_{\BR,1},\dotsc,\Bu_{\BR,r^\star}\}
	\subseteq \col(\BU_\BR),
\end{align}
where $\BU_\BR \triangleq [\Bu_{\BR,1},\dotsc,\Bu_{\BR,N_\Tx}]$, and $r^\star = \rank(\BP^\star) = \rank(\BQ^\star)$ denotes the pilot/precoder rank at the joint optimum $(\BP^\star,\BQ^\star)$ of Problem~\eqref{joint_problem_decomposed_revisited_inner}.
\footnote{Obviously, the rank $r^\star$ is not known {\em a priori} before solving the problem. The notation in \eqref{eigenbasis_alignment} is merely to indicate that $\col(\BU_\BP)$ and $\col(\BU_\BQ)$ should contain eigenvectors of $\BR$ corresponding to the {\em largest} eigenvalues of $\BR$.}
\end{mythm}
\begin{IEEEproof}
See Appendix~\ref{app:proof:thm:joint_basis}.
\end{IEEEproof}

Since Theorem~\ref{thm:joint_basis} holds irrespective of the value of the pair $(\mu_\mathcal{P},\mu_\mathcal{Q})$, it not only holds for the joint optimization {\em without} energy boost \eqref{joint_problem_decomposed_revisited_inner}, but as well for the joint optimization problem {\em with} energy boost \eqref{joint_problem_generic}/\eqref{joint_problem_generic_2}.

Consequently, and without loss of optimality, we will align the eigenbases of $\BP$ and $\BQ$ in conformity with \eqref{eigenbasis_alignment}. The scalars $[\Br]_i = r_i$, $[\Bp]_i = p_i$, and $[\Bq]_i = q_i$ shall denote the eigenvalues of $\BR$, $\BP$, and $\BQ$, respectively. Under such assumptions, all matrices involved in the expression of the effective SNR \eqref{S}, namely $\hat{\BR}$ and $\tilde{\BR}$ [cf.~\eqref{R_definitions}], as well as $\BQ$, acquire the same eigenbasis $\BU_\BR$. We can readily see from Expression~\eqref{S} that $\BS$ then inherits the (common) eigenvectors of $\BP$ and $\BQ$, i.e., $\col(\BU_\BS) = \col(\BU_\BP) = \col(\BU_\BQ) \subseteq \col(\BU_\BR)$, so that the profile $\Bs$ is given by [cf.~\eqref{S}]
\begin{align}   \label{s_vectorial}
	\Bs = \frac{\hat{\Br} \odot \Bq}{1 + \Bq^\Tr\tilde{\Br}}
\end{align}
and `$\odot$' denotes the componentwise product.
Here, the eigenvalue vectors $\tilde{\Br} = \tilde{\Br}(\Bp)$ and $\hat{\Br} = \hat{\Br}(\Bp) = \Br - \tilde{\Br}(\Bp)$ are functions of $\Bp$ and respectively have entries
\begin{align}   \label{r_hat_and_r_tilde}
	\tilde{r}_i(p_i) &= \frac{r_i}{1 + r_i p_i}, &
	\hat{r}_i(p_i) &= \frac{r_i^2 p_i}{1 + r_i p_i}.
\end{align}

Hereinforth, we will write $\Bs(\Bp,\Bq)$ instead of $\Bs(\BP,\BQ)$ whenever we implicitly assume that the eigenbases are optimally aligned according to \eqref{eigenbasis_alignment}. We do not impose any ordering of the eigenvalues $p_i$, $q_i$, and $r_i$. Instead we assume, without loss of generality, that they are arranged in such way that the $s_i$ are non-increasingly ordered.

\subsection{Pareto optimal allocation with energy boost}   \label{ssec:Pareto_optimal_allocation_with_average_power_constraint}

Upon optimally aligning the eigenbases as according to Theorem~\ref{thm:joint_basis}, we now consider the remaining problem that consists in jointly optimizing the allocation vector pair $(\Bp,\Bq)$, which belongs to a set that constrains the average power radiated by the transmitter array:
\begin{align}   \label{Gamma}
	\Gamma
	= \left\{ (\Bp,\Bq) \in \mathbb{R}_+^{2 N_\Tx} \middle| \myone^\Tr\Bp + (T-T_\tau) \myone^\Tr\Bq \leq T\mu \right\}.
\end{align}

\begin{figure}[ht]
\begin{center}
\includegraphics[width=.35\columnwidth]{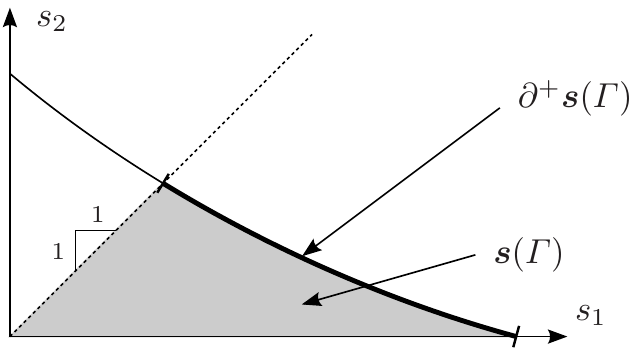}
\caption{Sketch of the typical shape of a set $\Bs(\Gamma)$ and its Pareto border $\partial^+\Bs(\Gamma)$ for $N_\Tx = 2$.}
\label{fig:overall_shape}
\end{center}
\end{figure}

By virtue of Theorem~\ref{thm:joint_basis}, we have $\Bs(\mathcal{PQ}) = \Bs(\Gamma)$. In the following, we will devise a procedure for computing the set of all allocations $(\Bp,\Bq)$ that yield points $\Bs(\Bp,\Bq)$ located on the Pareto border $\partial^+\Bs(\mathcal{PQ}) = \partial^+\Bs(\Gamma)$. Given the monotonicity of the function $\Bs(\Bp,\Bq)$ [cf.~\eqref{power_monotonicity}], we are certain that any Pareto optimal allocation $(\Bp,\Bq)$ will expend the full power budget, and thus belong to
\begin{align}
	\partial^+\Gamma
	= \left\{ (\Bp,\Bq) \in \Gamma \middle| \myone^\Tr\Bp + (T-T_\tau) \myone^\Tr\Bq = T\mu \right\}.
\end{align}
The joint problem \eqref{joint_problem_generic}/\eqref{joint_problem_generic_2} can thus be reformulated once more as
\begin{align}
	\max_{\Bs \in \partial^+\Bs(\partial^+\Gamma)} f(\Bs)
\end{align}
Now note that the search set $\partial^+\Bs(\partial^+\Gamma)$ is {\em not} equal to the set $\Bs(\partial^+\Gamma)$, meaning that it is not sufficient to simply choose some full-power allocation $(\Bp,\Bq)$ in order to obtain a Pareto optimal allocation. Instead, we have the proper inclusion
\begin{align}
	\partial^+\Bs(\partial^+\Gamma) \subsetneq \Bs(\partial^+\Gamma).
\end{align}
In fact, any Pareto optimal allocation is a full-power allocation, but the converse is not true. This becomes clear when counting dimensions: the vector $\Bs$ has $N_\Tx$ real entries, so any parametrization of the feasible set $\Bs(\Gamma)$ with minimal number of parameters will require at most $N_\Tx$ real parameters. However, the entries of the vector pair $(\Bp,\Bq)$ represent $2N_\Tx$ parameters. Even by replacing $\Gamma$ with $\partial^+\Gamma$, which implies the fulfillment of the linear constraint $\sum_i p_i + (T-T_\tau) \sum_i q_i = T\mu$, we only lose one parameter, which still leaves us with $2N_\Tx-1$ parameters. Thus, we are left with at least $N_\Tx-1$ redundant parameters that need to be eliminated. However, a direct elimination by working off the explicit expression of $\Bs(\Bp,\Bq)$ in \eqref{s_vectorial} does not seem possible.

The idea for reducing the parameter set so as to efficiently compute Pareto optimal allocations $(\Bp,\Bq)$ will be as follows: we choose some vector norm $\norm{\cdot}$, then fix a non-negative direction vector $\Be \geq \mynull$ that is normalized as $\norm{\Be} = 1$. This normalized vector points into the positive orthant of the $\Bs$ domain and defines a half-line departing from the origin. We then maximize the norm $\norm{\Bs(\Bp,\Bq)}$ with respect to the allocation $(\Bp,\Bq)$ under the constraint that $\Bs(\Bp,\Bq)$ points into the direction of $\Be$. In other terms, we determine the point from the set $\Bs(\mathcal{PQ}) = \Bs(\Gamma)$ which lies farthest away from the origin, and is located on the line running along $\Be$.

\begin{figure}[ht]
\begin{center}
\includegraphics[width=.35\columnwidth]{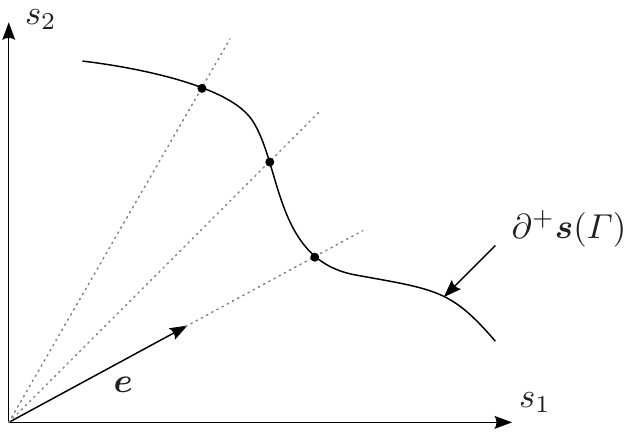}
\caption{Symbolic sketch of the procedure for computing Pareto border points from $\partial^+\Bs(\mathcal{P},\mathcal{Q})$. Said points are parametrized by a unit-norm direction vector $\Be$}
\label{fig:rays}
\end{center}
\end{figure}

Formally, the problem at hand can be stated as:
\begin{align}    \label{effective_snr_balancing_problem}
	\max_{(\Bp,\Bq) \in \Gamma} \ \nu \qquad \text{s.t.} \quad \Bs(\Bp,\Bq) = \nu \Be
\end{align}
where $\nu = \norm{\Bs(\Bp,\Bq)}$ stands for the norm of $\Bs$, while the function $\Bs(\Bp,\Bq)$ is given by \eqref{s_vectorial} as
\begin{align}   \label{s_vectorial_2}
	\Bs(\Bp,\Bq)
	= \frac{\hat{\Br} \odot \Bq}{1 + \tilde{\Br}^\Tr\Bq}
	= \frac{\hat{\Br} \odot \Bq}{1 + \Br^\Tr\Bq - \hat{\Br}^\Tr\Bq}
\end{align}
and $\Be$ is some normalized direction vector pointing into the positive orthant, i.e., $\Be \geq \mynull$ and $\norm{\Be} = 1$. As usual, the search set $\Gamma$ can be reduced to $\partial^+\Gamma$.

When we vary $\Be$, the set of all points $\nu_\maximum \Be$ that are determined by this maximization procedure constitute what we shall call a {\em front border}.

\begin{figure}[htb]
\begin{center}
\includegraphics[width=.3\columnwidth]{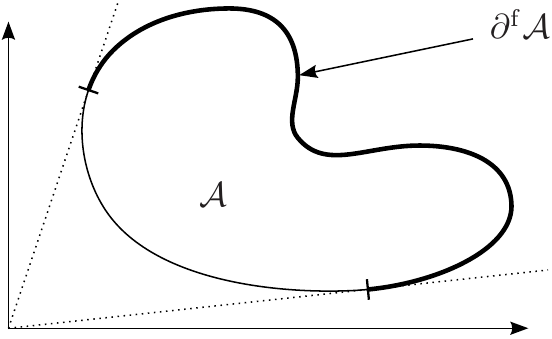}
\caption{Front border $\partial^\mathrm{f}\mathcal{A}$ of a closed set $\mathcal{A} \subset \mathbb{R}^2$}
\label{fig:front}
\end{center}
\end{figure}

The front border of a compact set $\mathcal{A} \subseteq \mathbb{R}_+^n$ shall be denoted as $\partial^\mathrm{f}\mathcal{A}$ and be formally defined as
\begin{align}   \label{front_border_definition}
	\partial^\mathrm{f}\mathcal{A}
	= \bigcup_{\begin{subarray}{c} \Be \geq \mynull \\ \norm{\Be} = 1 \end{subarray}} \argmax_{\begin{subarray}{c} \Ba \in \mathcal{A} \\ \Ba = \nu \Be \end{subarray}} \nu.
\end{align}
Note that certain directions $\Be$ may yield empty sets $\left\{ \Ba \in \mathcal{A} \middle| \Ba = \nu\Be \right\}$, so only non-trivial contributions (non-empty sets) should be retained when taking the union \eqref{front_border_definition}.
As we easily intuit from comparing Figures~\ref{fig:Pareto} and \ref{fig:front}, the {\em Pareto border} and {\em front border} of a compact set are not generally identical. However, according to the next Lemma, identity holds for the set $\Bs(\Gamma)$.

\begin{mylem}   \label{lem:front_border}
The Pareto border and the front border of the set $\Bs(\Gamma)$ coincide.
\end{mylem}
\begin{IEEEproof}
See Appendix~\ref{app:front_border}.
\end{IEEEproof}

As a consequence, we can compute the Pareto border by the above-mentioned technique. Let us choose the norm $\norm{\cdot}$ to be the $1$-norm $\norm{\Bs}_1 = \sum_i s_i$, as this will turn out to be a convenient choice. The quantity $\nu$ that is maximized in \eqref{effective_snr_balancing_problem} is the $1$-norm of the vector $\Bs(\Bp,\Bq)$, constrained to being colinear with $\Be$, i.e.,
\begin{align}   \label{s_colinearity}
	\Bs(\Bp,\Bq)
	&= \nu\Be
	&
	\norm{\Bs(\Bp,\Bq)}_1
	&= \frac{\eta}{1 + \Br^\Tr\Bq - \eta}
	= \nu
\end{align}
where $\eta$ stands for [cf.~\eqref{s_vectorial_2}]
\begin{align}   \label{eta}
	\eta = \norm{\hat{\Br} \odot \Bq}_1 = \hat{\Br}^\Tr\Bq.
\end{align}
Note that the colinearity constraint $\Bs(\Bp,\Bq) = \nu\Be$ implies the colinearity $\hat{\Br} \odot \Bq = \eta\Be$. Componentwise, the latter reads as [cf.~\eqref{r_hat_and_r_tilde}]
\begin{align}   \label{scaled_direction_vector}
	\frac{r_i^2 p_i q_i}{1 + r_i p_i}
	&= \eta e_i
\end{align}
Consider $\Be$ to be fixed. Then we see from \eqref{scaled_direction_vector} that, once $\eta$ is given, $p_i$ and $q_i$ are entirely determined from one another: given any value of $q_i \geq 0$, the corresponding value of $p_i \geq 0$ is uniquely determined (as long as $\frac{\eta e_i}{q_i} < r_i$), and conversely, given any value of $p_i \geq 0$, the value of $q_i \geq 0$ is uniquely determined. This allows us to effectuate a (one-to-one) change of parameters: we drop the $q_i$ and replace them by $e_i$, thus effectively replacing the parameter pair $(\Bp,\Bq) \in \Gamma$ by the new pair $(\Bp,\Be) \in \mathcal{D}(T\mu) \times \mathcal{D}(1)$. From \eqref{scaled_direction_vector}, the $q_i$ can now be expressed in terms of $p_i$ and $e_i$ as
\begin{align}   \label{q_i(p)}
	q_i(p_i,e_i)
	= \eta e_i \frac{1 + r_i p_i}{r_i^2 p_i}.
\end{align}
By summing \eqref{q_i(p)} up over $i$, and taking into account the energy conservation $\sum_i p_i + (T-T_\tau) \sum_i q_i = T\mu$, we obtain expressions of $\eta$ and of $q_i$ which are functions of $(\Bp,\Be)$:
\begin{align}
	\eta(\Bp,\Be)
	&= \frac{T\mu - \myone^\Tr\Bp}{T-T_\tau} \left( \sum_{i=1}^{N_\Tx} e_i \frac{1 + r_i p_i}{r_i^2 p_i} \right)^{-1}   \label{eta(p)} \\
	q_i(\Bp,\Be)
	&= \frac{T\mu - \myone^\Tr\Bp}{T-T_\tau} \frac{e_i \frac{1 + r_i p_i}{r_i^2 p_i}}{\sum_j e_j \frac{1 + r_j p_j}{r_j^2 p_j}}.   \label{q(p)}
\end{align}
Consequently, $\nu$ can itself be expressed as a function of $(\Bp,\Be)$ too [cf.~\eqref{effective_snr_balancing_problem}]:
\begin{align}
	\nu(\Bp,\Be)
	= \frac{\eta(\Bp,\Be)}{1 + \Br^\Tr\Bq(\Bp,\Be) - \eta(\Bp,\Be)}.   \label{nu(p)}
\end{align}
We can now dismiss the initial problem formulation \eqref{effective_snr_balancing_problem} in favor of the equivalent formulation
\begin{align}   \label{argmax_nu}
	\Bp^\star(\Be)
	= \argmax_{\Bp \in \mathcal{D}(T\mu)} \nu(\Bp,\Be)
\end{align}
with $\nu(\Bp,\Be)$ as given in \eqref{nu(p)}. Once the maximizer $\Bp^\star(\Be)$ is determined, we compute the corresponding $\Bq^\star(\Be)$ via
\eqref{q(p)} as
\begin{align}
	\Bq^\star(\Be)
	= \begin{bmatrix} q_1(\Bp^\star(\Be),\Be) \\ \vdots \\ q_{N_\Tx}(\Bp^\star(\Be),\Be) \end{bmatrix}.
\end{align}
The Pareto border $\partial^+\Bs(\mathcal{PQ}) = \partial^+\Bs(\Gamma)$ is described in its entirety by the union (see Figure~\ref{fig:rays})
\begin{align}   \label{entire_Pareto_1}
%	\partial^+\Bs(\mathcal{PQ})
	\partial^+\Bs(\Gamma)
	= \bigcup_{\begin{subarray}{c} \Be \geq \mynull \\ \norm{\Be}_1 = 1 \end{subarray}} \Bs(\Bp^\star(\Be),\Bq^\star(\Be)).
\end{align}

\begin{mydef}   \label{def:quasi-concavity}
A function $f \colon \mathcal{X} \mapsto \mathbb{R}$ is quasi-concave (resp. quasi-convex) on a convex and compact set $\mathcal{X} \subset \mathbb{R}^n$ if it can be represented as a concatenation
\begin{align}
	f(\Bx) = (g \circ h)(\Bx)
\end{align}
of a concave (resp. convex) function $h \colon \mathcal{X} \to \mathbb{R}$ and a non-decreasing function $g \colon \mathbb{R} \to \mathbb{R}$.
\end{mydef}

\begin{mylem}   \label{lem:nu_is_quasi_concave}
The function $\nu(\Bp,\Be)$ is quasi-concave in $\Bp$.
\end{mylem}

\begin{IEEEproof}
See Appendix~\ref{app:proof:lem:nu_is_quasi_concave}.
\end{IEEEproof}

This lemma renders \eqref{argmax_nu} a quasi-convex problem, which can be solved efficiently.

\begin{figure}[htb]
\begin{center}
\subfigure[$\nu(\Bp,\Be)$ as a function of $p_1$ and $p_2$]{
	\includegraphics[width=.45\columnwidth]{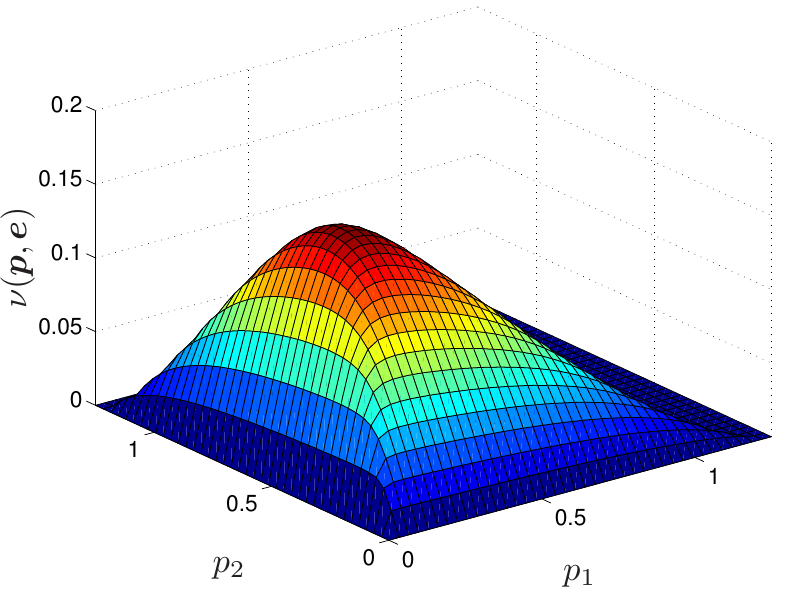}
	\label{fig:nu}
}
\subfigure[Contour plot of the function $\nu(\Bp,\Be)$ from Figure~\ref{fig:nu}]{
	\includegraphics[width=.35\columnwidth]{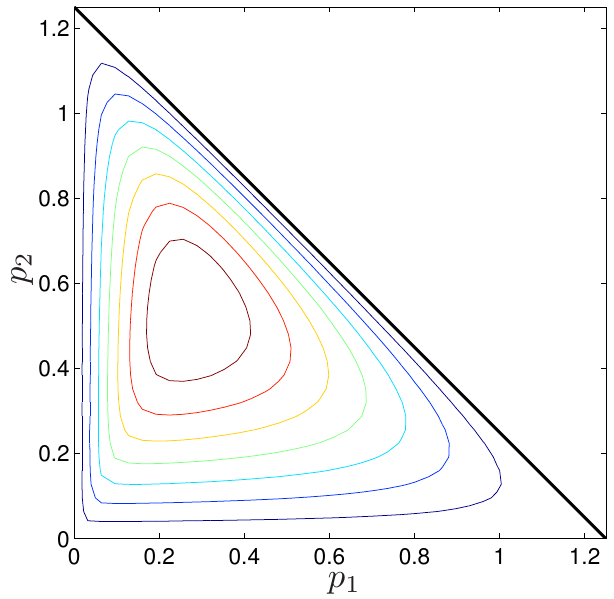}
	\label{fig:nu_contour}
}
\end{center}
\caption{Three-dimensional representation and corresponding contour plot of a function $\nu(\Bp,\Be)$}
\end{figure}

Figures~\ref{fig:nu} and \ref{fig:nu_contour} illustrate an example of a function $\nu(\Bp,\Be)$ for $N_\Tx = 2$ transmit antennas, channel coherence $T = 10$ and SNR $\mu = 1$, $\Br = \bigl[ 2/3 \ \ 1/3 \bigr]^\Tr$ and $\Be = \bigl[ 1/2 \ \ 1/2 \bigr]^\Tr$. The quasi-concavity (but non-concavity) can be well appreciated in said plot, since $\nu(\Bp,\Be)$ appears to be convex in $\Bp$ near the borders of its triangular domain $\mathcal{D}(T\mu)$, while it is concave in an inner region. Notwithstanding this change of curvature, the function is globally quasi-concave in $\Bp$, since all upper contour sets, as illustrated in Figure~\ref{fig:nu_contour}, are convex.

\subsection{Pareto optimal allocation without energy boost}    \label{ssec:Pareto_optimal_allocation_with_fixed_budgets}

The problem considered so far becomes a different one if the pilot energy $\myone^\Tr\Bp$ and the transmit power $\myone^\Tr\Bq$ are individually limited by fixed budgets (i.e., no energy boost is permitted). Instead of a sum energy constraint per fading block \eqref{energy_conservation} that is shared between the tasks of channel estimation and data transmission, let us consider a pair of constraints
\begin{align}   \label{fixed_budgets}
	\myone^\Tr\Bp
	&\leq \mu_\mathcal{P}
	&\myone^\Tr\Bq
	&\leq \mu_\mathcal{Q}
\end{align}
where the so-called {\em budgets} $\mu_\mathcal{P}$ and $\mu_\mathcal{Q}$ are two given constants. Repeating the approach taken in Subsection~\ref{ssec:Pareto_optimal_allocation_with_average_power_constraint}, we define a set
\begin{align}
	\bar{\Gamma}
	= \left\{ (\Bp,\Bq) \in \mathbb{R}_+^{2 N_\Tx} \middle| \myone^\Tr\Bp \leq \mu_\mathcal{P} \ \text{and} \ \myone^\Tr\Bq \leq \mu_\mathcal{Q} \right\}.
\end{align}
and seek to maximize the $1$-norm
\begin{align}
	\bar{\nu}
	= \norm{\Bs(\Bp,\Bq)}_1
	= \frac{\bar{\eta}}{1 + \Br^\Tr\Bq - \bar{\eta}}
\end{align}
over $\bar{\Gamma}$ under the colinearity constraint $\Bs(\Bp,\Bq) = \nu \Be$, and where $\bar{\eta} = \norm{\hat{\Br} \odot \Bq}_1$ [cf.~\eqref{s_colinearity}]. Similarly to \eqref{eta(p)}--\eqref{q(p)}, for a fixed direction $\Be$, the quantities $\bar{\eta}$ and $\Bq$ can be expressed as functions of $(\Bp,\Be)$, namely
\begin{align}
	\bar{\eta}(\Bp,\Be)
	&= \mu_\mathcal{Q} \left( \sum_{i=1}^{N_\Tx} e_i \frac{1 + r_i p_i}{r_i^2 p_i} \right)^{-1}   \label{bar_eta(p)} \\
	\bar{q}_i(\Bp,\Be)
	&= \mu_\mathcal{Q} \frac{e_i \frac{1 + r_i p_i}{r_i^2 p_i}}{\sum_j e_j \frac{1 + r_j p_j}{r_j^2 p_j}}.   \label{bar_q(p)}
\end{align}
With these two functions, $\bar{\nu}$ as well can be expressed as a function of $\Bp$:
\begin{align}
	\bar{\nu}(\Bp,\Be)
	= \frac{\bar{\eta}(\Bp,\Be)}{1 + \Br^\Tr\bar{\Bq}(\Bp,\Be) - \bar{\eta}(\Bp,\Be)}.   \label{bar_nu(p)}
\end{align}
Note that in constrast to \eqref{eta(p)}--\eqref{q(p)}, in \eqref{bar_eta(p)}--\eqref{bar_q(p)} the factor $\frac{T\mu - \myone^\Tr\Bp}{T-T_\tau}$ has been replaced by the constant $\mu_\mathcal{Q}$. This change makes the problem
\begin{align}   \label{argmax_nu_without_sharing}
	\bar{\Bp}^\star(\Be)
	= \argmax_{\Bp \in \partial^+\mathcal{D}(\mu_\mathcal{P})} \bar{\nu}(\Bp,\Be)
\end{align}
yet more amenable than its counterpart with energy boost \eqref{argmax_nu}. In fact, while \eqref{argmax_nu} is a quasi-convex problem according to Lemma~\ref{lem:nu_is_quasi_concave}, the problem \eqref{argmax_nu_without_sharing} is convex and admits a closed-form solution
\begin{align}   \label{pilot_allocation_without_sharing}
	\bar{p}_i^\star(\Be)
	= \mu_\mathcal{P} \frac{\frac{1}{r_i}\sqrt{e_i(1+\mu_\mathcal{Q}r_i)}}{\sum_j \frac{1}{r_j} \sqrt{e_j(1+\mu_\mathcal{Q}r_i)}}.
\end{align}
A detailed derivation is provided in Appendix~\ref{app:proof:pilot_allocation_without_sharing}.
Again, as in \eqref{entire_Pareto_1}, the entire Pareto border $\partial^+\Bs(\mathcal{P},\mathcal{Q}) = \partial^+\Bs(\bar{\Gamma})$ is parametrized by $\Be$ as
\begin{align}   \label{entire_Pareto_2}
	\partial^+\Bs(\bar{\Gamma})
	= \bigcup_{\begin{subarray}{c} \Be \geq \mynull \\ \norm{\Be}_1 = 1 \end{subarray}} \Bs(\bar{\Bp}^\star(\Be),\bar{\Bq}^\star(\Be)),
\end{align}
where $\bar{\Bq}^\star(\Be) = \bar{\Bq}(\Bp^\star(\Be),\Be) = [ \bar{q}_1(\Bp^\star(\Be),\Be) , \dotsc , \bar{q}_{N_\Tx}(\Bp^\star(\Be),\Be) ]^\Tr$.

It is worth mentioning that for $N_\Tx = 2$ transmit antennas, a simple closed-form parametrization of the Pareto border $\partial^+\Bs(\bar{\Gamma})$ was already reported in \cite[Equation (31)]{PaJoFo11_isit}, as the solution to an envelope equation. Said solution can be recovered by particularizing \eqref{entire_Pareto_2} to $N_\Tx = 2$ and an appropriate change of parameters.

\section{Iterative Joint Design}   \label{sec:iterative_joint_design}

Since we know from Theorem~\ref{thm:joint_basis} that the eigenbases are aligned as $\col(\BU_\BP) = \col(\BU_\BQ) \subseteq \col(\BU_\BR)$ at the joint optimum, it follows that we can align the eigenbases accordingly and keep working in the vector domain alone.

The interest of Problems \eqref{marginal_problem_Q}, \eqref{marginal_problem_P} is that, once we know how to compute their solutions with respect to some utility $f \in \mathcal{F}$, a natural way of tackling the joint design of $\Bp$ and $\Bq$ is by alternating between the two problems in the fashion of a block gradient ascent:
\begin{align}
	\begin{cases}
		\Bp_{n+1} = \Bp^\star(\Bq_n) \\
		\Bq_{n+1} = \Bq^\star(\Bp_{n+1})
	\end{cases}
& \text{or} &
	\begin{cases}
		\Bq_{n+1} = \Bq^\star(\Bp_n) \\
		\Bp_{n+1} = \Bp^\star(\Bq_{n+1})
	\end{cases}
\end{align}
This procedure converges monotonically toward a fixed-point of the iteration $(\Bp_{n+1},\Bq_{n+1}) = (\Bp^\star(\Bq_n),\Bq^\star(\Bp_n))$, or enters a cycle. However, this simple iteration is not sufficient to reach a global optimum of the joint problem, the reason being that there are too many parameters contained in the pair $(\Bp,\Bq)$, as already observed in Subsection~\ref{ssec:Pareto_optimal_allocation_with_average_power_constraint}.

Therefore, we need to insert an additional step in the iteration that readjusts the allocation $(\Bp,\Bq)$ so as to remain Pareto optimal. This step can be performed with the methods for computing the Pareto border, developed in Subsection~\ref{ssec:Pareto_optimal_allocation_with_average_power_constraint} when allowing energy boost, and in Subsection~\ref{ssec:Pareto_optimal_allocation_with_fixed_budgets} when precluding energy boost. Roughly speaking, the algorithm should cycle through the following three steps:
\begin{enumerate}
	\item	Optimize $\Bp$ for a prescribed $\Bq$
	\item	Optimize $\Bq$ for a prescribed $\Bp$
	\item	Adjust $(\Bp,\Bq)$ to be Pareto optimal
\end{enumerate}

\subsection{With energy boost}

If $\Bp$ and $\Bq$ are constrained by an average power constraint such as \eqref{energy_conservation}, i.e.,
\begin{align}
	\myone^\Tr\Bp + (T-T_\tau)\myone^\Tr\Bq \leq T\mu,
\end{align}
then the algorithm reads as follows.
\begin{algorithm}
\begin{algorithmic}[1]
\STATE	$\Bp_0 \gets \frac{T_\tau\mu}{N_\Tx} \myone_{N_\Tx}$
\STATE	$\Bq_0 \gets \frac{(T-T_\tau)\mu}{N_\Tx} \myone_{N_\Tx}$
\STATE	$n \gets 0$
\REPEAT
	\STATE	$\Bp' \gets \argmax_{\Bp \in \mathcal{D}(\myone^\Tr\Bp_n)} f(\Bs(\Bp,\Bq))$   \label{step:p_1}
	\STATE	$\Bq' \gets \argmax_{\Bq \in \mathcal{D}(\myone^\Tr\Bq_n)} f(\Bs(\Bp,\Bq))$   \label{step:q_1}
	\STATE	$\Be_{n+1} \gets \frac{\Bs(\Bp',\Bq')}{\norm{\Bs(\Bp',\Bq')}_1}$   \label{step:adjustment_start_1}
	\STATE	$\Bp_{n+1} \gets \argmax_{\Bp \in \mathcal{D}(T\mu)} \nu(\Bp,\Be_{n+1})$   \label{step:adjustment_1}
	\STATE	$\Bq_{n+1} \gets \Bq(\Bp_{n+1},\Be_{n+1})$
	\STATE	$\Bs_{n+1} \gets \Be_{n+1} \nu(\Bp_{n+1},\Be_{n+1})$   \label{step:adjustment_end_1}
	\STATE	$n \gets n+1$
\UNTIL $f(\Bs_n)-f(\Bs_{n-1}) \leq \epsilon$
\end{algorithmic}
\caption{Joint optimization with energy boost}
\label{alg1}
\end{algorithm}

For concave (resp. quasi-concave) utilities $f \in \mathcal{F}$, Steps \ref{step:p_1} and \ref{step:q_1} were shown to be convex (resp. hidden convex) optimizations in the findings of Section~\ref{sec:Pilot_Design_for_a_Prescribed_Precoder} and Subsection~\ref{ssec:Convexity_of_the_set_of_feasible_profiles_s_for_prescribed_P}, respectively. Computation of Steps \ref{step:adjustment_start_1} through \ref{step:adjustment_end_1} have been exposed in detail in Subsection~\ref{ssec:Pareto_optimal_allocation_with_average_power_constraint}, wherein Step \ref{step:adjustment_1} was shown to be a quasi-convex optimization.

\subsection{Without energy boost}

If $\Bp$ and $\Bq$ are constrained by two individual budget constraints $\myone^\Tr\Bp \leq \mu_\mathcal{P}$ and $\myone^\Tr\Bq \leq \mu_\mathcal{Q}$ as in \eqref{fixed_budgets} in Subsection~\ref{ssec:Pareto_optimal_allocation_with_fixed_budgets}, then the algorithm looks similar to the above Algorithm~\ref{alg1}, except for the fact that those steps that project an allocation $(\Bp_n,\Bq_n)$ onto the Pareto border (Steps \ref{step:adjustment_1} through \ref{step:adjustment_end_1}) make use of the barred functions $\bar{\nu}(\cdot,\cdot)$ and $\bar{\Bq}(\cdot,\cdot)$ instead of $\nu(\cdot,\cdot)$ and $\Bq(\cdot,\cdot)$, and that the search sets of the optimizations in Steps~\ref{step:p_1}, \ref{step:q_1} and \ref{step:adjustment_1} as well need to be changed accordingly. Moreover, the initial values for $\Bp_0$ and $\Bq_0$ need to be made consistent with the constraints $\myone^\Tr\Bp \leq \mu_\mathcal{P}$ and $\myone^\Tr\Bq \leq \mu_\mathcal{Q}$.
\begin{algorithm}
\begin{algorithmic}[1]
\STATE	$\Bp_0 \gets \frac{\mu_\mathcal{P}}{N_\Tx} \myone_{N_\Tx}$
\STATE	$\Bq_0 \gets \frac{\mu_\mathcal{Q}}{N_\Tx} \myone_{N_\Tx}$
\STATE	$n \gets 0$
\REPEAT
	\STATE	$\Bp' \gets \argmax_{\Bp \in \partial^+\mathcal{D}(\mu_\mathcal{P})} f(\Bs(\Bp,\Bq))$   \label{step:p_2}
	\STATE	$\Bq' \gets \argmax_{\Bq \in \partial^+\mathcal{D}(\mu_\mathcal{Q})} f(\Bs(\Bp,\Bq))$   \label{step:q_2}
	\STATE	$\Be_{n+1} \gets \frac{\Bs(\Bp',\Bq')}{\norm{\Bs(\Bp',\Bq')}_1}$   \label{step:adjustment_start_2}
	\STATE	$\Bp_{n+1} \gets \argmax_{\Bp \in \partial^+\mathcal{D}(\mu_\mathcal{P})} \bar{\nu}(\Bp,\Be_{n+1})$   \label{step:adjustment_2}
	\STATE	$\Bq_{n+1} \gets \bar{\Bq}(\Bp_{n+1},\Be_{n+1})$
	\STATE	$\Bs_{n+1} \gets \Be_{n+1} \bar{\nu}(\Bp_{n+1},\Be_{n+1})$   \label{step:adjustment_end_2}
	\STATE	$n \gets n+1$
\UNTIL $f(\Bs_n)-f(\Bs_{n-1}) \leq \epsilon$
\end{algorithmic}
\caption{Joint optimization without energy boost}
\label{alg2}
\end{algorithm}

Similarly to Algorithm~\ref{alg1}, we have that for concave (resp. quasi-concave) utilities $f \in \mathcal{F}$, Steps \ref{step:p_2} and \ref{step:q_2} are convex (resp. hidden convex) optimizations (cf.~Section~\ref{sec:Pilot_Design_for_a_Prescribed_Precoder} and Subsection~\ref{ssec:Convexity_of_the_set_of_feasible_profiles_s_for_prescribed_P}, respectively). Steps \ref{step:adjustment_start_2} through \ref{step:adjustment_end_2} are detailed in Subsection~\ref{ssec:Pareto_optimal_allocation_with_average_power_constraint} and can be solved in closed form.

\subsection{Optimal training duration}

We mentioned in the problem statement in Section~\ref{sec:problem_statement} that we could leave aside the problem of tuning the length $T_\tau$ of the training sequence, since its optimization amounts to an exhaustive search over the interval $\{1,\dotsc,N_\Tx\}$. Indeed, to tackle the full-fledged joint optimization problem stated in Equation~\eqref{joint_problem_full_fledged}, we simply need to wrap Algorithms~\ref{alg1} or \ref{alg2} into an extra loop. First, we need to internalize the time penalty into the utility function by redefining $f$ as
\begin{align}
	f(\Bs) \longleftarrow \frac{T-T_\tau}{T} f(\Bs).
\end{align}
Note that the functions $\nu(\cdot,\cdot)$ and $\bar{\nu}(\cdot,\cdot)$ are dependent on the parameter $T_\tau$, and should be updated accordingly with the loop count. The full-fledged algorithm then reads as follows:
\begin{algorithm}
\begin{algorithmic}[1]
	\STATE	$T_\tau \gets 1$
	\REPEAT
		\STATE	\text{\dots Algorithm~\ref{alg1} or \ref{alg2} \dots}
		\STATE	$(\Bp^\star(T_\tau),\Bq^\star(T_\tau)) \gets (\Bp_n,\Bq_n)$
		\STATE	$T_\tau \gets T_\tau + 1$
	\UNTIL{$T_\tau = N_\Tx$}
	\STATE	$(\Bp^\star,\Bq^\star) = \max_{i} f(\Bs(\Bp^\star(i),\Bq^\star(i)))$
\end{algorithmic}
\caption{Full-fledged joint optimization}
\label{alg3}
\end{algorithm}

\section{Simulations}

\begin{figure}[htb]
\begin{center}
\subfigure[convergence of the profile $\Bs$ towards the optimizer, located on the Pareto border (thick black line) of the feasible set $\Bs(\Gamma)$]{
\includegraphics[width=.45\columnwidth]{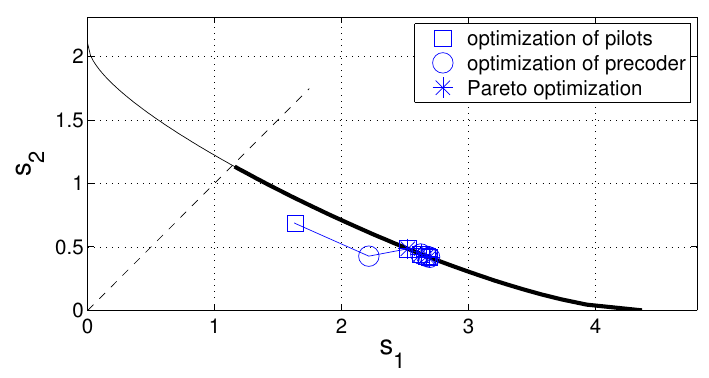}
\label{fig:iteration_in_s}
}
\subfigure[convergence of the utility function $I(\Bs)$ towards the optimum]{
\includegraphics[width=.45\columnwidth]{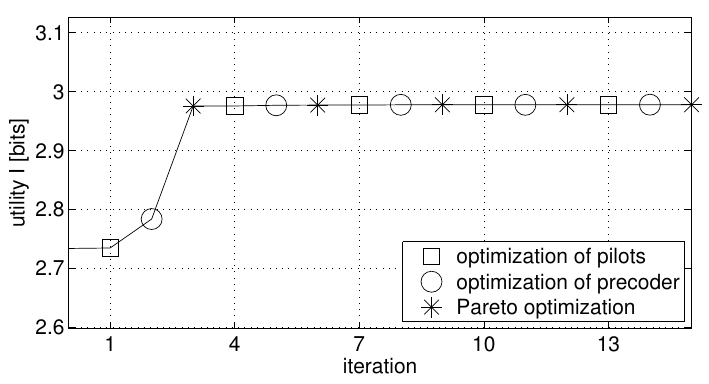}
\label{fig:iteration_in_f}
}
\end{center}
\caption{Convergence of Algorithm~\ref{alg1} at an SNR of $10 \dB$ ($\mu = 10$) for an exemplary $2 \times 2$ MIMO channel, both in the $\Bs$-domain [Fig.~\ref{fig:iteration_in_s}] and in terms of the utility value $I(\Bs)$ [Fig.~\ref{fig:iteration_in_f}]}
\label{fig:iteration_convergence}
\end{figure}

Figure~\ref{fig:iteration_convergence} shows how Algorithm~\ref{alg1} (with energy boost and for fixed $T_\tau = 2$) converges to the jointly optimal solution for the utility function $f(\Bs) = I(\Bs)$. The parameters chosen in this simulation are $T = 10$, $\mu = 10$ (i.e., $10\dB$), $(N_\Tx, N_\Rx) = (2,2)$, and $(r_1,r_2) = (\frac{2}{3},\frac{1}{3})$.

\begin{figure}[htb]
\begin{center}
\subfigure[achievable rate as a function of the SNR]{
	\includegraphics[width=.45\columnwidth]{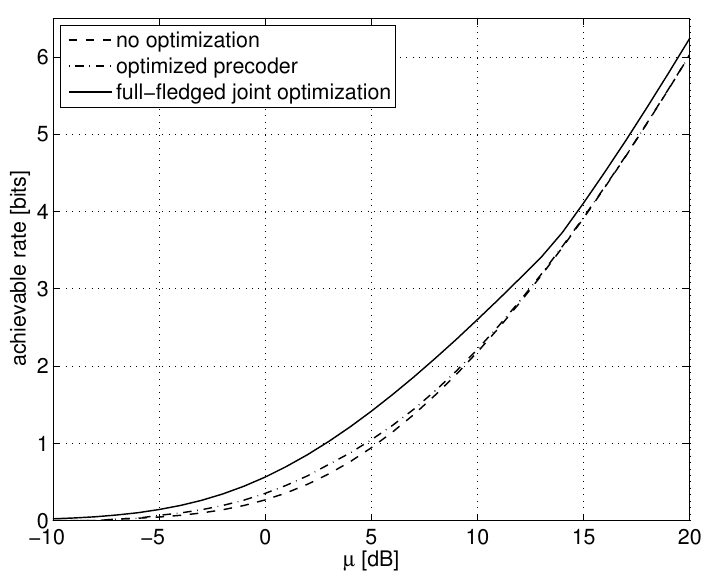}
	\label{fig:R_SNR}
}
\subfigure[achievable mean-square error as a function of the SNR]{
	\includegraphics[width=.45\columnwidth]{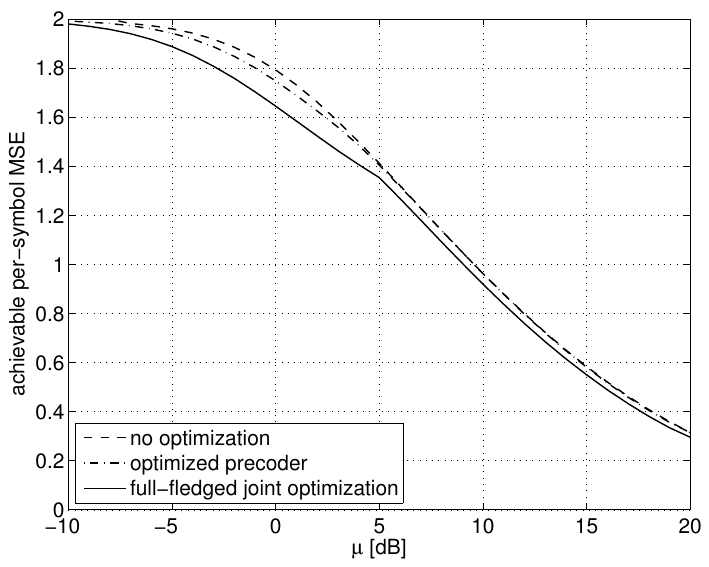}
	\label{fig:MMSE_SNR}
}
\end{center}
\caption{Two different utility functions against the SNR parameter $\mu$ for an examplary $2 \times 2$ MIMO system. For both utilities, the performance of full-fledged optimization is compared to a partial optimization (precoder only) and no optimization at all.}
\end{figure}

Figures~\ref{fig:R_SNR} and \ref{fig:MMSE_SNR} respectively show the quantities [cf.~\eqref{MMSE_upper}]
\begin{align}
	f_{\text{(a)}}(\Bs) &= \frac{T-T_\tau}{T}I(\Bs)
	&
	-f_{\text{(b)}}(\Bs) &= r - N_\Tx + \trace\Exp[(\myid+\hat{\BW}\BS\hat{\BW}^\He)^{-1}]
\end{align}
plotted against the SNR $\mu$ (in decibels) for the same $2 \times 2$ system as for Figure~\ref{fig:iteration_convergence}, i.e., $(r_1,r_2) = (\frac{2}{3},\frac{1}{3})$ and $T=10$. The former utility $f_{\text{(a)}}$ represents an achievable rate, since it is the mutual information $I(\Bs)$ weighted with a training overhead factor $\frac{T-T_\tau}{T}$ [see also formulation \eqref{joint_problem_full_fledged} of the full-fledged optimization problem]. The other utility $f_{\text{(b)}}$ is the negative of the bound on the per-symbol MMSE, derived in \eqref{MMSE_upper} [see also Utility~9 in Table~\ref{table:utilities}]. For this utility, we have fixed $r = T_\tau = 2$. In each of the two figures, three curves are plotted for comparison: the rate obtained with full-fledged joint optimization \eqref{joint_problem_full_fledged} computed with Algorithm~\ref{alg1} [including an exhaustive search over $T_\tau$ in the case of Figure~\ref{fig:R_SNR}, cf. Algorithm~\ref{alg3}]; the rate obtained in the case of precoder optimization alone; the rate obtained in the case of no optimization at all, i.e., $\BP_0 = \frac{T_\tau\mu}{N_\Tx}\myid_{N_\Tx}$ and $\BQ_0 = \frac{(T-T_\tau)\mu}{N_\Tx}\myid_{N_\Tx}$. The relative gains in mutual information are well noticeable especially for low and moderate SNR values. For higher SNR instead, these gains are far less significant. In fact, at high SNR, if the channel coherence $T$ is at least twice the number $N_\Tx$ of available transmit antennas, the optimal pilot-precoder pair tends toward the non-optimized pair $(\BP_0,\BQ_0)$.

\appendix
\subsection{Power monotonicity}   \label{app:proof:power_monotonicity}

We first prove that the function $(\BP,\BQ) \mapsto \BS(\BP,\BQ)$ is matrix-monotonic in the first argument, meaning that
\begin{align}   \label{strong_monotonicity}
	\mynull \preceq \BP \prec \BP'
	\Rightarrow \BS(\BP,\BQ) \prec \BS(\BP',\BQ).
\end{align}
Denote
\begin{align}
	\tilde{\BR} &= (\BR^{-1} + \BP)^{-1} &
	\tilde{\BR}' &= (\BR^{-1} + \BP')^{-1} \\
\intertext{and}
	\hat{\BR} &= \BR - \tilde{\BR} &
	\hat{\BR}' &= \BR - \tilde{\BR}'.
\end{align}
Then $\BP \prec \BP'$ obviously implies $\tilde{\BR} \succ \tilde{\BR}'$ and $\hat{\BR} \prec \hat{\BR}'$. This further implies $\trace(\BQ\tilde{\BR}) > \trace(\BQ\tilde{\BR}')$ and $\hat{\BR}^{\frac{1}{2}} \prec (\hat{\BR}')^{\frac{1}{2}}$, from which finally follows
\begin{align}
	\frac{\hat{\BR}^{\frac{1}{2}}\BQ\hat{\BR}^{\frac{1}{2}}}{1+\trace(\BQ\tilde{\BR})}
	\prec
	\frac{(\hat{\BR}')^{\frac{1}{2}}\BQ(\hat{\BR}')^{\frac{1}{2}}}{1+\trace(\BQ\tilde{\BR}')},
\end{align}
which is nothing else than $\BS(\BP,\BQ) \prec \BS(\BP',\BQ)$.
The monotonicity \eqref{strong_monotonicity} implies the weaker property
\begin{align}
	0 \leq k < k'
	\Rightarrow \BS(k\BP,\BQ) \prec \BS(k'\BP,\BQ).
\end{align}
Similarly we have
\begin{align}
	0 \leq k < k'
	\Rightarrow \BS(\BP,k\BQ) \prec \BS(\BP,k'\BQ).
\end{align}
This is because
\begin{align}
	\BS(\BP,k\BQ)
	&= \frac{k}{1+k\trace(\BQ\tilde{\BR})} \hat{\BR}^{\frac{1}{2}}\BQ\hat{\BR}^{\frac{1}{2}} \\
	&<
	\frac{k'}{1+k'\trace(\BQ\tilde{\BR})} \hat{\BR}^{\frac{1}{2}}\BQ\hat{\BR}^{\frac{1}{2}}
	= \BS(\BP,k'\BQ)
\end{align}
owing to the fact that $k \mapsto \frac{k}{1+k\trace(\BQ\tilde{\BR})}$ is monotonically increasing in $k$.

\subsection{Examples of utilities}   \label{app:examples_of_utilities}

\begin{table}[htb]
% increase table row spacing, adjust to taste
%\renewcommand{\arraystretch}{1.3}
% if using array.sty, it might be a good idea to tweak the value of
% \extrarowheight as needed to properly center the text within the cells
\caption{Examples of utilities from the class $\mathcal{F}$}
\label{table:utilities}
\centering
\begin{tabular}{|c|c|c|}
\hline
& Utility & Curvature in $\BS$ \\
\hline
\hline
1 & $I(\BS)$ & concave \\
\hline
2 & $\trace(\BS)$ & linear \\
\hline
3 & $\det(\BS)$ & log-concave \\
\hline
4 & $\trace(\BS^{-1})^{-1}$ & log-concave \\
\hline
5 & $\det(\myid + \nu \BS)$ with $\nu \geq 0$ & log-concave \\
\hline
\hline
6 & $\Exp\det(\myid + \hat{\BW}\BS\hat{\BW}^\He)$ & log-concave \\
\hline
7 & $\Exp\log\det(\hat{\BW}\BS\hat{\BW}^\He)$ for $N_\Tx \geq N_\Rx$ & concave \\
\hline
8 & $\Exp\det(\hat{\BW}\BS\hat{\BW}^\He)$ for $N_\Tx \geq N_\Rx$ & log-concave \\
\hline
9 & $-\trace\Exp\big\{ (\myid + \hat{\BW}\BS\hat{\BW}^\He)^{-1} \bigr\}$ & concave \\
\hline
10 & $\trace\Exp\big\{ (\BS^{-1} + \hat{\BW}^\He\hat{\BW})^{-1} \bigr\}$ for $\det(\BS) \neq 0$ & concave \\
\hline
\hline
11 & $\Prob\bigl(\det(\myid + \hat{\BW}\BS\hat{\BW}^\He) \geq \eta\bigr)$ & --/-- \\
\hline
12 & $\Prob\bigl(\log\det(\hat{\BW}\BS\hat{\BW}^\He) \geq \eta\bigr)$ for $N_\Tx \geq N_\Rx$ & --/-- \\
\hline
13 & $\Prob\bigl(\det(\hat{\BW}\BS\hat{\BW}^\He) \geq \eta\bigr)$ for $N_\Tx \geq N_\Rx$ & --/-- \\
\hline
14 & $\Prob\bigl(-\trace\big\{ (\myid + \hat{\BW}\BS\hat{\BW}^\He)^{-1} \bigr\} \geq \eta\bigr)$ & --/-- \\
\hline
15 & $\Prob\bigl(\trace\big\{ (\BS^{-1} + \hat{\BW}^\He\hat{\BW})^{-1} \bigr\} \geq \eta\bigr)$ & --/-- \\
\hline
\hline
16 & $\norm{\BS}_\Frob^2$ & convex \\
\hline
17 & $\lambda_\maximum(\BS)$ & convex \\
\hline
\end{tabular}
\end{table}

In the following, we provide a few examples illustrating from which bounds or approximations of the mutual information $I(\BS)$ the above utilities may arise.
\paragraph{Utility 2}
A simple upper bound on $I(\BS)$ is obtained using the fact that $\Exp[\hat{\BW}\BS\hat{\BW}^\He] = \trace(\BS) N_\Tx \myid$ and by applying Jensen's inequality to the concave $\log$-determinant:
\begin{align}   \label{concavity_upper_bound_1}
	I(\BS)
	&\leq \sum_{i=1}^{N_\Rx} \log(1 + \trace(\BS) N_\Tx).
\end{align}
\paragraph{Utility 5 with $\nu = N_\Tx N_\Rx$}
Using the determinant identity $\det(\myid+\BA\BB) = \det(\myid+\BB\BA)$ to write $I(\BS) = \Exp\log\det(\myid + \BS\hat{\BW}^\He\hat{\BW})$, and applying Jensen's inequality, we get another upper bound:
\begin{align}   \label{concavity_upper_bound_2}
	I(\BS)
	&\leq \log\det\left(\myid + N_\Tx N_\Rx \BS \right).
\end{align}
Here, we have used $\Exp[\hat{\BW}^\He\hat{\BW}] = N_\Tx N_\Rx$.
%The new objective is thus
%\begin{align}
%	\trace(\BS)
%	= \frac{\trace(\BQ\hat{\BR})}{1+\trace(\BQ\tilde{\BR})}
%	= \frac{\trace(\BQ\BA)}{\trace(\BQ\BB)}
%	= \frac{\Bq^\Tr\Balpha}{\Bq^\Tr\Bbeta}
%	%&= \frac{\sum_{i} q_i }{1+\sum_{i} q_i }.
%\end{align}
%where $\BA = \hat{\BR}$ and $\BB = \frac{1}{\Ptx}\myid+\tilde{\BR}$, and where the entries of $\Balpha$ and $\Bbeta$ are $\alpha_i = \bigl[\BU_\BQ^\He\BA\BU_\BQ\bigr]_{i,i}$ and $\beta_i = \bigl[\BU_\BQ^\He\BB\BU_\BQ]_{i,i}$, respectively. For fixed $\Balpha$ and $\Bbeta$ (i.e., for a fixed $\BP$), the optimal $\Bq$ has only one non-zero entry at the index $i$ corresponding to the largest fraction $\alpha_i/\beta_i$, except for possible multiplicities. Since this is true for every given $\BP$, the joint optimum has rank one (w.l.o.g?).
%\begin{align}
%	\max_{\mathcal{Q}} \trace(\BS) = \max_i \frac{\alpha_i}{\beta_i}.
%\end{align}
%Now, for such a fixed $\BQ$ of rank one, the pilot matrix $\BP$ maximizing $\trace(\BS)$ is the one minimizing $\trace(\BQ\tilde{\BR})$, since
%\begin{align}
%	\trace(\BS) = \frac{\trace(\BQ\BR) - \trace(\BQ\tilde{\BR})}{1 + \trace(\BQ\tilde{\BR})}.
%\end{align}

\paragraph{Utilities 6 and 11}
By applying Jensen's inequality to the concave $\log$ function, we get the upper bound
\begin{align}
	I(\BS)
	\leq \log\Exp[\det(\myid + \hat{\BW}\BS\hat{\BW}^\He)].
\end{align}

\paragraph{Utilities 3, 7 and 12}
We can lower bound $I(\BS)$ by removing the identity matrix inside the $\log$-determinant. Depending on the sizes of antenna arrays, this gives us a bound $I(\BS) \geq \underline{I}(\BS)$ with
\begin{align}
	\underline{I}(\BS)
	=
	\begin{cases}
		\Exp{\log\det\bigl(\hat{\BW}\BS\hat{\BW}^\He\bigr)} \quad \text{for $N_\Tx \geq N_\Rx$} \\
		\Exp{\log\det\bigl(\BS\hat{\BW}^\He\hat{\BW}\bigr)} \quad \text{for $N_\Tx \leq N_\Rx$},
	\end{cases}
\end{align}
The former case (i.e., $N_\Tx \geq N_\Rx$) justifies utilities 7 and 12. In the latter case (i.e., $N_\Tx \leq N_\Rx$), note that
\begin{align}
	\underline{I}(\BS)
	= \log\det(\BS) + \Exp{\log\det\bigl(\hat{\BW}^\He\hat{\BW}\bigr)}
\end{align}
leads to utility 3. Clearly, $\underline{I}(\BS)$ is good as an approximation of $I(\BS)$ at high SNR, and was used as such in \cite{GaYoCh00}, \cite{Gr02}. Let us also mention the tighter lower bound \cite{OyNaBoPa02}
\begin{align}   \label{Minkowski_lower}
	I(\BS)
	\geq N_\Rx \log\left( 1 + \exp\left( \frac{\log(\e)}{N_\Rx} \underline{I}(\BS) \right) \right),
\end{align}
the derivation of which makes use of the Minkowski inequality for determinants.

\subsection{Proof of Lemma~\ref{lem:linear_fractional}}   \label{proof:lem:linear_fractional}

Supposing we are in the first situation, i.e., $\BA$ has full column rank, then $\BA$ has a left pseudoinverse $\BA^\sharp = (\BA^\He\BA)^{-1}\BA^\He$ which can be used to define the inverse function $\varrho^{-1}$. Let $\BZ = \varrho(\BX)$ be the image of $\BX$. Given $\BZ$, one obtains $\BX$ by insulating it via left-multiplication with $\BA^\sharp$ and right-multiplication with $\BA^{\sharp\He}$, and appropriate scaling:
\begin{align}   \label{inverse_linear_fractional_1}
	\BA^\sharp\BZ\BA^{\sharp\He} (1+\trace(\BB\BX))
	= \BX
\end{align}
Left-multiplying \eqref{inverse_linear_fractional_1} with $\BB$ and taking the trace yields
\begin{align}
	\trace(\BB\BA^\sharp\BZ\BA^{\sharp\He})
	= \frac{\trace(\BB\BX)}{1 + \trace(\BB\BX)},
\end{align}
or equivalently,
\begin{align}   \label{inverse_linear_fractional_2}
	1+\trace(\BB\BX)
	= \frac{1}{1-\trace(\BB\BA^\sharp\BZ\BA^{\sharp\He})}
\end{align}
By combining \eqref{inverse_linear_fractional_1} with \eqref{inverse_linear_fractional_2}, we recover the pre-image $\BX = \varrho^{-1}(\BZ)$, and see that the inverse function $\varrho^{-1}$ is linear fractional with parameters $\BA^\sharp$ and $-\BA^{\sharp\He}\BB\BA^\sharp$.

We now suppose that we are in the second situation, i.e., $\BA$ has full row rank and $\mathcal{X}$ is such that $\range(\BX) = \range(\BA^\He)$ for every element of $\mathcal{X}$. Due to the latter constraint on the span of $\BX$, we can write any $\BX \in \mathcal{X}$ as $\BX = \BA^\He\hat{\BX}\BA$, with $\hat{\BX}$ given by the inverse relation $\hat{\BX} = \BA^{\flat\He}\BX\BA^\flat$, where $\BA^\flat = \BA^\He(\BA\BA^\He)^{-1}$ denotes the right pseudoinverse of $\BA$. The function $\varrho$ can be represented as
\begin{align}
	\varrho \colon \ 
%	\hat{\mathcal{X}} \ &\to \ \hat{\varrho}(\hat{\mathcal{X}}), \ 
	\BX \ \mapsto \ \frac{\BA\BX\BA^\He}{1 + \trace(\BB\BX)} = \frac{\hat{\BA}\hat{\BX}\hat{\BA}^\He}{1 + \trace(\hat{\BB}\hat{\BX})}
\end{align}
with abbreviations $\BA\BA^\He \triangleq \hat{\BA}$ and $\hat{\BB} \triangleq \BA\BB\BA^\He$. Since $\hat{\BA} = \BA\BA^\He$ has full rank (because $\BA$ has full row rank), the function $\varrho$ appears as an injective linear fractional function of $\hat{\BX}$ with parameters $\hat{\BA}$ and $\hat{\BB}$, whose inverse, according to findings above, is linear fractional with parameters $\hat{\BA}^\sharp$ and $-\hat{\BA}^{\sharp\He}\hat{\BB}\hat{\BA}^\sharp$. Denoting as $\BZ = \varrho(\BX)$ the image of $\BX$ under the function $\varrho$, we can thus recover the pre-image $\BX$ from $\BZ$ as
\begin{align}
	\BX
	= \BA^\He\hat{\BX}\BA
	&= \BA^\He \frac{\hat{\BA}^\sharp \BZ \hat{\BA}^{\sharp\He}}{1 - \trace(\hat{\BA}^{\sharp\He}\BB\hat{\BA}^\sharp\BZ)} \BA \nonumber\\
	&= \frac{\BA^\flat \BZ \BA^{\flat\He}}{1 - \trace(\BA^{\flat\He}\BB\BA^\flat\BZ)}.
\end{align}
Consequently, the inverse $\varrho^{-1}$ is linear fractional with parameters $\BA^\flat$ and $-\BA^{\flat\He}\BB\BA^\flat$.

\subsection{Proof of Theorem~\ref{thm:Q_rank}}   \label{proof:thm:Q_rank}

Let $\BP$ and $\BQ$ have $\rank(\BP) = r_\BP$ and $\rank(\BQ) = r_\BQ$ respectively, with $r_\BP,r_\BQ \in \{1,\dotsc,N_\Tx\}$. The absolute difference of ranks be $d = | r_\BP - r_\BQ |$. The pilot matrix and precoder have reduced spectral decompositions $\BP = \BU_\BP\BLambda_\BP\BU_\BP^\He$ and $\BQ = \BU_\BQ\BLambda_\BQ\BU_\BQ^\He$, respectively, where the eigenbases $\BU_\BP \in \mathbb{U}^{N_\Tx \times r_\BP}$ and $\BU_\BQ \in \mathbb{U}^{N_\Tx \times r_\BQ}$ are tall or square, whereas $\BLambda_\BP$ and $\BLambda_\BQ$ are diagonal and positive definite. Let $\BU_{\BP^\bot} \in \mathbb{U}^{N_\Tx \times (N_\Tx-r_\BP)}$ and $\BU_{\BQ^\bot} \in \mathbb{U}^{N_\Tx \times (N_\Tx-r_\BQ)}$ denote orthonormal bases of the nullspaces of $\BP$ and $\BQ$, respectively, so that $\BU_\BP^\He\BU_{\BP^\bot} = \mynull$ and $\BU_\BQ^\He\BU_{\BQ^\bot} = \mynull$.

The reduced eigendecomposition of $\hat{\BR}$ is consistently denoted as $\BU_{\hat{\BR}}\BLambda_{\hat{\BR}}\BU_{\hat{\BR}}^\He$, where $\BLambda_{\hat{\BR}} \in \mathbb{R}_+^{r_\BP \times r_\BP}$ is diagonal positive definite and of size $r_\BP \times r_\BP$, due to the rank equality \eqref{rank_equality} which states that $\rank(\hat{\BR}) = \rank(\BP)$. The orthonormal nullspace of $\hat{\BR}$ is denoted as $\BU_{\hat{\BR}^\bot} \in \mathbb{U}^{N_\Tx \times (N_\Tx-r_\BP)}$. We introduce the notation $\BL_{\BA \cap \BB}$ to designate an orthonormal basis of the intersection of range spaces $\range(\BA)$ and $\range(\BB)$. If it exists, $\BL_{\BA \cap \BB}$ is a matrix with the maximum number of columns defined as
\begin{align}   \label{L}
	\BL_{\BA \cap \BB} = \left\{ \BL \ \middle|
	\begin{array}{l}
		\BL^\He\BL = \myid, \\
		\forall \Bx \neq \mynull \colon \BA\BL\Bx \neq \mynull \text{ and } \BB\BL\Bx \neq \mynull
	\end{array}
	\right\}.
\end{align}

Assume that $\range(\BQ) \nsubseteq \range(\hat{\BR})$, so the matrix $\BL_{\BQ \cap \hat{\BR}^\bot}$ is defined and has at least one column. We define a new precoder $\BQ' \in \mathcal{Q}$ as
\begin{align}
	\BQ' = \BQ - \lambda_{r_\BQ}(\BQ) \BL_{\BQ \cap \hat{\BR}^\bot} \BL_{\BQ \cap \hat{\BR}^\bot}^\He,
\end{align}
where $\lambda_{r_\BQ}(\BQ)$ is the smallest non-zero eigenvalue of $\BQ$. First, we verify that $\BQ' \in \mathcal{Q}$. Clearly, since $\lambda_{r_\BQ}(\BQ) \geq 0$, we have $\BQ' \preceq \BQ$, and therefore $\trace(\BQ') \leq \trace(\BQ)$. What remains to prove is that $\BQ' \succeq \mynull$. The smallest eigenvalue of $\BQ'$ is
\begin{align*}
	\lambda_\minimum(\BQ')
	= \min_{\norm{\Bw} = 1} \Bw^\He \BQ' \Bw.
\end{align*}
But since by definition of $\BL_{\BQ \cap \hat{\BR}}$, the range space of $\BQ$ contains the range space of $\BL_{\BQ \cap \hat{\BR}} \BL_{\BQ \cap \hat{\BR}}^\He$, we have that $\Bw^\He\BQ'\Bw$ is equal to $\Bw^\He\BPi_\BQ^\He\BQ'\BPi_\BQ\Bw$, where $\BPi_\BQ = \BU_\BQ(\BU_\BQ^\He\BU_\BQ)^{-1}\BU_\BQ^\He$ is the projector from $\mathbb{C}^{N_\Tx \times N_\Tx}$ onto the basis $\BU_\BQ$. We thus have
%\begin{align}   %adjust
%	\lambda_\minimum(\BQ')
%	&= \lambda_\minimum \left( \BPi_\BQ^\He \Bigl( \BQ - \lambda_{r_\BQ}(\BQ) \BL_{\BQ \cap \hat{\BR}^\bot} \BL_{\BQ \cap \hat{\BR}^\bot}^\He \Bigr) \BPi_\BQ \right) \nonumber\\
%	&\geq \lambda_\minimum \Bigl( \BPi_\BQ^\He \BQ \BPi_\BQ \Bigr) \nonumber\\
%	&\qquad\qquad {} - \lambda_{r_\BQ}(\BQ) \lambda_\maximum \Bigl( \BPi_\BQ^\He \BL_{\BQ \cap \hat{\BR}^\bot} \BL_{\BQ \cap \hat{\BR}^\bot}^\He \BPi_\BQ \Bigr) \nonumber\\
%	&\geq \lambda_{r_\BQ}(\BQ) \Bigl( 1 - \lambda_\maximum \bigl( \BPi_\BQ^\He \BPi_\BQ  \bigr) \lambda_\maximum \bigl( \BL_{\BQ \cap \hat{\BR}^\bot}^\He \BL_{\BQ \cap \hat{\BR}^\bot} \bigr) \Bigr) \nonumber\\
%	&= 0.
%\end{align}
\begin{align}
	\lambda_\minimum(\BQ')
	&= \lambda_\minimum \left( \BPi_\BQ^\He \Bigl( \BQ - \lambda_{r_\BQ}(\BQ) \BL_{\BQ \cap \hat{\BR}^\bot} \BL_{\BQ \cap \hat{\BR}^\bot}^\He \Bigr) \BPi_\BQ \right) \nonumber\\
	&\geq \lambda_\minimum \Bigl( \BPi_\BQ^\He \BQ \BPi_\BQ \Bigr) - \lambda_{r_\BQ}(\BQ) \lambda_\maximum \Bigl( \BPi_\BQ^\He \BL_{\BQ \cap \hat{\BR}^\bot} \BL_{\BQ \cap \hat{\BR}^\bot}^\He \BPi_\BQ \Bigr) \nonumber\\
	&\geq \lambda_{r_\BQ}(\BQ) \Bigl( 1 - \lambda_\maximum \bigl( \BPi_\BQ^\He \BPi_\BQ  \bigr) \lambda_\maximum \bigl( \BL_{\BQ \cap \hat{\BR}^\bot}^\He \BL_{\BQ \cap \hat{\BR}^\bot} \bigr) \Bigr) \nonumber\\
	&= 0.
\end{align}
The second inequality holds because the spectral radius norm $\lambda_\maximum(\cdot)$ is sub-multiplicative, while the last equality holds because the projector $\BPi_\BQ$ and the (sub)unitary $\BL_{\BQ \cap \hat{\BR}^\bot}$ have a largest singular value of at most $1$. We infer that $\BQ' \in \mathcal{Q}$.

Notice that $\BQ'$ is purposely constructed so that $\BQ'\hat{\BR} = \BQ\hat{\BR}$. As compared to the matrix $\BS = \BS(\BP,\BQ)$ obtained with the precoder $\BQ$, the new matrix $\BS' = \BS(\BP,\BQ')$ thus reads as
\begin{align}
	\BS'
	&= \frac{\hat{\BR}^{\frac{1}{2}} \BQ' \hat{\BR}^{\frac{1}{2}}}{1 + \trace(\BQ'\tilde{\BR})} \nonumber\\
	&= \frac{\hat{\BR}^{\frac{1}{2}} \BQ \hat{\BR}^{\frac{1}{2}}}{1 + \trace(\BQ\tilde{\BR}) - \lambda_{r_\BQ}(\BQ) \trace\bigl(\BL_{\BQ \cap \hat{\BR}^\bot}^\He\tilde{\BR}\BL_{\BQ \cap \hat{\BR}^\bot}\bigr)} \nonumber\\
	&= k\BS
\end{align}
and thus turns out to be a scaled version of $\BS$, where the positive scalar $k$ is
\begin{align*}
	k = \frac{1 + \trace(\BQ\tilde{\BR})}{1 + \trace(\BQ\tilde{\BR}) - \lambda_{r_\BQ}(\BQ) \trace\bigl(\BL_{\BQ \cap \hat{\BR}^\bot}^\He\tilde{\BR}\BL_{\BQ \cap \hat{\BR}^\bot}\bigr)} > 1.
\end{align*}
Therefore, we have $\BS' \succ \BS$, so the precoder $\BQ$ is necessarily suboptimal, which means that $\range(\BQ) \nsubseteq \range(\hat{\BR})$ cannot hold for optimal $\BQ$. Instead, we must have $\range(\BQ) \subseteq \range(\hat{\BR})$ for optimality, which concludes the proof.

\subsection{Proof of Lemma~\ref{thm:prescribed_pilot_region}}   \label{proof:thm:prescribed_pilot_region}

As a consequence of Theorem~\ref{thm:Q_rank} and of the power monotonicity \eqref{power_monotonicity}, optimal precoders $\BQ$ will be elements of $\partial^+\mathcal{Q} \cap \range(\hat{\BR})$. By definition, this set can be parametrized by $r_\BP$ non-negative coefficients $\Bpsi = [\psi_1,\dotsc,\psi_{r_\BP}]^\Tr \in \mathbb{R}_+^{r_\BP}$ stored in a diagonal matrix $\BPsi = \diag(\Bpsi)$, and a tall or square (sub)unitary basis $\BUpsilon \in \mathbb{U}^{N_\Tx \times r_\BP}$ as follows:
\begin{align}   \label{Q_parametrization}
	\BQ_{\BPsi,\BUpsilon} = \bigl(\hat{\BR}^{\frac{1}{2}}\bigr)^+ \BUpsilon \BPsi \BUpsilon^\He \bigl(\hat{\BR}^{\frac{1}{2}}\bigr)^+,
\end{align}
where $(\bullet)^+$ denotes the Moore-Penrose pseudoinverse (cf. Section~\ref{sec:notation}), and where the parameter pair $(\BPsi,\BUpsilon)$ shall be subject to the four constraints
\begin{subequations}   \label{parametrization_constraints}
\begin{align}
	\Bpsi &\geq \mynull,   \label{first_parametrization_constraint} \\
	\trace(\BQ_{\BPsi,\BUpsilon}) &= \mu_\mathcal{Q},   \label{second_parametrization_constraint} \\
	\BUpsilon^\He\BUpsilon &= \myid,   \label{third_parametrization_constraint} \\
	\range(\BUpsilon) &= \range(\hat{\BR}).   \label{fourth_parametrization_constraint}
\end{align}
\end{subequations}
The first two constraints ensure that $\BQ_{\BPsi,\BUpsilon}$ belongs to $\partial^+\mathcal{Q}$, while the structure of Expression~\eqref{Q_parametrization} ensures that $\BQ_{\BPsi,\BUpsilon}$ belongs to $\range(\hat{\BR})$. The third and fourth constraints \eqref{third_parametrization_constraint}--\eqref{fourth_parametrization_constraint} are clearly not necessary to fulfill $\BQ_{\BPsi,\BUpsilon} \in \range(\hat{\BR}) \cap \partial^+\mathcal{Q}$, but they induce no loss of generality either and will turn out helpful later. The set $\partial^+\mathcal{Q}$ is thus entirely parametrized by the parameter pair $(\BPsi,\BUpsilon)$ subject to the constraints \eqref{parametrization_constraints}. Consider now the feasible vectors
\begin{align}
	\Bs(\BP,\BQ_{\BPsi,\BUpsilon})
	&= \Blambda\left(\frac{ \hat{\BR}^{\frac{1}{2}} \BQ_{\BPsi,\BUpsilon} \hat{\BR}^{\frac{1}{2}}}{1+\trace(\BQ_{\BPsi,\BUpsilon}\tilde{\BR})} \right) \nonumber\\
	&= \Blambda\left(\frac{ \BU_{\hat{\BR}}\BU_{\hat{\BR}}^\He \BUpsilon\BPsi\BUpsilon^\He \BU_{\hat{\BR}}\BU_{\hat{\BR}}^\He }{1+\trace(\BQ_{\BPsi,\BUpsilon}\tilde{\BR})} \right) \in \mathbb{R}_+^{N_\Tx}.
\end{align}
This vector has at most $r_\BP$ non-zero entries because $\BPsi$ is $r_\BP \times r_\BP$. Therefore, we define a vector $\bar{\Bs} \in \mathbb{R}_+^{r_\BP}$ of reduced dimension, which contains the $r_\BP$ topmost (i.e., largest) entries of $\Bs$. Since $\range(\BUpsilon)=\range(\hat{\BR})$ [cf.~\eqref{fourth_parametrization_constraint}], the matrix $\BU_{\hat{\BR}}^\He\BUpsilon$ is unitary, so we have
\begin{align}
	\bar{\Bs}(\BP,\BQ_{\BPsi,\BUpsilon})
	&= \frac{\Bpsi}{1+\trace(\BQ_{\BPsi,\BUpsilon}\tilde{\BR})} \nonumber\\
	&= \frac{\Bpsi}{1 + \trace\bigl(\BUpsilon^\He\bigl(\hat{\BR}^{\frac{1}{2}}\bigr)^+\tilde{\BR}\bigl(\hat{\BR}^{\frac{1}{2}}\bigr)^+\BUpsilon \BPsi\bigr)}.
\end{align}
Note that we have not assumed so far that the entries of $\Bpsi$ or $\bar{\Bs}$ are sorted in any specific order. For notational brevity, call $\Balpha$ the vector of entries $\alpha_i = \bigl[\BUpsilon^\He\bigl(\hat{\BR}^{\frac{1}{2}}\bigr)^+\tilde{\BR}\bigl(\hat{\BR}^{\frac{1}{2}}\bigr)^+\BUpsilon\bigr]_{i,i}$, then
\begin{align}   \label{varphi_parametrized}
	\bar{\Bs}(\BP,\BQ_{\BPsi,\BUpsilon}) =
	\frac{\Bpsi}{1 + \Balpha^\Tr\Bpsi}.
\end{align}
On the other hand, the second constraint \eqref{second_parametrization_constraint} translates to $\Bbeta^\Tr\Bpsi = \mu_\mathcal{Q}$, where $\Bbeta$ denotes the vector of diagonal entries of $\BUpsilon^\He\hat{\BR}^+\BUpsilon$, i.e., $\beta_i = [\BUpsilon^\He\hat{\BR}^+\BUpsilon]_{i,i}$. Together, this constraint and equation \eqref{varphi_parametrized} describe an affine plane of dimension $r_\BP-1$, because left-multiplying \eqref{varphi_parametrized} with $\frac{1}{\mu_\mathcal{Q}}\Bbeta^\Tr + \Balpha^\Tr$ leads to the affine equation
\begin{align}
	\left(\frac{1}{\mu_\mathcal{Q}}\Bbeta + \Balpha\right)^\Tr \bar{\Bs}(\BP,\BQ_{\BPsi,\BUpsilon})
	= 1.
\end{align}
%\begin{align}
%	\omega_i
%	&= \frac{1}{\mu_\mathcal{Q}} \beta_{\pi(i)} + \alpha_{\pi(i)} \nonumber\\
%	&= \bigl[ \BUpsilon^\He \bigl(\hat{\BR}^{\frac{1}{2}}\bigr)^+ \bigl(\mu_\mathcal{Q}^{-1}\myid + \tilde{\BR}\bigr) \bigl(\hat{\BR}^{\frac{1}{2}}\bigr)^+ \BUpsilon \bigr]_{\pi(i),\pi(i)}
%\end{align}
%where $\pi$ is a sorting permutation such that $\omega_1 \geq \dotso \geq \omega_{r_\BP}$. For a given value of $\BUpsilon$, the set of feasible vectors $\bar{\Bs}$ is determined by the plane equation \eqref{plane_equation} and by the constraint that $\bar{s}_i$ are sorted non-increasingly, i.e.,
%\begin{align}
%	\sum_{i=1}^{r_\BP} \omega_{\pi^{-1}(i)} \bar{s}_i &= 1 \nonumber\\
%	\bar{s}_1 \geq \dotso \geq \bar{s}_{r_\BP} &\geq 0.
%\end{align}
%\vspace{1mm}
%\hrule\hrule
%\hrule\hrule
%\vspace{1mm}
This affine equation, together with the non-negativity constraint $\Bpsi \geq \mynull$ [cf.~\eqref{first_parametrization_constraint}], thus delimit a $(r_\BP-1)$-dimensional simplex, whose elements fulfill
\begin{align}   \label{linear_equation}
	\left\{ \begin{array}{l}
		\displaystyle \sum_i \frac{\bar{s}_i}{\omega_i} = 1 \\
		\bar{\Bs} \geq \mynull
	\end{array} \right.
\end{align}
where
\begin{align}   \label{omega_i}
	\omega_i = \frac{1}{\frac{1}{\mu_\mathcal{Q}}\beta_i + \alpha_i}.
\end{align}
The $r_\BP$ vertices of the simplex described by \eqref{linear_equation} are the axis points $\omega_i \Be_i$.

Due to the symmetry property of utilities from the class $\mathcal{F}$, the ordering of the $\bar{s}_i$ does not influence the utility value. Assume that, for a given $\bar{\Bs} \geq \mynull$ fulfilling \eqref{linear_equation}, there exists an index permutation $\pi$ such that $\sum_i \frac{\bar{s}_{\pi(i)}}{\omega_i} < 1$, then $\bar{\Bs}$ is suboptimal, since there exists an $\bar{\Bs}' = k\BPi\bar{\Bs}$ with $k>1$ which also fulfills \eqref{linear_equation} and yields a larger utility value $f(\bar{\Bs}') = f(k\BPi\bar{\Bs}) = f(k\bar{\Bs}) > f(\bar{\Bs})$. Therefore, we can discard all $\bar{\Bs}$ for which some permutation $\pi$ yields $\sum_i \frac{\bar{s}_{\pi(i)}}{\omega_i} < 1$. This is equivalent to the requirement that the $\bar{s}_i$ and $\omega_i$ be ordered in the same way, i.e.,
\begin{align}
	\omega_i \leq \omega_j \Rightarrow \bar{s}_i \leq \bar{s}_j.
\end{align}
Hence, without loss of optimality, we will restrain the set of admissible $\bar{\Bs}$ to the following convex set, called $\mathcal{S}$:
\begin{align}   \label{set_S}
	\mathcal{S} = \left\{ \bar{\Bs} \in \mathbb{R}_+^{r_\BP} \ \middle| \ \sum_{i=1}^{r_\BP} \frac{\bar{s}_i}{\bar{\omega}_i} = 1, \forall j \colon \bar{s}_j \geq \bar{s}_{j+1} \right\},
\end{align}
where $\bar{\Bomega} = [\bar{\omega}_1,\dotsc,\bar{\omega}_{r_\BP}]^\Tr$ contains the entries of $\Bomega$ arranged in non-increasing order, i.e., $\bar{\omega}_1 \geq \dotso \geq \bar{\omega}_{r_\BP}$. Let us define $r_\BP$ special points pertaining to $\mathcal{S}$, which we shall denote as $\Bsigma^{(n)}$, and define as
\begin{align}   \label{vertices}
	\forall n = 1,\dotsc,r_\BP \colon \quad
	\Bsigma^{(n)}
	= \mathcal{H}(\bar{\omega}_1,\dotsc,\bar{\omega}_n) \sum_{j=1}^n \Be_j,
\end{align}
where $\mathcal{H}(\cdot,\dotsc,\cdot)$ and $\Be_j$ are defined in the statement of Lemma~\ref{thm:prescribed_pilot_region}. In fact, it is easy to see that the $\Bsigma^{(n)}$ have non-increasing entries and fulfill $\sum_{i=1}^{r_\BP} \frac{\sigma_i^{(n)}}{\bar{\omega}_i} = 1$, and thus belong to $\mathcal{S}$.
Now, we will show that the set $\mathcal{S}'$ of all convex combinations of the $\Bsigma^{(n)}$, i.e,
\begin{align}
	\mathcal{S}' = \left\{ \sum_{n=1}^{r_\BP} \nu_n \Bsigma^{(n)} \middle| \sum_n \nu_n = 1, \forall n \colon \nu_n \geq 0 \right\},
\end{align}
is the same as the set $\mathcal{S}$. We know that $\mathcal{S}'$ is a subset of the convex set $\mathcal{S}$, for being a convex combination of a collection of points from $\mathcal{S}$, hence $\mathcal{S}' \subseteq \mathcal{S}$. Next, we argue that, if we assume that some particular point $\tilde{\Bsigma}$ belongs to $\mathcal{S}\setminus\mathcal{S}'$, this implies that $\tilde{\Bsigma}$ does not lie in $\mathcal{S}$ because it would fail to comply with some constraint from the definition \eqref{set_S} of $\mathcal{S}$. Therefrom, it will follow that $\mathcal{S} = \mathcal{S}'$.

Since the $\Bsigma^{(n)}$ pertain to $\mathcal{S}$ and are $r_\BP$ linearly independent vectors, they define the $(r_\BP-1)$-dimensional affine plane described by $\sum_{i=1}^{r_\BP} \frac{\bar{s}_i}{\bar{\omega}_i} = 1$. Therefore, to prove the equality $\mathcal{S}=\mathcal{S}'$, it will be sufficient to take some point $\tilde{\Bsigma} = [\tilde{\sigma}_1,\dotsc,\tilde{\sigma}_{r_\BP}]^\Tr$ to lie on said plane, and show that an infringement of an inequality $\tilde{\sigma}_i \geq \tilde{\sigma}_{i+1}$ implies that $\tilde{\Bsigma} = \sum_{n=1}^{r_\BP} \tilde{\nu}_n \Bsigma^{(n)}$ with coefficients $\tilde{\nu}_n$ such that either $\sum_n \tilde{\nu}_n \neq 1$ or $\tilde{\nu}_n < 0$ for some index $n$. So, assume that $\tilde{\sigma}_i < \tilde{\sigma}_{i+1}$ for a given $i$. There exist unique coefficients $\tilde{\nu}_n$ such that $\tilde{\Bsigma} = \sum_{n=1}^{r_\BP} \tilde{\nu}_n \Bsigma^{(n)}$. The inequality $\tilde{\sigma}_i < \tilde{\sigma}_{i+1}$ can thus be written as
\begin{align}
	\sum_{n=1}^{r_\BP} \tilde{\nu}_n \Be_i^\Tr\Bsigma^{(n)}
	< \sum_{n=1}^{r_\BP} \tilde{\nu}_n \Be_{i+1}^\Tr\Bsigma^{(n)}.
\end{align}
By inserting \eqref{vertices} into the latter inequality, we get
\begin{align}
	\sum_{n=i}^{r_\BP} \tilde{\nu}_n \mathcal{H}(\bar{\omega}_1,\dotsc,\bar{\omega}_n)
	< \sum_{n=i+1}^{r_\BP} \tilde{\nu}_n \mathcal{H}(\bar{\omega}_1,\dotsc,\bar{\omega}_n),
\end{align}
which boils down to $\tilde{\nu}_i < 0$. This concludes the proof that $\mathcal{S}=\mathcal{S}'$. Also, this simplex $\mathcal{S}$ contains only Pareto border points, in the sense that $\mathcal{S} = \partial^+\mathcal{S}$. In fact, any point $\bar{\Bs}''$ dominating some point $\bar{\Bs}' \in \mathcal{S}$ would fulfill $\sum_{i=1}^{r_\BP} \bar{s}_i''/\bar{\omega}_i > 1$ and thus lie outside $\mathcal{S}$.

Now that we have fully characterized the set of Pareto border points $\bar{\Bs} = \bar{\Bs}(\BP,\BQ_{\BPsi,\BUpsilon})$ for a fixed $\BUpsilon$ as a simplex set $\mathcal{S}$, we ask what the best choice for $\BUpsilon$ is under the constraints \eqref{third_parametrization_constraint}--\eqref{fourth_parametrization_constraint}. Clearly, if there exists one single $\BUpsilon^\star$ that simultaneously maximizes all vertices $\Bsigma^{(n)}$ in the sense that for any $\BUpsilon$, we have
\begin{align}
	\Bsigma^{(n)}(\BUpsilon^\star) \geq \Bsigma^{(n)}(\BUpsilon), \qquad n = 1,\dotsc,r_\BP
\end{align}
then this $\BUpsilon^\star$ is optimal. Here, $\Bsigma^{(n)}(\BUpsilon)$ denotes the value of $\Bsigma^{(n)}$, as defined in \eqref{vertices}, with the $\bar{\omega}_i$ interpreted as functions of $\BUpsilon$. Next, we show that such $\BUpsilon^\star$ is well-defined and characterize it.

We state the multiobjective optimization problem
\begin{align}   \label{moo}
	\forall n = 1,\dotsc,r_\BP \colon \max_{\substack{\BUpsilon \in \mathbb{U}^{N_\Tx \times r_\BP} \\ \range(\BUpsilon) = \range(\hat{\BR})}} \mathcal{H}(\bar{\omega}_1,\dotsc,\bar{\omega}_n).
\end{align}
Omitting the range space constraint on $\BUpsilon$, we have that, with the definition \eqref{omega_i} of the coefficients $\omega_i$ together with the definitions of $\alpha_i$ and $\beta_i$, this multiobjective problem reads as
\begin{align}   \label{moo_2}
	\forall n \colon \min_{\BUpsilon \in \mathbb{U}^{N_\Tx \times r_\BP}} \sum_{i=1}^n \textstyle \Bigl[ \BUpsilon^\He \bigl(\hat{\BR}^{\frac{1}{2}}\bigr)^+ \bigl( \frac{\myid}{\mu_\mathcal{Q}} + \tilde{\BR} \bigr) \bigl(\hat{\BR}^{\frac{1}{2}}\bigr)^+ \BUpsilon \Bigr]_{\pi(i),\pi(i)},
\end{align}
where $\pi$ denotes the permutation which orders the diagonal entries of the matrix between square brackets so as to be non-decreasingly ordered. If $\BW\BD\BW^\He$ denotes the spectral decomposition of $\bigl(\hat{\BR}^{\frac{1}{2}}\bigr)^+ \bigl( \frac{\myid}{\mu_\mathcal{Q}} + \tilde{\BR} \bigr) \bigl(\hat{\BR}^{\frac{1}{2}}\bigr)^+$ where $\BW \in \mathbb{U}^{N_\Tx \times r_\BP}$, and $\BD$ has non-increasingly ordered, positive diagonal entries, then it is well known from majorization theory that the solution of \eqref{moo_2} is $\BUpsilon^\star = \BW$, up to a column permutation (e.g., \cite[Theorem~4.3.26]{HoJo90}). It turns out as well that $\range(\BUpsilon^\star) = \range(\BW) = \range(\hat{\BR})$, so the range space constraint is systematically fulfilled. The columns of $\BUpsilon^\star$ contain the eigenvectors $\Bupsilon_i$ of the generalized eigenvalue problem
\begin{align}
	\hat{\BR}\Bupsilon_i = \omega_i \bigl( \tfrac{1}{\mu_\mathcal{Q}}\myid + \tilde{\BR} \bigr) \Bupsilon_i,
\end{align}
corresponding to the $r_\BP$ largest generalized eigenvalues $\omega_i$.

\subsection{Proof of Theorem~\ref{thm:P_rank}}   \label{proof:thm:P_rank}

Assume that $r_\BP > r_\BQ$. Similarly as for the proof of Theorem~\ref{thm:Q_rank} in Appendix~\ref{proof:thm:Q_rank}, we will proceed by constructing another pilot matrix in $\mathcal{P}$ which strictly outperforms $\BP$. Recall that the covariance of the channel estimate is $\hat{\BR} = \BR-\tilde{\BR}$, as usual, and $\tilde{\BR} = (\BR^{-1} + \BP)^{-1}$ is the estimation error covariance. We construct $\BP'$ as
\begin{align}
	\BP' = \Bigl[ \tilde{\BR} + \lambda_{r_\BP}(\hat{\BR})\BL_{\BQ^\bot \cap \hat{\BR}}\BL_{\BQ^\bot \cap \hat{\BR}}^\He \Bigr]^{-1} - \BR^{-1}.
\end{align}
The subunitary matrix $\BL_{\BQ^\bot \cap \hat{\BR}}$ is defined as in the proof of Theorem~\ref{thm:Q_rank}. It exists and has at least $d = r_\BP - r_\BQ$ columns.
First, we verify that $\BP' \in \mathcal{P}$. In fact, $\BP'$ can be written out as
\begin{align}
	\BP' = \Bigl[ \BR - \hat{\BR} + \lambda_{r_\BP}(\hat{\BR}) \BL_{\BQ^\bot \cap \hat{\BR}} \BL_{\BQ^\bot \cap \hat{\BR}}^\He \Bigr]^{-1} - \BR^{-1},
\end{align}
where it becomes clear that $\BP' \succeq \mynull$, because $\hat{\BR} - \lambda_{r_\BP}(\hat{\BR}) \BL_{\BQ^\bot \cap \hat{\BR}} \BL_{\BQ^\bot \cap \hat{\BR}}^\He \succeq \mynull$. On the other hand, the trace of $\BP'$ is upper-bounded as
\begin{align}
	\trace(\BP')
	&= \trace \left( \Bigl[ \tilde{\BR} + \lambda_{r_\BP}(\hat{\BR}) \BL_{\BQ^\bot \cap \hat{\BR}} \BL_{\BQ^\bot \cap \hat{\BR}}^\He \Bigr]^{-1} \right) - \trace(\BR^{-1}) \nonumber\\
	&< \trace (\tilde{\BR}^{-1}) - \trace(\BR^{-1}) \nonumber\\
	&= \trace (\BP).
\end{align}
If we write $\tilde{\BR} = \tilde{\BR}(\BP)$ to stress that it is essentially a function of $\BP$, then we notice that $\BP'$ is designed so as to leave the product
\begin{align}
	\BQ \tilde{\BR}(\BP')
	&= \BQ \bigl( \tilde{\BR} + \lambda_{r_\BP}(\hat{\BR})\BL_{\BQ^\bot \cap \hat{\BR}}\BL_{\BQ^\bot \cap \hat{\BR}}^\He \bigr) \nonumber\\
	&= \BQ\tilde{\BR}(\BP)
\end{align}
unchanged, irrespective of whether the pilots are $\BP$ or $\BP'$. The same is true for $\Bs = \Blambda(\BS)$, which is left unchanged when replacing $\BP$ by $\BP'$, because $\Bs$ depends on $\BP$ only via the product $\BQ \tilde{\BR}(\BP)$, as seen from the relationship
\begin{align}
	\Bs
	&= \frac{\Blambda\bigl( \hat{\BR}^{\frac{1}{2}}\BQ\hat{\BR}^{\frac{1}{2}} \bigr)}{1 + \trace(\BQ\tilde{\BR})}
	= \frac{\Blambda\bigl( \BQ\BR - \BQ\tilde{\BR} \bigr)}{1 + \trace(\BQ\tilde{\BR})}.
\end{align}
We have thus constructed alternative pilots $\BP'$ which yield the same utility value $f(\Bs)$, yet saving on the training energy, since $\trace(\BP') < \trace(\BP)$. We generate another pilot matrix $\BP'' = \kappa \BP'$ with $\kappa = \trace(\BP)/\trace(\BP')$. The new pilots $\BP''$ spend the same amount of training energy as $\BP$, but yield a strictly larger $\BS'' = \BS(\BP'',\BQ) \succ \BS(\BP,\BQ)$. Hence, $\BP$ is suboptimal.

\subsection{Convexity of the set of feasible $\hat{\BR}$}   \label{app:R_hat_convexity}

Showing the convexity of the set of feasible $\hat{\BR}$ is equivalent to showing the convexity of the set of feasible $\tilde{\BR}$, because $\hat{\BR} = \BR - \tilde{\BR}$ is merely $\tilde{\BR}$ scaled with $-1$ and summed with a constant matrix $\BR$.
Therefore, we show that the set
\begin{align}
	\bigl\{\tilde{\BR} = (\BR^{-1} + \BP)^{-1} \big| \BP \in \mathcal{P} \bigr\}
\end{align}
is convex. For any pair $(\BP_1,\BP_2) \in \mathcal{P}^2$, there exists a $\BP_3 \in \mathcal{P}$ and a $\alpha \in [0;1]$ such that
\begin{align}   \label{R_tilde_convexity}
	\alpha \tilde{\BR}_1 + (1-\alpha) \tilde{\BR}_2 = \tilde{\BR}_3,
\end{align}
where $\tilde{\BR}_i = (\BR^{-1} + \BP_i)^{-1}$ for $i = 1,2,3$. By isolating $\BP_3$ in \eqref{R_tilde_convexity}, the pilot Gram $\BP_3$ is given by
\begin{align}
	\BP_3
	= \bigl[ \alpha \tilde{\BR}_1 + (1-\alpha) \tilde{\BR}_2 \bigr]^{-1} - \BR^{-1}.
\end{align}
Obviously, since $\tilde{\BR}_i \preceq \BR$ for $i=1,2$, we have $\BP_3 \succeq \mynull$. What remains to prove is that $\trace(\BP_3) \leq \mu_\mathcal{P}$. Knowing that the function $\BX \mapsto \trace(\BX^{-1})$ is convex on the positive cone $\BX \succ \mynull$, we have
\begin{align}
	\trace(\BP_3)
	&\leq \alpha \trace(\tilde{\BR}_1^{-1}) + (1-\alpha) \trace(\tilde{\BR}_2^{-1}) - \trace(\BR^{-1}) \nonumber\\
	&= \alpha \trace(\BP_1) + (1-\alpha) \trace(\BP_2) \nonumber\\
	&\leq \mu_\mathcal{P} .
\end{align}
Hence, the set of feasible $\hat{\BR}$ is convex, and so is Problem \eqref{marginal_problem_P}.

\subsection{Proof of Theorem~\ref{thm:joint_basis}}   \label{app:proof:thm:joint_basis}

%We will show that for Problem~\eqref{joint_problem_decomposed_revisited_inner}, there is no loss of optimality in setting
%\begin{align}
%	\col(\BU_\BP)
%	= \col(\BU_\BQ)
%	= \{\Bu_{\BR,1},\dotsc,\Bu_{\BR,r^\star}\}
%	\subseteq \col(\BU_\BR),
%\end{align}
%where $r^\star = \rank(\BP^\star) = \rank(\BQ^\star)$ denotes the pilot/precoder rank at the joint optimum $(\BP^\star,\BQ^\star)$ of Problem~\eqref{joint_problem_decomposed_revisited_inner}.

We will proceed by showing that, in Problem \eqref{joint_problem_decomposed_revisited_inner}, for any given value of the pair $(\mu_\mathcal{P},\mu_\mathcal{Q})$, the search set $\Bs(\mathcal{P},\mathcal{Q})$---and thus its Pareto border $\partial^+\Bs(\mathcal{P},\mathcal{Q})$---is left unchanged whether we allow $(\BP,\BQ)$ to take {\em any} value within $\mathcal{P} \times \mathcal{Q}$, or whether we restrict the choice of the basis $\BU_\BP$ such that $\col(\BU_\BP) = \{\Bu_{\BR,1},\dotsc,\Bu_{\BR,r^\star}\}$, where $r^\star$ denotes the number of non-zero entries of the $\Bs^\star$ . With a consequence of Theorem~\ref{thm:prescribed_pilot_region}, we will eventually conclude on the desired result $\col(\BU_\BP) = \col(\BU_\BQ) = \{\Bu_{\BR,1},\dotsc,\Bu_{\BR,r^\star}\}$.

To begin with, note that the set $\Bs(\mathcal{P},\mathcal{Q})$ can be represented as the union
\begin{align}
	\Bs(\mathcal{P},\mathcal{Q}) = \bigcup_{\BP \in \mathcal{P}} \Bs(\BP,\mathcal{Q}).
\end{align}
As a consequence of the rank equality \eqref{rank_equality}, the elements of $\Bs(\BP,\mathcal{Q})$ have at most $r_\BP$ non-zero entries, because $\rank(\BS) = \rank(\hat{\BR}^{\frac{1}{2}}\BQ\hat{\BR}^{\frac{1}{2}}) \leq \rank(\hat{\BR}) = \rank(\BP) = r_\BP$. They can thus be written as $\Bs(\BP,\BQ) = \bigl[ \bar{\Bs}(\BP,\BQ)^\Tr \ \ \mynull^\Tr \bigr]^\Tr$ with $\bar{\Bs}(\BP,\BQ) \in \mathbb{R}_+^{r_\BP}$ of reduced size. According to Theorem~\ref{thm:prescribed_pilot_region}, this set of reduced-size vectors $\bar{\Bs}(\BP,\mathcal{Q})$ is the simplex given by the convex hull of the points
\begin{align}
	\Bsigma^{(0)}
	&= \mynull
	&
	\Bsigma^{(n)}
	&= \mathcal{H}(\omega_1,\dotsc,\omega_n) \sum_{j=1}^n \Be_j,
	\quad n \in \{1,\dotsc,r_\BP\}.
\end{align}
Every such simplex is entirely described by $\Bomega = [\omega_1,\dotsc,\omega_{r_\BP}]^\Tr$, the vector non-increasingly ordered eigenvalues of the matrix $\hat{\BR}\bigl(\mu_\mathcal{Q}^{-1}\myid + \tilde{\BR}\bigr)^{-1}$, which is a function of $\BP$ alone (not of $\BQ$). Consistently with the notation used so far, $\Bomega(\mathcal{P})$ shall denote the set of feasible $\Bomega$ given that $\BP$ belongs to $\mathcal{P}$. To prove Theorem~\ref{thm:joint_basis}, we will first show that the set of Pareto border points $\partial^+\Bomega(\mathcal{P})$ is still achievable under the restriction $\col(\BU_\BP) \subseteq \{\Bu_{\BR,1},\dotsc,\Bu_{\BR,r^\star}\}$. Recalling that $\BR = \hat{\BR} + \tilde{\BR}$ and $\tilde{\BR} = (\BR^{-1}+\BP)^{-1}$, we write out $\Bomega$ as
\begin{align*}
	\Bomega
	&= \Blambda\left(\hat{\BR}\bigl(\tfrac{1}{\mu_\mathcal{Q}}\myid + \tilde{\BR}\bigr)^{-1}\right) \nonumber\\
	&= \Blambda\left( \Bigl(\BR - (\BR^{-1}+\BP)^{-1}\Bigr) \Bigl(\tfrac{1}{\mu_\mathcal{Q}}\myid + (\BR^{-1}+\BP)^{-1}\Bigr)^{-1}\right).
\end{align*}
Let us denote $\BP' = \BR^{\frac{1}{2}}\BP\BR^{\frac{1}{2}}$, then using the property $\Blambda(\BA\BB) = \Blambda(\BB\BA)$, the last expression can be rewritten as
\begin{align*}
	\Bomega
	&= \Blambda\left( \Bigl(\myid - (\myid+\BP')^{-1}\Bigr) \Bigl((\mu_\mathcal{Q}\BR)^{-1} + (\myid+\BP')^{-1}\Bigr)^{-1}\right).
\end{align*}
Let us now denote $\BP'' = \BU_\BR^\He\BP'\BU_\BR$, so that the last expression becomes
\begin{align}   \label{omega_with_Xi}
	\Bomega
	&= \Blambda\left( \Bigl(\myid - (\myid+\BP'')^{-1}\Bigr) \Bigl((\mu_\mathcal{Q}\BLambda_\BR)^{-1} + (\myid+\BP'')^{-1}\Bigr)^{-1}\right).
%	&= \Blambda\left( \BP'' \Bigl((\mu_\mathcal{Q}\BLambda_\BR)^{-1}(\myid+\BP'') + \myid\Bigr)^{-1}\right) \nonumber\\
%	&= \Blambda\left( \BP'' \Bigl((\mu_\mathcal{Q}\BLambda_\BR)^{-1}\BP'' + \BXi \Bigr)^{-1}\right)
\end{align}
%with $\BXi = (\mu_\mathcal{Q}\BLambda_\BR)^{-1} + \myid$.
%(\texttt{Wir hatten einen Ausdruck dieser Bauart an der Tafel hergeleitet. Er ist auch deutlich h\"ubscher als der Ausdruck in der ersten Zeile in Gleichung \eqref{omega_with_Xi}. Leider gelingt es mir aber nicht, ihn effizient f\"ur den Beweis einzusetzen, weil er nicht von der Form $\Blambda(\BA\BB)$ mit zwei positiv definiten Matrizen $\BA$ und $\BB$ ist. Diese spezielle Struktur, die in der ersten Gleichungszeile noch erhalten ist, verschwindet in der zweiten und dritten. Ich ben\"otige sie aber f\"ur Gleichung \eqref{successive_maximizations}.})
Let us write out the mutual relations linking $\BP$ and $\BP''$ in full:
\begin{subequations}
\begin{align}
	\BP'' &= \BLambda_\BR^{\frac{1}{2}}\BU_\BR^\He\BU_\BP\diag(\Bp)\BU_\BP^\He\BU_\BR\BLambda_\BR^{\frac{1}{2}}   \label{P_double_prime_function_of_P} \\
	\BP &= \BU_\BR\BLambda_\BR^{-\frac{1}{2}}\BU_{\BP''}\diag(\Bp'')\BU_{\BP''}^\He\BLambda_\BR^{-\frac{1}{2}}\BU_\BR^\He.   \label{P_function_of_P_double_prime}
\end{align}
\end{subequations}
Regarding the (non-reduced) eigendecomposition $\BP'' = \BU_{\BP''}\BLambda_{\BP''}\BU_{\BP''}^\He$ with $\BU_{\BP''} \in \mathbb{U}^{N_\Tx \times N_\Tx}$ and $\BLambda_{\BP''} = \diag(\Bp'') = \diag(p_1'',\dotsc,p_{N_\Tx}'')$, one can say that, if $\BP$ is drawn from $\mathcal{P}$, then the corresponding eigenvalue profile $\Bp'' = \Blambda(\BP'') = \Blambda\bigl(\BLambda_{\BR}^{\frac{1}{2}} \BU_\BR^\He \BP \BU_\BR \BLambda_{\BR}^{\frac{1}{2}}\bigr) = \Blambda(\BP\BR)$ [cf.~\eqref{P_double_prime_function_of_P}] is drawn from a feasible set which we shall call $\Bp''(\mathcal{P})$, a notation which emphasizes its direct dependence on the domain $\mathcal{P}$. As to the eigenbasis $\BU_{\BP''}$, it obviously belongs to $\mathbb{U}^{N_\Tx \times N_\Tx}$ by definition, yet in general, we must presume that not all pairs $(\Bp'',\BU_{\BP''}) \in \Bp''(\mathcal{P}) \times \mathbb{U}^{N_\Tx \times N_\Tx}$ are jointly feasible, since the eigenbasis $\BU_{\BP''}$ and the eigenvalues $\Bp''$ cannot be chosen independently of each other, due to the special structure of Expression~\eqref{P_double_prime_function_of_P}. Instead, $\BU_\BP''$ belongs to a feasible set $\BU_{\BP''}(\Bp'') \subseteq \mathbb{U}^{N_\Tx \times N_\Tx}$ (which depends on $\Bp''$), so the overall set of feasible pairs $(\Bp'',\BU_{\BP''})$ forms a {\em subset} of the Cartesian product $\Bp''(\mathcal{P}) \times \mathbb{U}^{N_\Tx \times N_\Tx}$.
%Hence, the set of feasible pairs $(\Bp'',\BU_{\BP''})$ is
%\begin{align}
%	\bigcup_{\Bp'' \in \Bp''(\mathcal{P})} (\Bp'',\BU_{\BP''}(\Bp''))
%\end{align}
%and the set $\Bomega(\mathcal{P})$ may be written out as a double union [cf.~\eqref{conditioned_feasible_set_identity}]
%\begin{align}   \label{double_union}
%	\Bomega(\mathcal{P})
%	&= \bigcup_{\Bp'' \in \Bp''(\mathcal{P})} \quad \bigcup_{\BU_{\BP''} \in \BU_{\BP''}(\Bp'')} \Bomega(\Bp'',\BU_{\BP''}(\Bp'')) \nonumber\\
%	&= \bigcup_{\Bp'' \in \Bp''(\mathcal{P})} \Bomega\bigl(\Bp'',\BU_{\BP''}(\Bp'')\bigr),
%\end{align}
%wherein [cf.~\eqref{P_function_of_P_double_prime}]
%\begin{align}
%	\Bomega(\Bp'',\BU_{\BP''})
%	\triangleq \Bomega( \BU_\BR\BLambda_\BR^{-\frac{1}{2}}\BU_{\BP''}\diag(\Bp'')\BU_{\BP''}^\He\BLambda_\BR^{-\frac{1}{2}}\BU_\BR^\He ).
%\end{align}

However, suppose for a while that $\Bp''$ and $\BU_{\BP''}$ can be drawn {\em independently} of each other from their respective domains $\Bp''(\mathcal{P})$ and $\mathbb{U}^{N_\Tx \times N_\Tx}$. This assumption then corresponds to a {\em relaxation} of the original problem, as it possibly extends the overall set of feasible $\BP''$, and consequently, of feasible $\Bomega$. The resulting set of achievable vectors $\Bomega$ under this relaxation shall be denoted $\bar{\Bomega}(\mathcal{P}) \supseteq \Bomega(\mathcal{P})$ and is formally defined as
\begin{align}   \label{relaxation}
	\bar{\Bomega}(\mathcal{P})
	= \left\{ \Bomega(\Bp'',\BU_{\BP''}) \middle|
	(\Bp'',\BU_{\BP''}) \in \Bp''(\mathcal{P}) \times \mathbb{U}^{N_\Tx \times N_\Tx}
	\right\},
\end{align}
wherein the two-argument notation $\Bomega(\bullet,\bullet)$ is defined as [cf.~\eqref{P_function_of_P_double_prime}]
\begin{align}   \label{two_argument_notation}
	\Bomega(\Bp'',\BU_{\BP''})
	\triangleq \Bomega( \BU_\BR\BLambda_\BR^{-\frac{1}{2}}\BU_{\BP''}\diag(\Bp'')\BU_{\BP''}^\He\BLambda_\BR^{-\frac{1}{2}}\BU_\BR^\He ).
\end{align}
The set $\bar{\Bomega}(\mathcal{P})$ can be represented as a double union
%We will eventually prove that this relaxation does not actually increase the feasible set of simplices
%{\color{red} (\texttt{
%Diese Aussage hat sich leider als \textbf{falsch} herausgestellt. Diese Relaxation vergr\"oßert sowohl die Dom\"ane $\Bomega$ als auch die Dom\"ane der erreichbaren Simplizes. Schade!
%})
%}
%{\color{blue} (\texttt{
%Alles was jetzt in blau kommt, ist nun leider M\"ull.
%})
%, which means that $\Bomega(\mathcal{P}) = \bar{\Bomega}(\mathcal{P})$.
\begin{align}   \label{double_union_relaxed}
	\bar{\Bomega}(\mathcal{P})
	&= \bigcup_{\Bp'' \in \Bp''(\mathcal{P})} \quad \bigcup_{\BU_{\BP''} \in \mathbb{U}^{N_\Tx \times N_\Tx}} \Bomega(\Bp'',\BU_{\BP''}) \nonumber\\
	&= \bigcup_{\Bp'' \in \Bp''(\mathcal{P})} \Bomega\bigl(\Bp'',\mathbb{U}^{N_\Tx \times N_\Tx}\bigr).
\end{align}
As seen from expression \eqref{omega_with_Xi}, $\Bomega(\Bp'',\BU_{\BP''})$ is monotonic in the eigenvalues $p_i''$, meaning that
\begin{align*}
	\forall \Bd \geq \mynull \colon
	\Bomega(\Bp'' + \Bd,\BU_{\BP''}) \geq \Bomega(\Bp'',\BU_{\BP''}).
\end{align*}
Hence, since we are essentially interested in the Pareto border $\partial^+\bar{\Bomega}(\mathcal{P})$ of the set $\bar{\Bomega}(\mathcal{P})$, we can restrict our further analysis to the set\footnote{Note that $\bar{\Bomega}^+(\mathcal{P})$ is generally not the Pareto border of $\bar{\Bomega}(\mathcal{P})$, but rather a superset thereof.}
\begin{align}   \label{double_union_pareto}
	\bar{\Bomega}^+(\mathcal{P})
%	= \bigcup_{\Bp'' \in \partial^+\Bp''(\mathcal{P})} \quad \bigcup_{\BU_{\BP''} \in \mathbb{U}^{N_\Tx \times N_\Tx}} \Bomega(\Bp'',\BU_{\BP''}).
	&= \bigcup_{\Bp'' \in \partial^+\Bp''(\mathcal{P})} \Bomega\bigl(\Bp'',\mathbb{U}^{N_\Tx \times N_\Tx}\bigr).
\end{align}

The remainder of the proof of Theorem~\ref{thm:joint_basis} is completed in four successive steps, each of which is detailed in a separate paragraph, for the sake of a clearer structure: first, we specify a method for constructing a particular Pareto border point of the set $\Bomega(\Bp'',\mathbb{U}^{N_\Tx \times N_\Tx})$ given a particular value of the vector $\Bp''$, where we show that this construction requires the alignment $\col(\BU_\BP) \subseteq \col(\BU_\BR)$; second, we show that the point constructed this way, besides yielding a Pareto border point of the relaxed set $\bar{\Bomega}(\mathcal{P})$, is also contained in the smaller (non-relaxed) set $\Bomega(\mathcal{P})$, so it must be a Pareto border point of $\Bomega(\mathcal{P})$ as well; third, we show that, by varying the eigenvalues $\Bp''$ over the feasible set $\Bp''(\mathcal{P})$, with the aforementioned method of constructing particular Pareto border points, we reach the whole Pareto border of $\Bomega(\mathcal{P})$; fourth, we show that the alignment $\col(\BU_\BP) \subseteq \col(\BU_\BR)$ implies that $\BU_\BQ$ must as well be aligned such that $\col(\BU_\BQ) = \col(\BU_\BP)$ to reach the whole feasible set $\Bs(\mathcal{P},\mathcal{Q})$, and conclude.

\indent {\itshape 1)} %{-\baselineskip} %{0.5\baselineskip}
Let the orthonormal eigenbasis $\BU_{\BP''} = [\Bu_1,\dotsc,\Bu_{N_\Tx}]$ be spanned by unit vectors $\Bu_i$, where the $i$-th vector $\Bu_i$ is associated to the $i$-th largest eigenvalue $p_i''$.
%We intend to show that
%\begin{align}
%	\bigcup_{\BU_{\BP''} \in \mathbb{U}^{N_\Tx \times N_\Tx}} \Bomega(\BP)
%	= \bigcup_{\BU_{\BP''} \in \mathbb{P}_{N_\Tx}} \Bomega(\BP).
%\end{align}
Given a fixed value of $\Bp'' \in \Bp''(\mathcal{P})$, we construct a particular point of the Pareto border $\partial^+\Bomega(\Bp'',\mathbb{U}^{N_\Tx \times N_\Tx})$ by solving the sequence of optimization problems:
\begin{align}   \label{successive_maximization_1}
	\forall i \in \{1,\dotsc,N_\Tx\} &\colon& \BU^{(i)} &= \argmax_{\BU \in \mathbb{U}^{N_\Tx \times N_\Tx}} \omega_i\bigl(\Bp'',\BU\bigr) \nonumber\\
	\text{s.t. } \ \forall \ell \in \{1,\dotsc,i-1\} &\colon& \quad \omega_\ell &= \omega_\ell\bigl(\Bp'',\BU^{(i-1)}\bigr).
\end{align}
Clearly, $\BU^{(N_\Tx)}$ will yield a Pareto optimal point, that is,
\begin{align}
	\Bomega(\Bp'',\BU^{(N_\Tx)}) \in \partial^+\Bomega(\Bp'',\mathbb{U}^{N_\Tx \times N_\Tx}).
\end{align}
%This is because the sequence of optimization problems \eqref{successive_maximization_1} constructs a sequence of bases $\BU^{(i)} \in \mathbb{U}^{N_\Tx \times N_\Tx}$ such that the last basis $\BU^{(N_\Tx)}$ simultaneously maximizes of $\omega_1$, the largest possible value of $\omega_2$, etc.
Next, we will show by induction that $\BU^{(N_\Tx)} = \myid$. For this purpose, let us explicitly solve the first problem ($i=1$) of \eqref{successive_maximization_1}, i.e.,
\begin{align}
	\BU^{(1)}
	= \argmax_{\BU \in \mathbb{U}^{N_\Tx \times N_\Tx}} \omega_1\bigl(\Bp'',\BU\bigr)
\end{align}
With Expression~\eqref{omega_with_Xi}, this reads as
\begin{align}   \label{1_optimization}
	&\max_{\BU \in \mathbb{U}^{N_\Tx \times N_\Tx}} \max_{\norm{\Bv_1}=1} \left[ \frac{\Bv_1^\He \bigl(\myid - (\myid+\BU\BLambda_{\BP''}\BU^\He)^{-1}\bigr) \Bv_1}{\Bv_1^\He \bigl((\mu_\mathcal{Q}\BLambda_\BR)^{-1} + (\myid+\BU\BLambda_{\BP''}\BU^\He)^{-1}\bigr) \Bv_1} \right] \nonumber\\
	&\leq \max_{\BU \in \mathbb{U}^{N_\Tx \times N_\Tx}} \left[ \frac{ 1 - \lambda_\minimum\bigl((\myid+\BU\BLambda_{\BP''}\BU^\He)^{-1}\bigr) }{ \lambda_\minimum\bigl((\mu_\mathcal{Q}\BLambda_\BR)^{-1}\bigr) + \lambda_\minimum\bigl((\myid+\BU\BLambda_{\BP''}\BU^\He)^{-1}\bigr) } \right] \nonumber\\
	&= \frac{ 1 - \frac{1}{1+\lambda_\maximum(\BLambda_{\BP''})} }{ (\mu_\mathcal{Q}\lambda_\maximum(\BLambda_\BR))^{-1} + \frac{1}{1+\lambda_\maximum(\BLambda_{\BP''})} } \nonumber\\
	&= \frac{ 1 - \frac{1}{1+p_1''} }{ (\mu_\mathcal{Q}r_1)^{-1} + \frac{1}{1+p_1''} }.
\end{align}
This upper bound is tight and achieved if and only if $\Bv_1 = \Be_1$, and when $\BU$ is of the form
\begin{align}   \label{1_solution}
	\BU^{(1)}
	&=
	\begin{bmatrix}
		1       & \mynull \\
		\mynull & \BW^{(1)}
	\end{bmatrix}
\end{align}
with some arbitrary $\BW^{(1)} \in \mathbb{U}^{(N_\Tx-1) \times (N_\Tx-1)}$. To prove the induction step, we will show that if for a certain $i \geq 1$, all maximizers $\BU^{(i)}$ are of the form
\begin{align}   \label{U_structure}
	\BU^{(i)}
	&=
	\begin{bmatrix}
		\myid_i & \mynull \\
		\mynull & \BW^{(i)}
	\end{bmatrix}
\end{align}
with some arbitrary $\BW^{(i)} \in \mathbb{U}^{(N_\Tx-i) \times (N_\Tx-i)}$, and $\forall \ell = 1,\dotsc,i \colon \Bv_\ell = \Be_\ell$, then $\BU^{(i+1)}$ also has the above block structure \eqref{U_structure}, with an identity matrix $\myid_{i+1}$ top left and an arbitrary rotation matrix $\BW^{(i+1)}$ bottom right.
After solving the $i$-th problem, we know that all solutions thereof are of the form \eqref{U_structure}, which implies that the equality constraints for the $(i+1)$-th problem [cf.~\eqref{successive_maximization_1}] can only be fulfilled if $\BU^{(i+1)}$ has the same structure as $\BU^{(i)}$, i.e.,
\begin{align}
	\BU^{(i+1)}
	&=
	\begin{bmatrix}
		\myid_i & \mynull \\
		\mynull & \tilde{\BW}^{(i)}
	\end{bmatrix}
\end{align}
with some unitary matrix $\tilde{\BW}^{(i)} \in \mathbb{U}^{(N_\Tx-i) \times (N_\Tx-i)}$ to be determined. According to a straightforward adaptation of the Courant-Fisher Theorem \cite[Theorem~4.2.11]{HoJo90}, the non-increasingly ordered eigenvalues $\lambda_i(\BA\BB^{-1})$ with corresponding eigenvectors $\Bv_i$ of a product of two Hermitian matrices $\BA$ and $\BB^{-1}$ can be expressed as
\begin{align}
	\lambda_i(\BA\BB^{-1}) = \max_{\Bv \perp \BB\Bv_{i-1},\dotsc,\BB\Bv_1} \frac{\Bv^\He\BA\Bv}{\Bv^\He\BB\Bv}.
\end{align}
The $(i+1)$-th optimization problem reads as
\begin{align}   \label{i+1_optimization}   %adjust
	\BU^{(i+1)}
	&= \argmax_{\BU \in \mathbb{U}^{N_\Tx \times N_\Tx}} \left\{ \max_{\Bv_{i+1} \perp \BB\Bv_i,\dotsc,\BB\Bv_1} \frac{\Bv_{i+1}^\He\BA\Bv_{i+1}}{\Bv_{i+1}^\He\BB\Bv_{i+1}} \right\}
	\qquad\qquad \text{s.t.} \quad
	\BU
	=
	\begin{bmatrix}
		\myid_i & \mynull \\
		\mynull & \tilde{\BW}^{(i)}
	\end{bmatrix}
\end{align}
with $\BA = \myid - (\myid+\BU\BLambda_{\BP''}\BU^\He)^{-1}$ and $\BB = (\mu_\mathcal{Q}\BLambda_\BR)^{-1} + (\myid+\BU\BLambda_{\BP''}\BU^\He)^{-1}$ [cf.~\eqref{omega_with_Xi}], and $\forall \ell = 1,\dotsc,i \colon \Bv_\ell = \Be_\ell$.
When writing out $\BB$, the vectors involved in the orthogonality constraints $\Bv_i \perp \BB\Be_{i-1},\dotsc,\BB\Be_1$ read as
\begin{align}
	\BB\Be_\ell
	&= \Bigl[ (\mu_\mathcal{Q}\BLambda_\BR)^{-1} + (\myid+\BU\BLambda_{\BP''}\BU^\He)^{-1} \Bigr] \Be_\ell \nonumber\\
	&= \frac{\Be_\ell}{\mu_\mathcal{Q}r_\ell} + \begin{bmatrix} \bigl(\myid \! + \! \BLambda_{\BP''}^{[i]}\bigr)^{\!-1} & \!\!\!\! \mynull \\ \mynull & \!\!\!\! \tilde{\BW}^{(i)} \bigl(\myid \! + \! \bar{\BLambda}_{\BP''}^{[i]}\bigr)^{\!-1} \bigl(\tilde{\BW}^{(i)}\bigr)^{\!\He} \end{bmatrix} \! \Be_\ell \nonumber\\
	&= \Bigl[ \frac{1}{\mu_\mathcal{Q}r_\ell} + \frac{1}{1+p_\ell''} \Bigr] \Be_\ell
	\qquad \forall \ell = 1,\dotsc,i
\end{align}
where $\BLambda_{\BP''}^{[i]} = \diag(p_1'',\dotsc,p_i'')$ and $\bar{\BLambda}_{\BP''}^{[i]} = \diag(p_{i+1}'',\dotsc,p_{N_\Tx}'')$. Thus, the orthogonality constraints simply translate into $\Bv_{i+1} \perp \Be_i,\dotsc,\Be_1$. In other terms, the first $i$ entries of $\Bv_{i+1}$ must be zero. Thus, we can define matrices $(N_\Tx-i) \times (N_\Tx-i)$ matrices $\breve{\BA}$ and $\breve{\BB}$ as
\begin{align}
	\breve{\BA} &= \myid - \bigl(\myid + \tilde{\BW}^{(i)}\bar{\BLambda}_{\BP''}^{[i]}\bigl(\tilde{\BW}^{(i)}\bigr)^\He \bigr)^{-1} \nonumber\\
	\breve{\BB} &= (\mu_\mathcal{Q}\bar{\BLambda}_\BR^{(i)})^{-1} + \bigl(\myid + \tilde{\BW}^{(i)}\bar{\BLambda}_{\BP''}^{[i]}\bigl(\tilde{\BW}^{(i)}\bigr)^\He \bigr)^{-1},
\end{align}
so that the optimization problem \eqref{i+1_optimization} boils down to solving
\begin{align}   \label{i+1_optimization_2}
	\tilde{\BW}^{(i)}
	&= \argmax_{\BW \in \mathbb{U}^{(N_\Tx-i) \times (N_\Tx-i)}} \left\{ \max_{\breve{\Bv}_{i+1}} \frac{\breve{\Bv}_{i+1}^\He\breve{\BA}\breve{\Bv}_{i+1}}{\breve{\Bv}_{i+1}^\He\breve{\BB}\breve{\Bv}_{i+1}} \right\}.
\end{align}
This problem is fully equivalent in structure to the first optimization problem ($i=1$) as written out in Equation~\eqref{1_optimization} and has the same solution, i.e., [cf.~\eqref{1_solution}]
\begin{align}
	\tilde{\BW}^{(i)}
	&=
	\begin{bmatrix}
		1       & \mynull \\
		\mynull & \BW^{(i+1)}
	\end{bmatrix}.
\end{align}
Consequently, $\BU^{(i+1)}$ has indeed the structure \eqref{U_structure}, which concludes the induction proof. We infer that $\BU^{(N_\Tx)} = \myid$, and thus
\begin{align}
	\Bomega(\Bp'',\myid) \in \partial^+\Bomega(\Bp'',\mathbb{U}^{N_\Tx \times N_\Tx})
	\subset \bar{\Bomega}^+(\mathcal{P}).
\end{align}
Thus, we have specified a method to construct specific Pareto optimal points of the inner union in \eqref{double_union_relaxed}.

\indent {\itshape 2)}
Recalling how $\BP''$ is obtained from $\BP \in \mathcal{P}$, namely [cf.~\eqref{P_double_prime_function_of_P}]
\begin{align*}
	\BP'' &= \BLambda_\BR^{\frac{1}{2}}\BU_\BR^\He\BU_\BP\diag(\Bp)\BU_\BP^\He\BU_\BR\BLambda_\BR^{\frac{1}{2}},
\end{align*}
we can leverage Theorem~\ref{thm:prescribed_pilot_region} (although with other variables) to characterize the set $\Bp''(\mathcal{P})$ of vectors of feasible, non-increasingly sorted eigenvalues of the above matrix. First note that $\BP''$ has the same eigenvalues as $\BU_\BR\BP''\BU_\BR^\He$, so that the set $\Bp''(\mathcal{P})$ may be defined as [compare with \eqref{simplex_set_written_out}]
\begin{align}
	\Bp''(\mathcal{P})
	= \left\{ \Blambda\left( \BR^{\frac{1}{2}}\BP\BR^{\frac{1}{2}} \right) \ \middle| \ \BP \in \mathbb{C}_+^{N_\Tx \times N_\Tx}, \ \trace(\BP) \leq \mu_\mathcal{P} \right\}
\end{align}
Now, Theorem~\ref{thm:prescribed_pilot_region} can be applied upon replacing $\hat{\BR}$, $\tilde{\BR}$, $\BQ$, $\mathcal{Q}$ and $\mu_\mathcal{Q}$ (as they appear in the formulation of said theorem) with $\BR$, $\mynull$, $\BP$, $\mathcal{P}$ and $\mu_\mathcal{P}$ respectively. This leads to $\Bp''(\mathcal{P})$ being characterized as the convex hull of the points ${\Bsigma''}^{(n)}, n=0,\dotsc,N_\Tx$ defined as
\begin{align}
	{\Bsigma''}^{(0)} &= \mynull
	&
	{\Bsigma''}^{(n)} &= \mu_\mathcal{P} \cdot \mathcal{H}(r_1,\dotsc,r_n) \sum_{\ell=1}^n \Be_\ell, \quad n = 1,\dotsc,N_\Tx.
\end{align}
It can be readily verified that all points of this convex hull can be reached when setting $\col(\BU_\BP) \subseteq \col(\BU_\BR)$.
%We then have $\Bp'' = \Bp \odot \Br$, where the entries of $\Bp$ and $\Br$ are arranged such that the entries of $\Bp''$ are in non-increasing order.
When doing so, the eigenbasis of $\BP''$ is precisely $\BU_{\BP''} = \myid$. But remember that the choice $\BU_{\BP''} = \myid$ was required in the previous paragraph for constructing a Pareto optimal point of $\partial^+\Bomega(\Bp'',\mathbb{U}^{N_\Tx \times N_\Tx})$. Consequently, this Pareto optimal point is also contained in the subset $\Bomega(\Bp'',\BU_{\BP''}(\Bp'')) \subseteq \Bomega(\Bp'',\mathbb{U}^{N_\Tx \times N_\Tx})$, and is thus necessarily a Pareto optimal point of $\Bomega(\Bp'',\BU_{\BP''}(\Bp''))$, too.
%What is more, if $\Bp'' \in \partial^+\Bp''(\mathcal{P})$, then $\Bomega(\Bp'',\myid)$ lies on the Pareto boundary not only of $\Bomega(\Bp'',\BU_{\BP''}(\Bp''))$, but also of the overall set (union) $\Bomega(\mathcal{P}) = \bigcup_{\Bp'' \in \Bp''({\mathcal{P}})} \Bomega(\Bp'',\BU_{\BP''}(\Bp''))$.

\indent {\itshape 3)}
We now ask whether all points of the overall Pareto border $\partial^+\Bomega(\mathcal{P})$ are attained by the construction method specified above, i.e., whether
\begin{align}
	\partial^+\Bomega(\mathcal{P}) \subseteq \bigcup_{\Bp'' \in \partial^+\Bp''(\mathcal{P})} \Bomega(\Bp'',\myid).
\end{align}
%To each point $\Bp''$ from the $(r_\BP-1)$-simplex $\partial^+\Bp''(\mathcal{P})$, we associate one point $\Bomega(\Bp'',\myid)$ lying on the Pareto boundary $\partial^+\Bomega(\mathcal{P})$ by means of the method proposed above [cf.~\eqref{successive_maximization_1}--\eqref{successive_maximization_2}], which requires $\BU_\BP = \BU_\BR$ and thus $\BU_{\BP''} = \myid$. Said Pareto optimal point has the expression
%Let us write out $\Bomega(\Bp'',\myid)$ [see \eqref{two_argument_notation} with $\BU_{\BP''} = \myid$]:
%\begin{align}   \label{omega_aligned}
%	\Bomega(\Bp'',\myid)
%	&= \Bomega\bigl( \BU_\BR\diag(\Br^{-1} \odot \Bp'')\BU_\BR^\He \bigr).
%\end{align}
Let us write out $\Bomega(\Bp'',\myid)$ by means of \eqref{omega_with_Xi} as
\begin{align}
	\BPi \Bomega(\Bp'',\myid)
	&= \Bp'' \odot \bigl( (\mu_\mathcal{Q}\Br)^{-1} \Bp'' + \Bxi \bigr)^{-1},
\end{align}
where $\BPi \in \mathbb{P}^{N_\Tx}$ is a sorting permutation, `$\odot$' denotes componentwise multiplication, $\Br^{-1}$ denotes the vector of entries $r_i^{-1}$ (i.e., componentwise reciprocal), and $\diag(\Bxi) = \BXi = \myid + (\mu_\mathcal{Q}\BLambda_\BR)^{-1}$. The mapping $\Bp'' \mapsto \Bp'' \odot \bigl( (\mu_\mathcal{Q}\Br)^{-1} \Bp'' + \Bxi \bigr)^{-1}$ is clearly injective, since $\Bxi > \mynull$. Additionally, it has the property that for any real unit-norm vector $\Be \geq \mynull$, there exists a scalar $\epsilon > 0$ and a single feasible vector $\Bp'' \in \partial^+\Bp''(\mathcal{P})$ such that
\begin{align}   \label{omega_balancing}
	\Bp'' \odot \bigl( (\mu_\mathcal{Q}\Br)^{-1} \Bp'' + \Bxi \bigr)^{-1}
	&= \epsilon \Be.
\end{align}
To see this, we first rewrite Expression~\eqref{omega_balancing} as
\begin{align}   \label{epsilon_mapsto_p_pp}
	\Bp'' = \epsilon\Be \odot \Bxi \odot \bigl( \myone - \epsilon\Be \odot (\mu_\mathcal{Q}\Br)^{-1} \bigr)^{-1}.
\end{align}
Since $\Bp'' \geq \mynull$, the scalar $\epsilon$ must lie in the semi-open interval $\epsilon \in [0;\min_i \mu_\mathcal{Q} r_i / e_i[$. From taking the Euclidian norm of Expression~\eqref{epsilon_mapsto_p_pp}, we obtain a function $\epsilon \mapsto \norm{\Bp''}_2$ which bijectively maps $[0;\min_i \mu_\mathcal{Q} r_i / e_i[$ onto $\mathbb{R}_+$. Since any $\Bp'' \in \partial^+\Bp''(\mathcal{P})$ has finite norm, there must necessarily exist one single value of $\epsilon$ fulfilling
\begin{align}
	\epsilon\Be \odot \Bxi \odot \bigl( \myone - \epsilon\Be \odot (\mu_\mathcal{Q}\Br)^{-1} \bigr)^{-1} \in \partial^+\Bp''(\mathcal{P}).
\end{align}
Consequently, all Pareto optimal points $\partial^+\Bomega(\mathcal{P})$ can be reached by the construction method from paragraphs 2) and 3), so that we may write
\begin{align}
	\partial^+\Bomega(\mathcal{P})
	= \Bomega\bigl(\partial^+\Bp''(\mathcal{P}),\myid\bigr).
\end{align}

\indent {\itshape 4)}
Now that we have established that the Pareto border $\partial^+\Bomega(\mathcal{P})$ can be reached by setting $\col(\BU_\BP) \subseteq \col(\BU_\BR)$, we have that $\tilde{\BR} = (\BR^{-1} + \BP)^{-1}$ and $\hat{\BR} = \BR - \tilde{\BR}$ acquire the same eigenbasis, up to a column permutation. Specifically, we have that the alignment $\col(\BU_\BP) \subseteq \col(\BU_\BR)$ implies $\col(\BU_\BP) = \col(\BU_{\hat{\BR}}) \subseteq \col(\BU_{\tilde{\BR}})$. But as a consequence of Theorem~\ref{thm:prescribed_pilot_region}, the alignment $\col(\BU_{\hat{\BR}}) \subseteq \col(\BU_{\tilde{\BR}})$ leads to [cf.~\eqref{alignment_of_Q}]
\begin{align}
	\col(\BU_\BQ) \subseteq \col(\BU_{\hat{\BR}}).
\end{align}
Hence, we obtain $\col(\BU_\BQ) \subseteq \col(\BU_\BP) \subseteq \col(\BU_\BR)$. Since we know from Section~\ref{sec:Number of pilot symbols and number of streams} that $\rank(\BP^\star) = \rank(\BQ^\star)$ at any joint optimum $(\BP^\star,\BQ^\star)$, we get the desired alignment property
\begin{align}   \label{inclusion_property}
	\col(\BU_\BP) = \col(\BU_\BQ) \subseteq \col(\BU_\BR).
\end{align}
Obviously, in case \eqref{inclusion_property} is a strict inclusion, the eigenbases of $\BP$ and $\BQ$ should contain the eigenvectors of $\BR$ associated to the largest eigenvalues of $\BR$, hence
\begin{align}
	\col(\BU_\BP) = \col(\BU_\BQ) = \{\Bu_{\BR,1},\dotsc,\Bu_{\BR,r^\star}\} \subseteq \col(\BU_\BR),
\end{align}
which concludes the proof of Theorem~\ref{thm:joint_basis}.

\subsection{Proof of Lemma~\ref{lem:front_border}}   \label{app:front_border}

For any set $\mathcal{A} \subseteq \mathbb{R}_+^n$, the Pareto border $\partial^+\mathcal{A}$ is a subset of the front border $\partial^\mathrm{f}\mathcal{A}$. In fact, if it were not so, then there would exist a Pareto optimal point, say $\Ba' \in \partial^+\mathcal{A}$, which would not be the solution to
\begin{align}
	\max_{\begin{subarray}{c} \Ba \in \mathcal{A} \\ \Ba = \nu \Ba' \end{subarray}} \nu
\end{align}
in that another $\Ba'' \in \mathcal{A}$ colinear with $\Ba'$ would exist that would have larger norm, i.e., $\norm{\Ba''} > \norm{\Ba'}$. Yet this is impossible by the definition of $\partial^+\mathcal{A}$, because $\Ba''$ would dominate $\Ba'$ in the sense $\Ba'' \geq \Ba'$, hence the contradiction.

It thus suffices to prove that $\partial^\mathrm{f}\Bs(\Gamma) \subseteq \partial^+\Bs(\Gamma)$ in order to conclude on set equality $\partial^\mathrm{f}\Bs(\Gamma) = \partial^+\Bs(\Gamma)$. For this purpose, take $\Bs'$ to be some point of the front border $\partial^\mathrm{f}\Bs(\Gamma)$. Assume that there would exist another point $\Bs'' \in \Bs(\Gamma)$ different from $\Bs'$ that dominates $\Bs'$, that is, $\Bs'' \geq \Bs'$. For belonging to the set $\Bs(\Gamma)$, which is the union
\begin{align}
	\Bs(\Gamma)
	= \bigcup_{\begin{subarray}{c} (\mu_\mathcal{P},\mu_\mathcal{Q}) \\ \mu_\mathcal{P} + (T-T_\tau)\mu_\mathcal{Q} \leq T\mu \end{subarray}} \bigcup_{\BP \in \mathcal{P}} \Bs(\BP,\mathcal{Q}),
\end{align}
the point $\Bs''$ would be contained in at least one of the sets $\Bs(\BP,\mathcal{Q})$. Call $\BP'' \in \mathcal{P}$ a pilot Gram of rank $r_{\BP''}$ such that $\Bs''$ lies in $\Bs(\BP'',\mathcal{Q})$. According to Theorem~\ref{thm:prescribed_pilot_region}, the set $\Bs(\BP'',\mathcal{Q})$ is a simplex consisting of all convex combinations of $r_{\BP''}+1$ points $\Bsigma^{(n)}, n=0,\dotsc,r_{\BP''}$, with $\Bsigma^{(0)} = \mynull$ and [cf.~\eqref{simplex_vertices}]
\begin{align}
	\Bsigma^{(n)}
	= \mathcal{H}(\omega_1,\dotsc,\omega_n) \sum_{j=1}^n \Be_j,
	\quad n \in \{1,\dotsc,r_{\BP''}\},
\end{align}
where $\omega_i$ are the non-increasingly ordered eigenvalues of the generalized eigenvalue problem [cf.~\eqref{generalized_evp}]
\begin{align}
	\hat{\BR}''\Bv_i
	= \omega_i (\mu_\mathcal{Q}^{-1}\myid + \tilde{\BR}'')\Bv_i
\end{align}
with $\tilde{\BR}'' = (\BR^{-1}+\BP'')^{-1}$ and $\hat{\BR}'' = \BR-\tilde{\BR}''$.
Notice that the linearly independent vectors $\Bsigma^{(n)}, n=1,\dotsc,r_{\BP''}$, when linearly combined with non-negative coefficients, span the linear subspace of $\mathbb{R}_+^{N_\Tx}$ of vectors having non-increasingly sorted entries on positions $1$ through $r_{\BP''}$, and zero entries on positions $r_{\BP''}+1$ through $N_\Tx$.
%That is because
%\begin{align}
%	\mathcal{H}(\omega_1)
%	> \mathcal{H}(\omega_1,\omega_2)
%	> \dotso
%	> \mathcal{H}(\omega_1,\dotsc,\omega_{r_{\BP''}}).
%\end{align}
Consequently, both $\Bs'$ and $\Bs''$, which by definition have non-increasing non-negative entries, can be written as linear combinations
\begin{align}
	\Bs'
	&= \sum_{n=1}^{r_{\BP''}} \nu_n' \Bsigma^{(n)}
	&\Bs''
	&= \sum_{n=1}^{r_{\BP''}} \nu_n'' \Bsigma^{(n)}
\end{align}
with unique non-negative coefficients $\nu_n'$ and $\nu_n''$. Since $\Bs'' \in \Bs(\BP'',\mathcal{Q})$, the coefficients $\nu_n''$ sum up to $\sum_n \nu_n'' \leq 1$. Now, since $\Bs'$ and $\Bs''$ are distinct, and $\Bs' \leq \Bs''$ by assumption, we must have
\begin{align}
	\sum_{n=1}^{r_{\BP''}} \nu_n' < \sum_{n=1}^{r_{\BP''}} \nu_n'' \leq 1.
\end{align}
Therefore $\Bs'$ lies in the interior of $\Bs(\BP'',\mathcal{Q})$. Consequently, for a small enough $\epsilon > 0$, the point $(1+\epsilon)\Bs'$ is element of $\Bs(\BP'',\mathcal{Q})$, and thus of $\Bs(\Gamma)$, which contradicts the initial assumption that $\Bs' \in \partial^\mathrm{f}\Bs(\Gamma)$. Hence $\partial^\mathrm{f}\Bs(\Gamma) = \partial^+\Bs(\Gamma)$.

\subsection{Proof of Lemma~\ref{lem:nu_is_quasi_concave}}   \label{app:proof:lem:nu_is_quasi_concave}

Clearly, maximizing $\nu(\Bp,\Be)$ as defined in \eqref{nu(p)} is equivalent to minimizing the function
\begin{align}
	\breve{\nu}(\Bp)
	= \frac{1}{\nu(\Bp,\Be)} + 1
	= \frac{1 + \Br^\Tr\Bq(\Bp,\Be)}{\eta(\Bp,\Be)},
\end{align}
where contrary to $\nu(\Bp,\Be)$, the direction vector $\Be$ is omitted in the notation of the function $\breve{\nu}(\Bp)$.
Writing the latter function out in full with help of definitions \eqref{eta(p)} and \eqref{q(p)} yields
\begin{align}
	\breve{\nu}(\Bp)
	&= \frac{1 + \frac{T\mu - \sum_i p_i}{T-T_\tau} \left( \sum_i e_i \frac{1 + r_i p_i}{r_i^2 p_i} \right)^{-1} \left( \sum_i e_i \frac{1 + r_i p_i}{r_i p_i} \right)}{\frac{T\mu - \sum_i p_i}{T-T_\tau} \left( \sum_i e_i \frac{1 + r_i p_i}{r_i^2 p_i} \right)^{-1}} \nonumber\\
	&= \frac{T-T_\tau}{T\mu - \sum_i p_i} \left( \sum_i e_i \frac{1 + r_i p_i}{r_i^2 p_i} \right) + \sum_i e_i \frac{1 + r_i p_i}{r_i p_i} \nonumber\\
	&\triangleq (T-T_\tau) \bigl( \breve{\nu}_1(\Bp) + \breve{\nu}_2(\Bp) \bigr) + \breve{\nu}_3(\Bp),
\end{align}
the three functions $\breve{\nu}_1(\Bp)$, $\breve{\nu}_2(\Bp)$, $\breve{\nu}_3(\Bp)$ in the last line being
\begin{subequations}
\begin{align}
	\breve{\nu}_1(\Bp)
	&= \sum_j e_j r_j^{-2} \frac{1}{p_j(T\mu - \sum_i p_i)} \\
	\breve{\nu}_2(\Bp)
	&= \sum_j e_j r_j^{-1} \frac{1}{T\mu - \sum_i p_i} \\
	\breve{\nu}_3(\Bp)
	&= \sum_i e_i \frac{1 + r_i p_i}{r_i p_i}.
\end{align}
\end{subequations}
We will now show that the three functions $\breve{\nu}_1(\Bp)$, $\breve{\nu}_2(\Bp)$ and $\breve{\nu}_3(\Bp)$ are all convex functions of $\Bp$ on the interior of $\mathcal{D}(T\mu) \subset (0;\infty)^{N_\Tx}$, which we shall denote as $\interior(\mathcal{D}(T\mu))$. It is easy to see that $\breve{\nu}_3$ is essentially a linear combination (plus a constant) of functions $1/p_i$ that are convex on the entire open orthant $(0,\infty)^{N_\Tx}$, and thus on $\interior(\mathcal{D}(T\mu)) \subset (0;\infty)^{N_\Tx}$. Similarly, $\breve{\nu}_2$ is convex on the open half-space $\sum_i p_i < T\mu$, and thus on the subset $\interior(\mathcal{D}(T\mu))$ thereof. Finally, $\breve{\nu}_1$ is a linear combination of functions $\frac{1}{p_j}\frac{1}{T\mu-\sum_i p_i}$, each of which is convex in $\Bp$ on $\interior(\mathcal{D}(T\mu))$. This can be shown as follows: take a pair of points $\bigl(\Bp^{(1)},\Bp^{(2)}\bigr) \in \interior(\mathcal{D}(T\mu))^2$, then for any $\theta \in [0;1]$,
%\begin{align}   \label{convexity_in_theta_to_show}   %adjust
%	&\frac{1}{\theta p_j^{(1)} + (1-\theta) p_j^{(2)}} \frac{1}{T\mu-\sum_i \bigl( \theta p_i^{(1)} + (1-\theta) p_i^{(2)} \bigr)} \nonumber\\
%	&\leq \theta \frac{1}{p_j^{(1)}}\frac{1}{T\mu-\sum_i p_i^{(1)}} + (1-\theta) \frac{1}{p_j^{(2)}}\frac{1}{T\mu-\sum_i p_i^{(2)}},
%\end{align}
\begin{align}   \label{convexity_in_theta_to_show}
	\frac{1}{\theta p_j^{(1)} + (1-\theta) p_j^{(2)}} \frac{1}{T\mu-\sum_i \bigl( \theta p_i^{(1)} + (1-\theta) p_i^{(2)} \bigr)}
	\leq \theta \frac{1}{p_j^{(1)}}\frac{1}{T\mu-\sum_i p_i^{(1)}} + (1-\theta) \frac{1}{p_j^{(2)}}\frac{1}{T\mu-\sum_i p_i^{(2)}},
\end{align}
because the left-hand side of the latter inequality is convex in $\theta \in [0;1]$, since it is of the form
\begin{align}   \label{product_of_reciprocals}
	A \frac{1}{1+B\theta} \frac{1}{1+C\theta}
\end{align}
with constants $A = \frac{1}{p_j^{(2)}(T\mu-\sum_i p_i^{(2)})} \geq 0$, $B = \frac{p_j^{(1)}-p_j^{(2)}}{p_j^{(2)}}$, $C = \frac{\sum_i (p_i^{(2)}-p_i^{(1)})}{T\mu-\sum_i p_i^{(2)}}$, and $1+B\theta \geq 0$ and $1+C\theta \geq 0$ by construction. The convexity of \eqref{product_of_reciprocals} is best seen by differentiating twice:
%\begin{align*}   %adjust
%	&\frac{\diffd^2}{\diffd \theta^2} \left[ \frac{1}{1+B\theta} \frac{1}{1+C\theta} \right]
%	= \frac{2}{(1+B\theta)(1+C\theta)} \times \nonumber\\
%	&\quad {} \times \left[ \frac{B^2}{(1+B\theta)^2} + \frac{C^2}{(1+C\theta)^2} + \frac{BC}{(1+B\theta)(1+C\theta)} \right].
%\end{align*}
\begin{align}
	\frac{\diffd^2}{\diffd \theta^2} \left[ \frac{1}{1+B\theta} \frac{1}{1+C\theta} \right]
	= \frac{2}{(1+B\theta)(1+C\theta)}
	\left( \frac{B^2}{(1+B\theta)^2} + \frac{C^2}{(1+C\theta)^2} + \frac{BC}{(1+B\theta)(1+C\theta)} \right).
\end{align}
The above expression is obviously positive if $BC \geq 0$. Otherwise, if $BC \leq 0$, then the expression between square brackets on the right-hand side of the last equality is lower bounded by
%\begin{align}   %adjust
%	&\frac{B^2}{(1+B\theta)^2} + \frac{C^2}{(1+C\theta)^2} + \frac{2BC}{(1+B\theta)(1+C\theta)} \nonumber\\
%	&\qquad\qquad\qquad\qquad {} = \left[ \frac{B}{(1+B\theta)} + \frac{C}{(1+C\theta)} \right]^2 \geq 0.
%\end{align}
\begin{align}
	\frac{B^2}{(1+B\theta)^2} + \frac{C^2}{(1+C\theta)^2} + \frac{2BC}{(1+B\theta)(1+C\theta)}
	= \left[ \frac{B}{(1+B\theta)} + \frac{C}{(1+C\theta)} \right]^2 \geq 0.
\end{align}
Hence \eqref{convexity_in_theta_to_show}, and all the three functions $\breve{\nu}_1$, $\breve{\nu}_2$ and $\breve{\nu}_3$ are convex in $\Bp$ on the open set $\interior(\mathcal{D}(T\mu))$. Thus, $\breve{\nu}(\Bp)$ is convex on $\interior(\mathcal{D}(T\mu))$. Therefore $\nu(\Bp,\Be) = 1/(\breve{\nu}(\Bp)-1)$, which is a decreasing function of $\breve{\nu}(\Bp) > 1$, is quasi-concave in $\Bp$ on $\interior(\mathcal{D}(T\mu))$, according to Definition~\ref{def:quasi-concavity}. Since $\nu(\Bp,\Be)$ vanishes on the boundary of $\mathcal{D}(T\mu)$ and is continuous in the vicinity of this boundary, we conclude that $\Bp \mapsto \nu(\Bp,\Be)$ is quasi-concave on the closure $\mathcal{D}(T\mu)$.

\subsection{Derivation of \eqref{pilot_allocation_without_sharing}}   \label{app:proof:pilot_allocation_without_sharing}

Rather than maximizing $\bar{\nu}(\Bp,\Be)$, we minimize its reciprocal
\begin{align}   \label{nu_reciprocal}
	\frac{1}{\bar{\nu}(\Bp,\Be)}
	&= \frac{1 + \Br^\Tr\bar{\Bq}(\Bp,\Be)}{\bar{\eta}(\Bp,\Be)} - 1 \nonumber\\
	&= \mu_\mathcal{Q}^{-1} \left( \sum_i e_i \frac{1 + r_i p_i}{r_i^2 p_i} \right) + \sum_i e_i \frac{1 + r_i p_i}{r_i p_i} - 1 \nonumber\\
	&= \sum_i \frac{e_i\bigl(\mu_\mathcal{Q}^{-1} + r_i\bigr)}{r_i^2} \cdot \frac{1}{p_i} + \mu_\mathcal{Q}^{-1} \sum_i \frac{e_i}{r_i}.
\end{align}
In the last equality we have made use of the normalization $\sum_i e_i = 1$.
The above expression is a non-negatively weighted sum of reciprocals of $p_i$ (plus a positive constant), and thus a convex function of $\Bp$. Therefore, we are dealing with a convex problem, for which a Lagrange approach yields necessary and sufficient optimality conditions.
To minimize this convex function under the (convex) sum constraint $\sum_i p_i \leq \mu_\mathcal{P}$, we define the Lagrangian
\begin{align}
	L(\Bp,\lambda) = \frac{1}{\bar{\nu}(\Bp,\Be)} + \lambda(\myone^\Tr\Bp - \mu_\mathcal{P})
\end{align}
and four associated Karush-Kuhn-Tucker conditions
\begin{align*}
	\left. \frac{\partial L}{\partial \Bp} \right|_{\Bp = \bar{\Bp}^\star} &= \mynull
	&\lambda(\myone^\Tr\bar{\Bp}^\star - \mu_\mathcal{P}) &= 0 \\
	\lambda &\geq 0
	&\myone^\Tr\bar{\Bp}^\star - \mu_\mathcal{P} &\leq 0.
\end{align*}
With \eqref{nu_reciprocal}, the stationarity condition $\left. \partial L/\partial \Bp \right|_{\Bp = \Bp^\star} = \mynull$ reads as
\begin{align}
	-\frac{e_i\bigl(\mu_\mathcal{Q}^{-1} + r_i\bigr)}{r_i^2} \cdot \frac{1}{(\bar{p}_i^\star)^2} + \lambda = 0,
\end{align}
whence
\begin{align}   \label{optimal_p_i}
	\bar{p}_i^\star
	= \frac{\sqrt{e_i\bigl(\mu_\mathcal{Q}^{-1} + r_i\bigr)}}{r_i \sqrt{\lambda}}.
\end{align}
Since we must have $\lambda > 0$, the complementary slackness condition $\lambda(\myone^\Tr\Bp - \mu_\mathcal{P}) = 0$ requires that the inequality constraint be fulfilled with equality, i.e., $\myone^\Tr\Bp = \mu_\mathcal{P}$, hence $\sqrt{\lambda} = \sum_i \frac{1}{\mu_\mathcal{P}r_i} \sqrt{e_i\bigl(\mu_\mathcal{Q}^{-1} + r_i\bigr)}$, so the solution \eqref{optimal_p_i} reads as
\begin{align}
	\bar{p}_i^\star
	= \mu_\mathcal{P} \frac{\frac{1}{r_i}\sqrt{e_i\bigl(\mu_\mathcal{Q}^{-1} + r_i\bigr)}}{\sum_j \frac{1}{r_j} \sqrt{e_j\bigl(\mu_\mathcal{Q}^{-1} + r_j\bigr)}}
	= \mu_\mathcal{P} \frac{\frac{1}{r_i}\sqrt{e_i(1+\mu_\mathcal{Q}r_i)}}{\sum_j \frac{1}{r_j} \sqrt{e_j(1+\mu_\mathcal{Q}r_i)}}.
\end{align}

\bibliographystyle{IEEEtran}
\bibliography{custom/IEEEabrv,../Shared/references,../Shared/references_own}

% Generated by IEEEtran.bst, version: 1.12 (2007/01/11)
\begin{thebibliography}{10}
\providecommand{\url}[1]{#1}
\csname url@samestyle\endcsname
\providecommand{\newblock}{\relax}
\providecommand{\bibinfo}[2]{#2}
\providecommand{\BIBentrySTDinterwordspacing}{\spaceskip=0pt\relax}
\providecommand{\BIBentryALTinterwordstretchfactor}{4}
\providecommand{\BIBentryALTinterwordspacing}{\spaceskip=\fontdimen2\font plus
\BIBentryALTinterwordstretchfactor\fontdimen3\font minus
  \fontdimen4\font\relax}
\providecommand{\BIBforeignlanguage}[2]{{%
\expandafter\ifx\csname l@#1\endcsname\relax
\typeout{** WARNING: IEEEtran.bst: No hyphenation pattern has been}%
\typeout{** loaded for the language `#1'. Using the pattern for}%
\typeout{** the default language instead.}%
\else
\language=\csname l@#1\endcsname
\fi
#2}}
\providecommand{\BIBdecl}{\relax}
\BIBdecl

\bibitem{HaHo03}
B.~Hassibi and B.~Hochwald, ``How much training is needed in multiple-antenna
  wireless links?'' \emph{IEEE Transactions on Information Theory}, vol.~49,
  no.~4, pp. 951--963, Apr. 2003.

\bibitem{PaJoFo11_isit}
A.~Pastore, M.~Joham, and J.~Fonollosa, ``Joint pilot and precoder design for
  optimal throughput,'' in \emph{Proc. IEEE International Symposium on
  Information Theory}, Aug. 2011, pp. 371--375.

\bibitem{ToSaDo04}
L.~Tong, B.~M. Sadler, and M.~Dong, ``Pilot-assisted wireless transmissions:
  general model, design criteria, and signal processing,'' \emph{IEEE Signal
  Processing Magazine}, vol.~21, no.~6, pp. 12--25, Nov. 2004.

\bibitem{BiGe06}
M.~Biguesh and A.~Gershman, ``Training-based {MIMO} channel estimation: a study
  of estimator tradeoffs and optimal training signals,'' \emph{IEEE
  Transactions on Signal Processing}, vol.~54, no.~3, pp. 884--893, Mar. 2006.

\bibitem{LaSh02}
A.~Lapidoth and S.~Shamai, ``Fading channels: how perfect need `perfect side
  information' be?'' \emph{IEEE Transactions on Information Theory}, vol.~48,
  no.~5, pp. 1118--1134, May 2002.

\bibitem{WeStSh04}
H.~Weingarten, Y.~Steinberg, and S.~Shamai, ``Gaussian codes and weighted
  nearest neighbor decoding in fading multiple-antenna channels,'' \emph{IEEE
  Transactions on Information Theory}, vol.~50, no.~8, pp. 1665--1686, Aug.
  2004.

\bibitem{Me00}
M.~M\'edard, ``The effect upon channel capacity in wireless communications of
  perfect and imperfect knowledge of the channel,'' \emph{IEEE Transactions on
  Information Theory}, vol.~46, no.~3, pp. 933--946, May 2000.

\bibitem{BaFoMe01}
J.~Baltersee, G.~Fock, and H.~Meyr, ``Achievable rate of {MIMO} channels with
  data-aided channel estimation and perfect interleaving,'' \emph{IEEE Journal
  on Selected Areas in Communications}, vol.~19, no.~12, pp. 2358--2368, Dec.
  2001.

\bibitem{YoYoGo04}
T.~Yoo, E.~Yoon, and A.~Goldsmith, ``{MIMO} capacity with channel uncertainty:
  Does feedback help?'' in \emph{Proc. IEEE Global Telecommunications
  Conference}, vol.~1, Dec. 2004, pp. 96--100.

\bibitem{MuDoNaAg05}
L.~Musavian, M.~Dohler, M.~Nakhai, and A.~Aghvami, ``Transmitter design in
  partially coherent antenna systems,'' in \emph{Proc. IEEE International
  Conference on Communications}, vol.~4, May 2005, pp. 2261--2265.

\bibitem{Lo08}
A.~Lozano, ``Interplay of spectral efficiency, power and {D}oppler spectrum for
  reference-signal-assisted wireless communication,'' \emph{IEEE Transactions
  on Wireless Communications}, vol.~7, no.~12, pp. 5020--5029, Dec. 2008.

\bibitem{SoUl10a}
A.~Soysal and S.~Ulukus, ``Joint channel estimation and resource allocation for
  {MIMO} systems--{P}art {I}: Single-user analysis,'' \emph{IEEE Transactions
  on Wireless Communications}, vol.~9, no.~2, pp. 624--631, Feb. 2010.

\bibitem{DiBl10}
M.~Ding and S.~Blostein, ``Maximum mutual information design for {MIMO} systems
  with imperfect channel knowledge,'' \emph{IEEE Transactions on Information
  Theory}, vol.~56, no.~10, pp. 4793--4801, Oct. 2010.

\bibitem{VoSc06}
A.~Vosoughi and A.~Scaglione, ``On the effect of receiver estimation error upon
  channel mutual information,'' \emph{IEEE Transactions on Signal Processing},
  vol.~54, no.~2, pp. 459--472, Feb. 2006.

\bibitem{YoGo06}
T.~Yoo and A.~Goldsmith, ``Capacity and power allocation for fading {MIMO}
  channels with channel estimation error,'' \emph{IEEE Transactions on
  Information Theory}, vol.~52, no.~5, pp. 2203--2214, May 2006.

\bibitem{PaJo09_asilomar}
A.~Pastore and M.~Joham, ``Mutual information bounds for {MIMO} channels under
  imperfect receiver {CSI},'' in \emph{Proc. 43rd Asilomar Conference on
  Signals, Systems and Computers}, Nov. 2009, pp. 1456--1460.

\bibitem{BoVa04}
S.~Boyd and L.~Vandenberghe, \emph{Convex Optimization}.\hskip 1em plus 0.5em
  minus 0.4em\relax New York, NY, USA: {Cambridge University Press}, Mar. 2004.

\bibitem{GaYoCh00}
E.~Gauthier, A.~Yonga\c{c}oglu, and J.~Chouinard, ``Capacity of multiple
  antenna systems in {R}ayleigh fading channels,'' in \emph{Canadian Conference
  on Electrical and Computer Engineering}, vol.~1, 2000, pp. 275--279.

\bibitem{Gr02}
A.~Grant, ``Rayleigh fading multi-antenna channels,'' \emph{EURASIP Journal of
  Applied Signal Processing}, vol. 2002, no.~1, pp. 316--329, Jan. 2002.

\bibitem{OyNaBoPa02}
O.~Oyman, R.~Nabar, H.~B\"{o}lcskei, and A.~Paulraj, ``Tight lower bounds on
  the ergodic capacity of {R}ayleigh fading {MIMO} channels,'' vol.~2, Nov.
  2002, pp. 1172--1176.

\bibitem{HoJo90}
R.~A. Horn and C.~R. Johnson, \emph{Matrix Analysis}.\hskip 1em plus 0.5em
  minus 0.4em\relax {Cambridge University Press}, Feb. 1990.

\end{thebibliography}

\end{document}